\documentclass[a4paper,11pt,final]{article}

\usepackage{graphicx,amsmath,amssymb,algorithm,algorithmic,textcomp,tikz-cd} % Add all your packages here
\usepackage{epsfig}

\begin{document}

% paper title: Must keep \ \\ \LARGE\bf in it to leave enough margin.
\title{\ \\ \LARGE\bf Properties of interaction networks, structure coefficients, and benefit--to--cost ratios}

\author{Hendrik Richter \\
HTWK Leipzig University of Applied Sciences \\ Faculty of
Electrical Engineering and Information Technology\\
        Postfach 301166, D--04251 Leipzig, Germany. \\ Email: 
hendrik.richter@htwk-leipzig.de. }

\maketitle

\begin{abstract}
%% Text of abstract
In structured populations the spatial arrangement of cooperators and defectors on the interaction graph together with the structure of the graph itself determines the game dynamics and particularly whether or not fixation of cooperation (or defection) is favored. For a single cooperator (and a single defector) and a network described by a regular graph the question of  fixation can be addressed by a single parameter, the structure coefficient.  As this quantity is generic for any regular graph, we may call it the generic structure coefficient. For two and more cooperators (or several defectors) fixation properties can also be assigned by structure coefficients. These structure coefficients, however, depend  on  
the arrangement of cooperators and defectors which we may interpret as a configuration of the game. Moreover, the coefficients are specific for a given interaction network modeled as regular graph, which is why we may call them specific  structure coefficients. In this paper, we study how specific  structure coefficients vary over interaction graphs and link the distributions obtained over different graphs
 to spectral properties of interaction networks. 
We also discuss implications for the benefit--to--cost ratios of donation games.

\end{abstract}

\section{Introduction}
Assessing probabilities for the fixation of a strategy, given that one or several players start to use this strategy, is a fundamental question in evolutionary game theory,~\cite{nowak06,broom13}. However,
calculating fixation probabilities  exactly  in structured populations is computationally intractable, see for example~\cite{hinder16,ibs15,vor13}. The computational costs increase exponentially  with the number of players (and coplayers) in evolutionary (and coevolutionary) games. Recently, \cite{chen16} devised a method for calculating conditions of weak selection favoring fixation that can be done in polynomial time for regular graphs. The method employs modeling the strategies used by players and coplayers as configurations. By such a configuration--based model, conditions that favor fixation have been studied for birth--death (BD) and death--birth (DB) updating. In another recent work, a method for linking configurations with graph--theoretical properties of interaction networks  has been proposed which uses the framework of dynamic game landscapes,~\cite{rich16,rich17,rich18}.~Game landscapes connect co--evolutionary games with the framework of dynamic fitness landscapes,~\cite{foster13,richengel14}.  The method also involves modeling interaction networks by regular graph.  This paper combines these two approaches. 

There is a considerable amount of work on dynamics of games with $N$ players on $d$--regular graphs. In particular, it was shown that for a game with two strategies (for instance cooperating and defecting) and for weak selection the conditions for one strategy to be favored over the other can be described by a single parameter, the structure coefficient $\sigma$,~\cite{taylor07,ohts06,tarnita09, allen14}. For instance, in a well-mixed population ($d=N-1$) and a wide class of updating rules, there is $\sigma=(N-2)/N$,~\cite{traulsen08,antal09}, while for games on cycles ($d=2$), we find $\sigma=(3N-8)/N$,~\cite{ohtsnow06} and for generally structured populations, low mutation, large $N$ and $N\gg d$, we have $\sigma=(d+1)/(d-1)$~\cite{ohts06}. Generalizing these results for population size $N$ and any $2 \leq d \leq N-1$, the structure coefficient $\sigma=((d+1)N-4d)/((d-1)N)$ has been obtained,~\cite{taylor07,lehmann07,tarnita09}. We may interpret all these structure coefficients as generic for a given $N$ and $d$ and say that  $\sigma=\sigma(N,d)$ are generic structure coefficients. They strictly apply only for a randomly--placed single cooperator (or defector). 

The generic description, however,  neither takes into account how for structured populations the game dynamics may be modified by several players starting to use a certain strategy,
nor considers how the structure of the interaction network itself influences the outcome. 
To include these variables into the framework of structure coefficients, \cite{chen16} recently suggested structure coefficients that can be defined for any configuration of cooperators and defectors \emph{and} for any network of interaction modeled by regular graphs.  We may interpret these structure coefficients as specific for a configuration $\pi$ and an interaction graph $A_I$ and say that   $\sigma(\pi)=\sigma(N,d,\pi,A_I)$ are specific  structure coefficients. 

In this paper, we analyze the specific structure coefficients over configurations and interaction networks. In particular, it is shown that while for configurations with a single cooperator (and a single defector) the results reproduce those for generic coefficients, for other configurations of cooperators and defectors, the specific structure coefficients vary over interaction networks. 
A frequently studied question is to what extend different networks of interaction contribute to differences in the fixation properties of games.
We approach this question be analyzing how network properties relate to specific structure coefficients. Therefore, 
the distributions obtained by these variations are studied and linked to spectral properties of the interaction networks, thus proposing a spectral analysis of evolutionary graphs.

The paper is organized as follows. In Sec.~\ref{sec:results} the main results and experimental findings are presented, where games with $N=\{8,10,12,14\}$ and $2 \leq d \leq N-1$ are considered. We focus on DB updating as BD updating appears to be less interesting in terms of game dynamics since it never favors cooperators and is always opposed to the emergence of cooperation. 
The results are discussed and possible implications are addressed in Sec.~\ref{sec:dis}.  Finally, in Sec.~\ref{sec:meth} methods are given and particularly coevolutionary games, configurations and structure coefficients as well as spectral graph measures of interaction networks are reviewed.

\section{Results} \label{sec:results}

Fixation properties of general $2 \times 2$ games for any configuration $\pi$ of cooperators and defectors on any regular graph can be determined by the structure coefficient $\sigma(\pi)$, see \ref{sec:meth}. Methods.  Recently, is was shown by \cite{chen16}  that for weak selection and DB as well as BD updating, the structure coefficient can be calculated with time complexity  $\mathcal{O}(d^2N)$, where $N$ is the number of players in the game (and also the number of vertices of the interaction graph $A_I$) and $d$ is the number of coplayers (and also the degree of the graph). In particular, it was shown that for DB updating we have  \begin{equation} 
\sigma(\pi)=\frac{N\left(1+1/d \right) \overline{\omega^1} \cdot \overline{\omega^0}-2\overline{\omega^{10}}-\overline{\omega^1 \omega^0} }{N\left(1-1/d \right) \overline{\omega^1} \cdot \overline{\omega^0}+\overline{\omega^1 \omega^0}},  \label{eq:sigma}
\end{equation} which depends (apart from $N$ and $d$) on four local frequencies:  $\overline{\omega^1}$, $\overline{\omega^0}$, $\overline{\omega^{10}}$ and $\overline{\omega^1 \omega^0}$. For these local frequencies, the following probabilistic interpretation may be given,~\cite{chen16}. Suppose on a given interaction graph $A_I$ a random walk is carried out with the starting 
vertex chosen uniformly--at--random. Then the local frequency $\overline{\omega^1}$ (or $\overline{\omega^0}=1-\overline{\omega^1}$) is the probability that for a configuration $\pi$ the player at the first step of the walk is a cooperator (or defector). The local frequency  $\overline{\omega^{10}}$ is the probability that the player at the first step is  a cooperator and at the second step it is a defector. Finally, if we carry out two random walks independent of each other, the local frequency   $\overline{\omega^1 \omega^0}$ is the probability that the player at the first step on the first walk is a cooperator, but a defector on the second walk. In \ref{sec:meth}. Methods the calculation of the local frequencies is given in detail for $N=4$.

\begin{figure}[tb]
\includegraphics[trim = 0mm 0mm 0mm 0mm,clip,width=8.25cm, height=5.9cm]{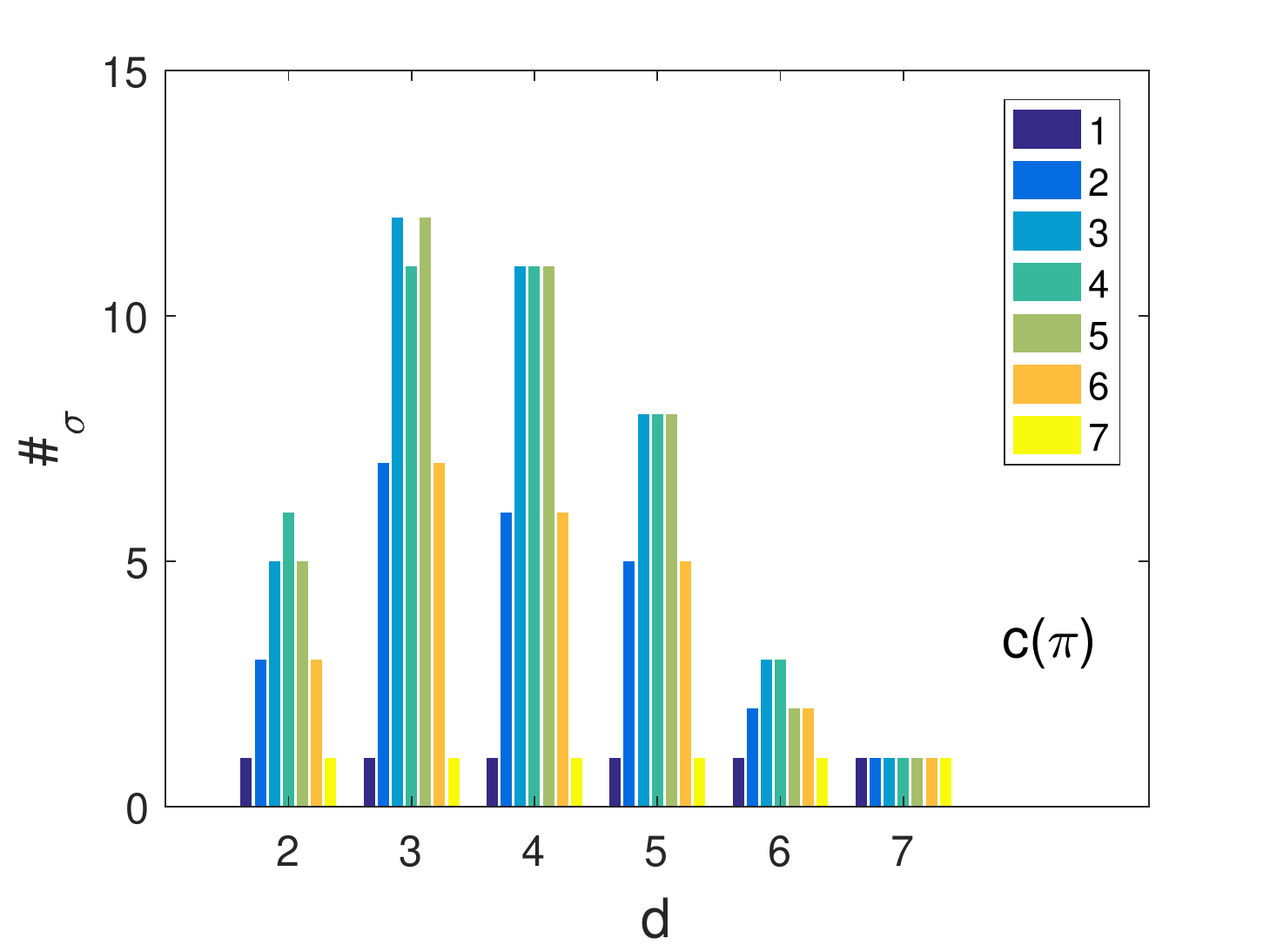}
\includegraphics[trim = 0mm 0mm 0mm 0mm,clip,width=8.25cm, height=5.9cm]{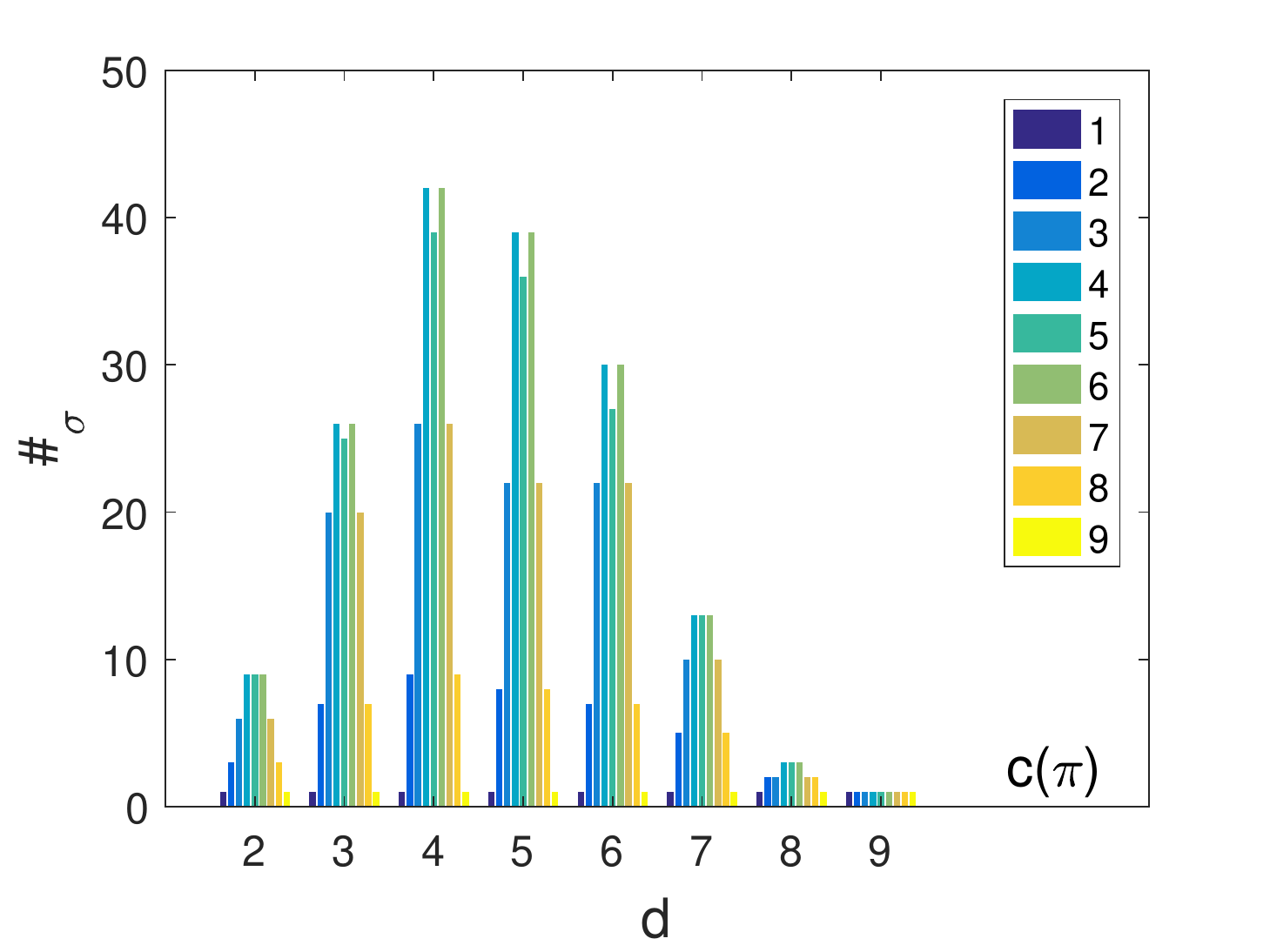}

\hspace{1cm}a) $N=8$  \hspace{7cm} b) $N=10$

\includegraphics[trim = 0mm 0mm 0mm 0mm,clip,width=8.25cm, height=5.9cm]{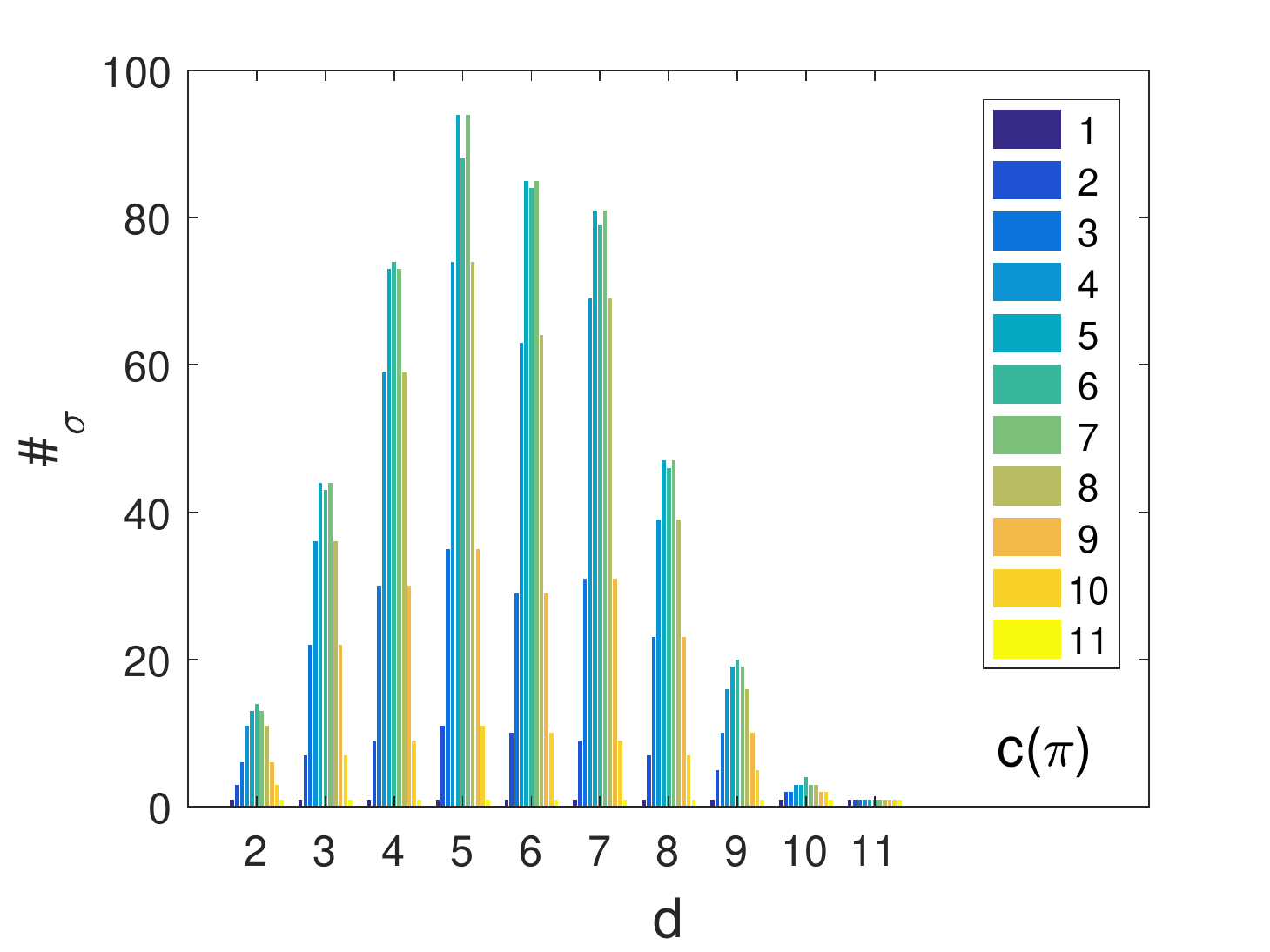}
\includegraphics[trim = 0mm 0mm 0mm 0mm,clip,width=8.25cm, height=5.9cm]{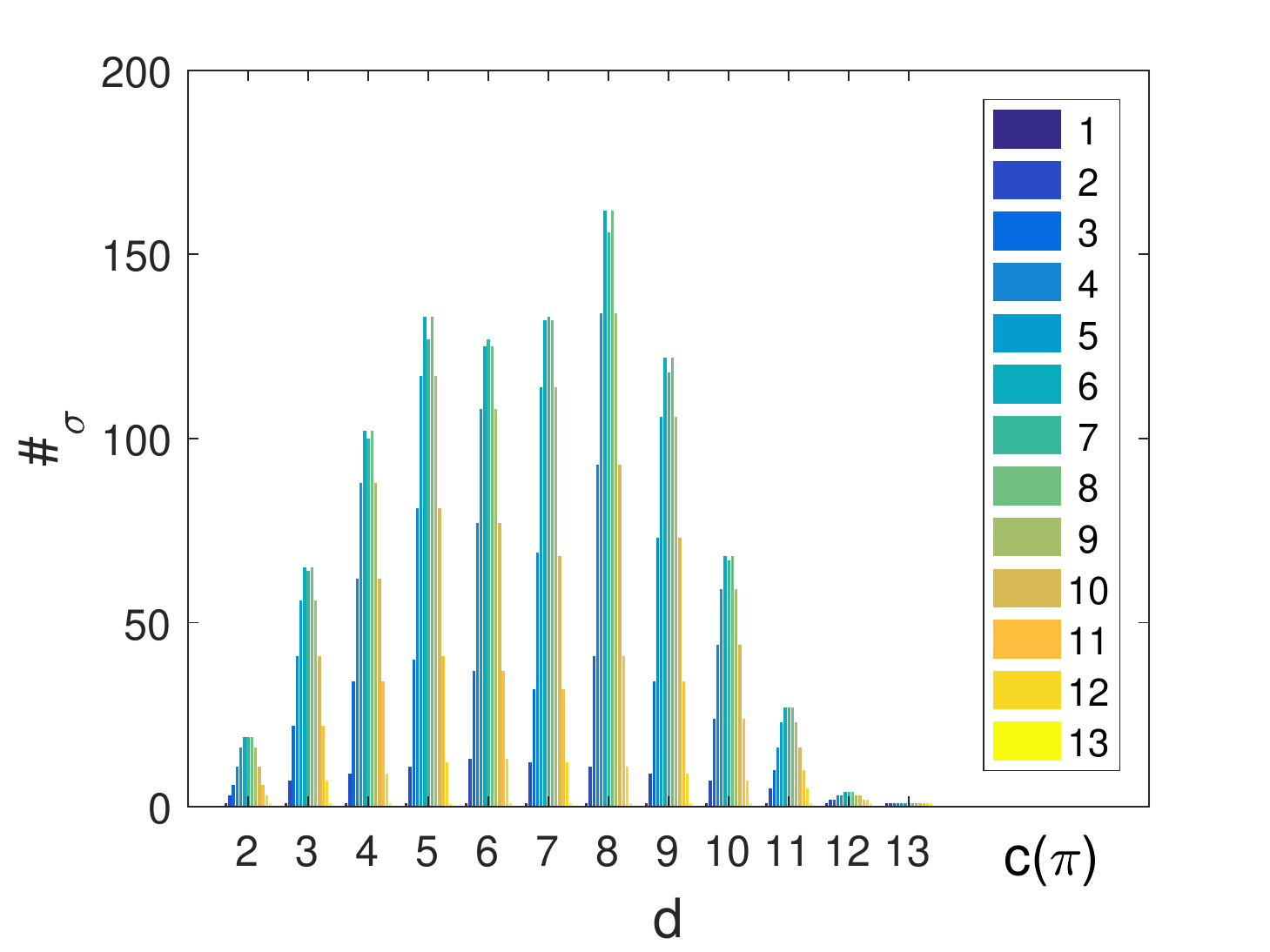}

\hspace{1cm}c) $N=12$  \hspace{7cm} d) $N=14$

\caption{The number $\#_\sigma$ of different values of the structure coefficient $\sigma(\pi)$, Eq. (\ref{eq:sigma}), for players $N$ and coplayers $2 \leq d \leq N-1$. The results are grouped with respect to the number of cooperators $c(\pi)$, $1 \leq c(\pi) \leq N-1$ according to the color code on the right--hand side of the figures.  We see that the number of different values of the structure coefficient $\sigma(\pi)$ increases with $N$, except for $d=N-1$ and that different values are most numerous for intermediate values of $d$.}
\label{fig:sigma_abs_8_14}
\end{figure}

\begin{figure}[tb]
\includegraphics[trim = 0mm 0mm 0mm 0mm,clip,width=8.25cm, height=5.9cm]{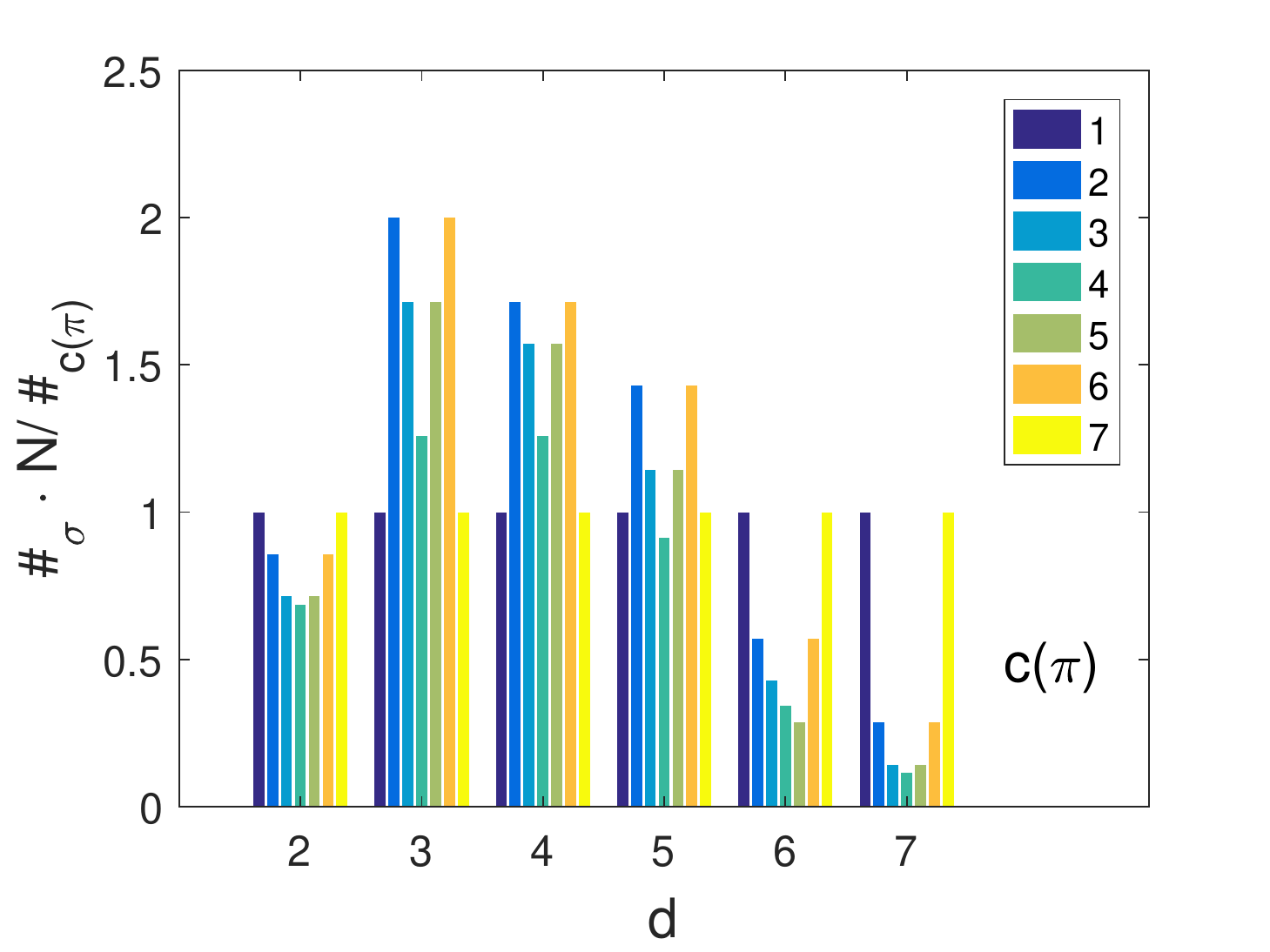}
\includegraphics[trim = 0mm 0mm 0mm 0mm,clip,width=8.25cm, height=5.9cm]{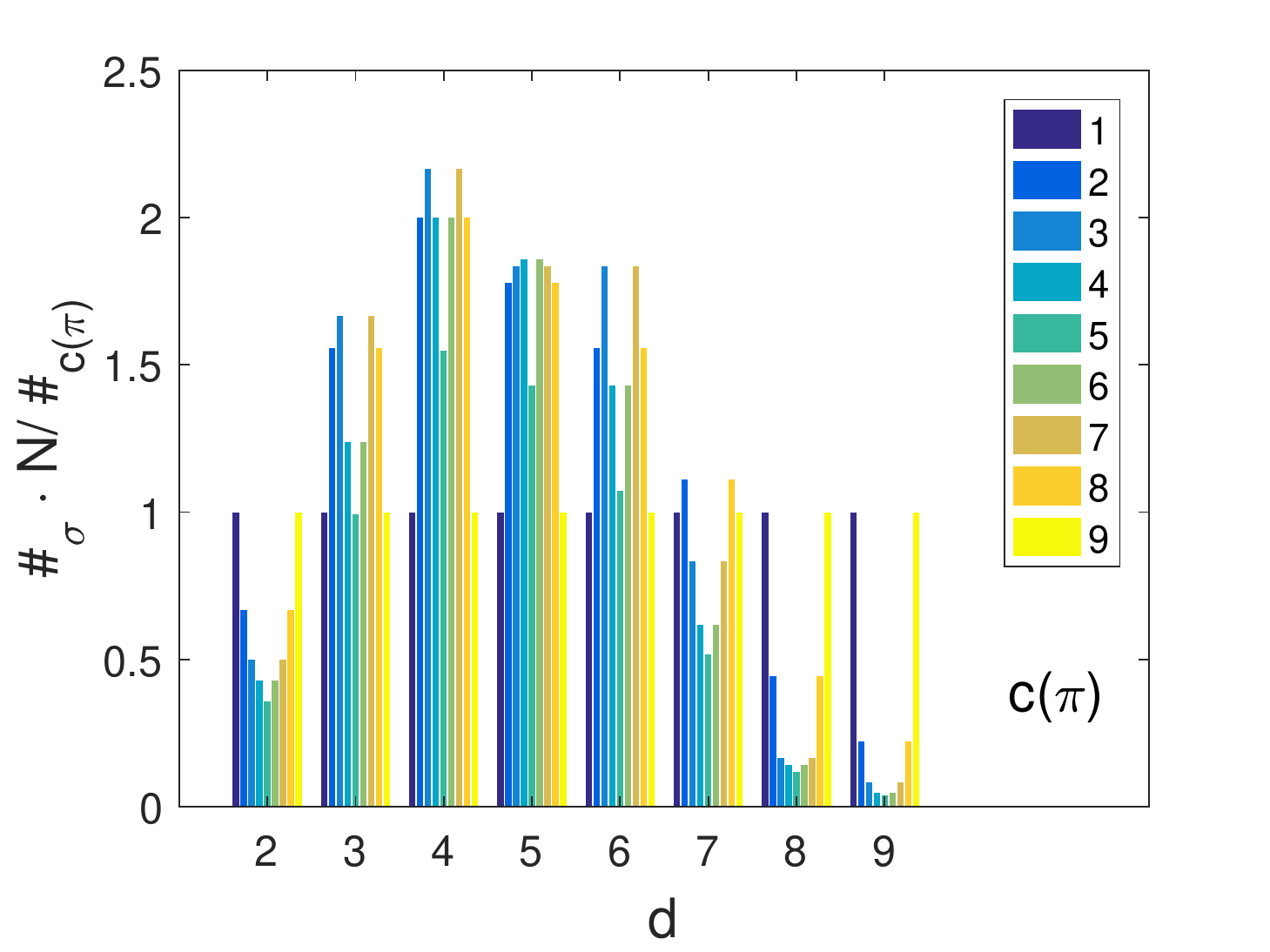}

\hspace{1cm}a) $N=8$  \hspace{7cm} b) $N=10$

\includegraphics[trim = 0mm 0mm 0mm 0mm,clip,width=8.25cm, height=5.9cm]{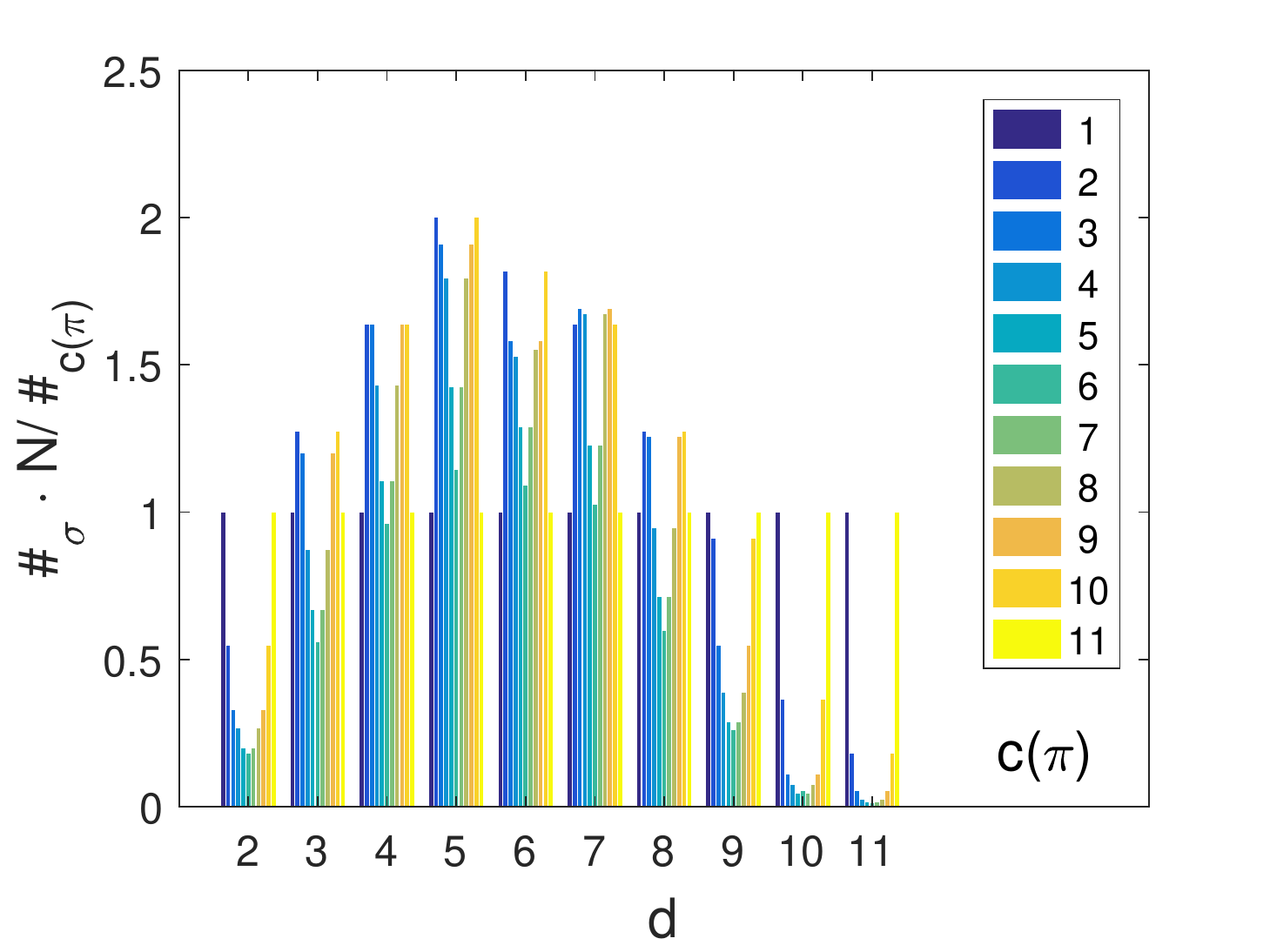}
\includegraphics[trim = 0mm 0mm 0mm 0mm,clip,width=8.25cm, height=5.9cm]{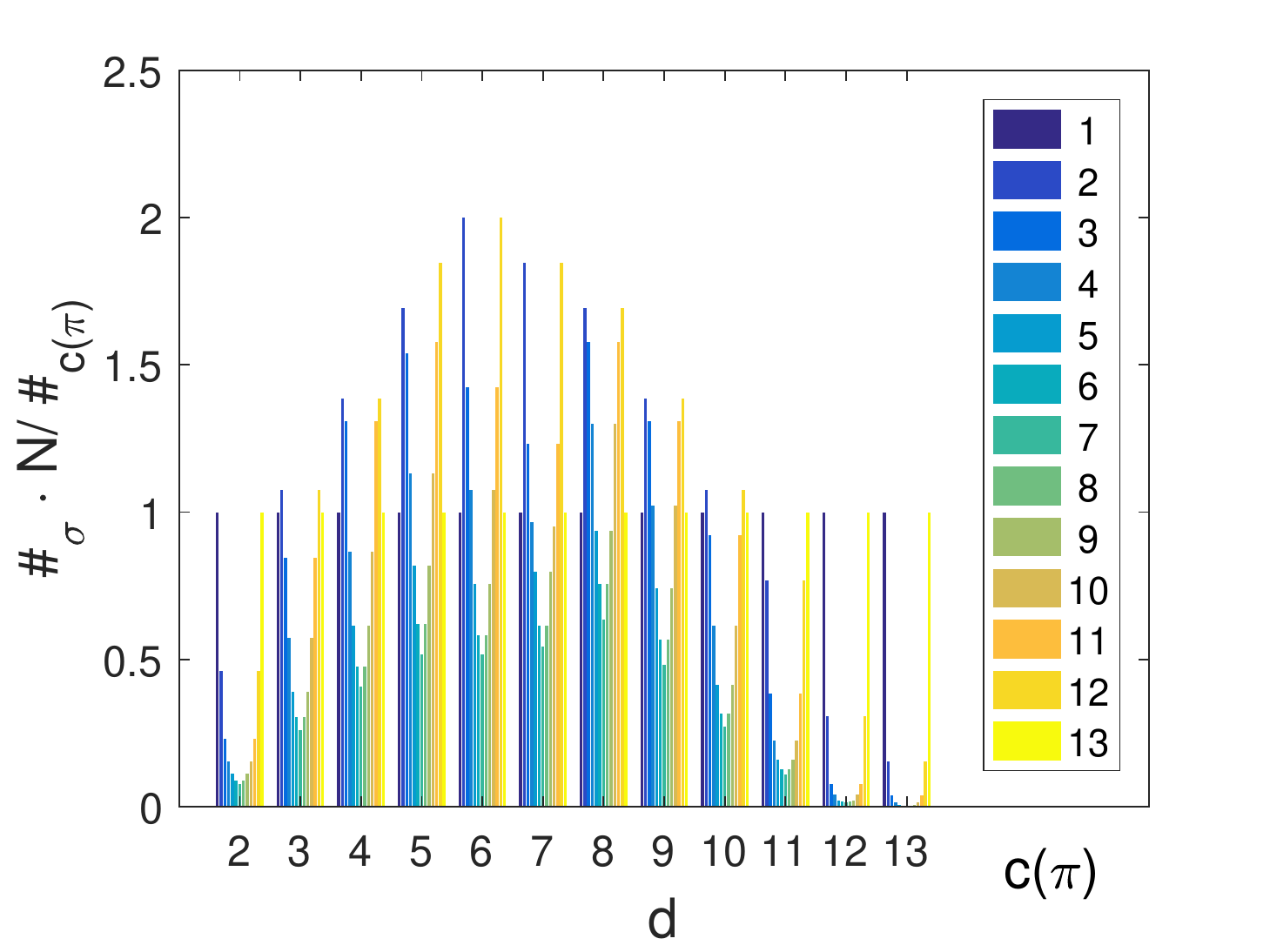}

\hspace{1cm}c) $N=12$  \hspace{7cm} d) $N=14$
\caption{The normalized  number $\frac{\#_\sigma}{\#_{c(\pi)}} N$ of different values of the structure coefficient $\sigma(\pi)$, Eq. (\ref{eq:sigma}), for players $N$ and coplayers $2 \leq d \leq N-1$. The results are grouped with respect to the number of cooperators $c(\pi)$, $1 \leq c(\pi) \leq N-1$ according to the color code on the right--hand side of the figures. }
\label{fig:sigma8_14}
\end{figure}

For a given $N$ and $d$, we may obtain varying  local frequencies in Eq. (\ref{eq:sigma})  for each network of interaction and possibly varying $\sigma(\pi)$, see also the discussion in  \ref{sec:meth}. Methods. For numerical experiments we analyze $N=\{8,10,12,14\}$, $2 \leq d \leq N-1$, and use a set of interaction networks described by $d$--regular graphs for which the adjacency matrices $A_I$ are generated algorithmically,~\cite{bay10,blitz11}.  The number $\mathcal{L}_d(N)$ of different interaction networks may be very large, even for a moderate number of players. For instance, for $N=12$ players with  $d=2$ coplayers each, we have $\mathcal{L}_2(12)=34.944.085$ different networks,~\cite{rich17}. Thus, as it is not feasible to numerically evaluate all networks, we bound  the magnitude of the set  of $A_I$ taken into account from above  by $G=10.000$. In case of $\mathcal{L}_d(N)<G$ for any $N$ and $d$ considered, we take the whole set.  To summarize, we analyze for each configuration $\pi$ a set $\sigma(\pi)= \left (\sigma_1(\pi),\sigma_2(\pi),\ldots,\sigma_I(\pi)  \right)$ with 
$I=\min{(G,\mathcal{L}_d(N))}$ for a set of adjacency matrices  $A_I(k)$ with $k=1,2,\ldots,I$. The variable $k$ can be interpreted as a temporal counter, similar to a discrete dynamical system. Furthermore, we understand the set $\sigma(\pi)$  to form a distribution over   adjacency matrices  $A_I(k)$.
Our central questions in the analysis are. How do the structure coefficients (\ref{eq:sigma}) vary over interaction networks? For what configurations $\pi$ and  what number of cooperators $c(\pi)$ does variation happen? Are there interaction networks that produces larger (or smaller) values of $\sigma(\pi)$ for a given configuration $\pi$ and the same $N$ and $d$? And if so, are these interaction networks also distinct with respect to spectral graph measures?

In a first series of experiments we analyze the distribution of $\sigma(\pi)$ over interaction networks. We count how many different values of $\sigma(\pi)$ there are for a given configuration $\pi$ and interaction graph $A_I$. This distribution is based on $\ell=2^N-2$ configurations for which Eq. (\ref{eq:sigma})  gives structure coefficients for each $N$, $d$ and $A_I$.  The values of $\sigma(000\ldots 0)$ and $\sigma(111\ldots 1)$ are not defined as they represent absorbing configurations. The distributions can be grouped according to the Hamming weight (or bitcount) of a configuration $\pi$ which equals the number of cooperators $c(\pi)$. Fig. \ref{fig:sigma_abs_8_14} gives the  number $\#_\sigma$ of how many  different values of $\sigma(\pi)$ there are for each number of cooperators $c(\pi)$ over players $N$ and coplayers $d$. 
 The
results  show  substantial structure and symmetry over all $N$ and $d$ tested.   
To begin with, the number of different $\sigma(\pi)$ is countable and the count is rather low. 
If every network of interaction were to produce a different value of $\sigma(\pi)$, the count $\#_\sigma$ could be as high as the number of interaction graphs tested (up to $G=10.000$ according to our experimental setup) for each configuration, but the actual results are much lower. A second property is symmetry with respect to the number of cooperators $c(\pi)$, which follows from the symmetry under conjugation $\sigma(\pi)=\sigma(1-\pi)$, \cite{chen16}. The values of $\#_\sigma$ generally increase with the number of cooperators $c(\pi)$ until cooperator and defectors are exactly (or almost)  balanced, before symmetrically decreasing for $c(\pi)$ getting even larger. A third result is $\#_\sigma=1$ for all $c(\pi)=1$ and $c(\pi)=N-1$. In other words, the $N$ configurations each with either only a single cooperator (and $N-1$ defectors) or $N-1$ cooperators (and a single defector) all produce the same $\sigma(\pi)$ for all $A_I$ tested. Lastly, we obtain   $\#_\sigma=1$  for all $c(\pi)$ and $d=N-1$. This relates to the finding  that for a well--mixed population there is exactly one network of interaction which consequently produces one $\sigma(\pi)$. The results in Fig. \ref{fig:sigma_abs_8_14} also show differences over $N$ and $d$. There is no real symmetry with respect to $d$, even if the well--mixed case $d=N-1$ is discarded. However, we obtain the largest values of $\#_\sigma$ for intermediate values of $d$, that is $d=\{3,4\}$ for $N=8$, $d=\{4,5\}$ for $N=10$, $d=\{4,5,6,7\}$ for $N=12$, and $d=\{5,6,7,8,9\}$ for $N=14$, and lower values for smaller and larger $d$. Moreover, the values of $\#_\sigma$ generally increase with the number of players, but if we scale the number of different values of $\sigma(\pi)$ with the number of configurations, the results in Fig.  \ref{fig:sigma_abs_8_14}  suggest $\frac{\#_\sigma}{2^N-2} \rightarrow 0$ for $N$ getting larger. This means as $N$ increases it becomes rare to find a configuration $\pi$ for which $\sigma(\pi)$ varies over interaction networks.  

An interesting aspect of the distribution of $\sigma(\pi)$ is variety and symmetry over the number of cooperators $c(\pi)$, which can be analyzed by normalizing  the values of $\#_\sigma$ by
$ N / \#_{c(\pi)}$, see Fig. \ref{fig:sigma8_14}.  The number $\#_{c(\pi)}$ of configurations with the same number of cooperators $c(\pi)$ can be calculated as binomials: $\#_{c(\pi)}=\left(\begin{array}{c} N \\ c(\pi) \end{array} \right)$ for $1 \leq c(\pi) \leq N-1$.  
There are two results that are invariant under the number of players $N$. First, the normalized number of different structure coefficients $\sigma(\pi)$ yields $\frac{\#_\sigma}{\#_{c(\pi)}}N=1$ for  all $c(\pi)=1$ and $c(\pi)=N-1$, which follows from $\#_1=\left(\begin{array}{c} N \\ 1 \end{array} \right)=\#_{N-1}=N$.  The other constant values over varying $N$  are obtained for $c(\pi)=2$ and $c(\pi)=N-2$. We get  $\frac{\#_\sigma}{\#_{c(\pi)}}N =2$ for intermediate numbers of coplayers, for which also $\#_\sigma$ has large values, see Fig. \ref{fig:sigma_abs_8_14}. The number of coplayers for which this can be observed follows $d=N/2-1$. For $N$ increasing, the value of $\#_\sigma$ grows linearly with $\#_{c(\pi)}$ and $N$ for $c(\pi)=2$ and $c(\pi)=N-2$.
As $\#_2=\left(\begin{array}{c} N \\ 2 \end{array} \right)=\#_{N-2}=N(N-1)/2$,   we have $\#_\sigma=N-1$. For all other numbers of coplayers (except  $c(\pi)=\{1,2,N-2,N-1\}$),  the number of configurations with the same number of cooperators $\#_{c(\pi)}$ grows faster than the number of configurations with different structure coefficients, which can be seen by  $\frac{\#_\sigma}{\#_{c(\pi)}}N$ getting smaller and smaller for $N$ increasing. A possible interpretation is that if we intend to find at random a configuration $\pi$ that yields varying structure coefficients $\sigma(\pi)$ over interaction networks specified by $A_I$, the chances are best for $c(\pi)=2$ or $c(\pi)=N-2$ cooperators and an intermediate number of coplayers $d$ with $d \approx N/2$. Moreover, as the number of configurations with the same number of cooperators $\#_{c(\pi)}$  grows quadratic with $N$ for $c(\pi)=2$ and $c(\pi)=N-2$, finding this configuration by enumeration is numerically feasible. For $c(\pi)\geq 3$ and $c(\pi) \leq N-3$, the absolute number of configurations with varying $\sigma(\pi)$ increases, but finding them becomes harder as they get rare. The number of configurations with the same $c(\pi)$, $\#_{c(\pi)}$, grows polynomial with the degree depending on  $c(\pi)$. For $c(\pi)=\frac{N}{2}$, the growth is even exponential.

\begin{figure}[tb]
\includegraphics[trim = 0mm 0mm 0mm 0mm,clip,width=8.25cm, height=5.9cm]{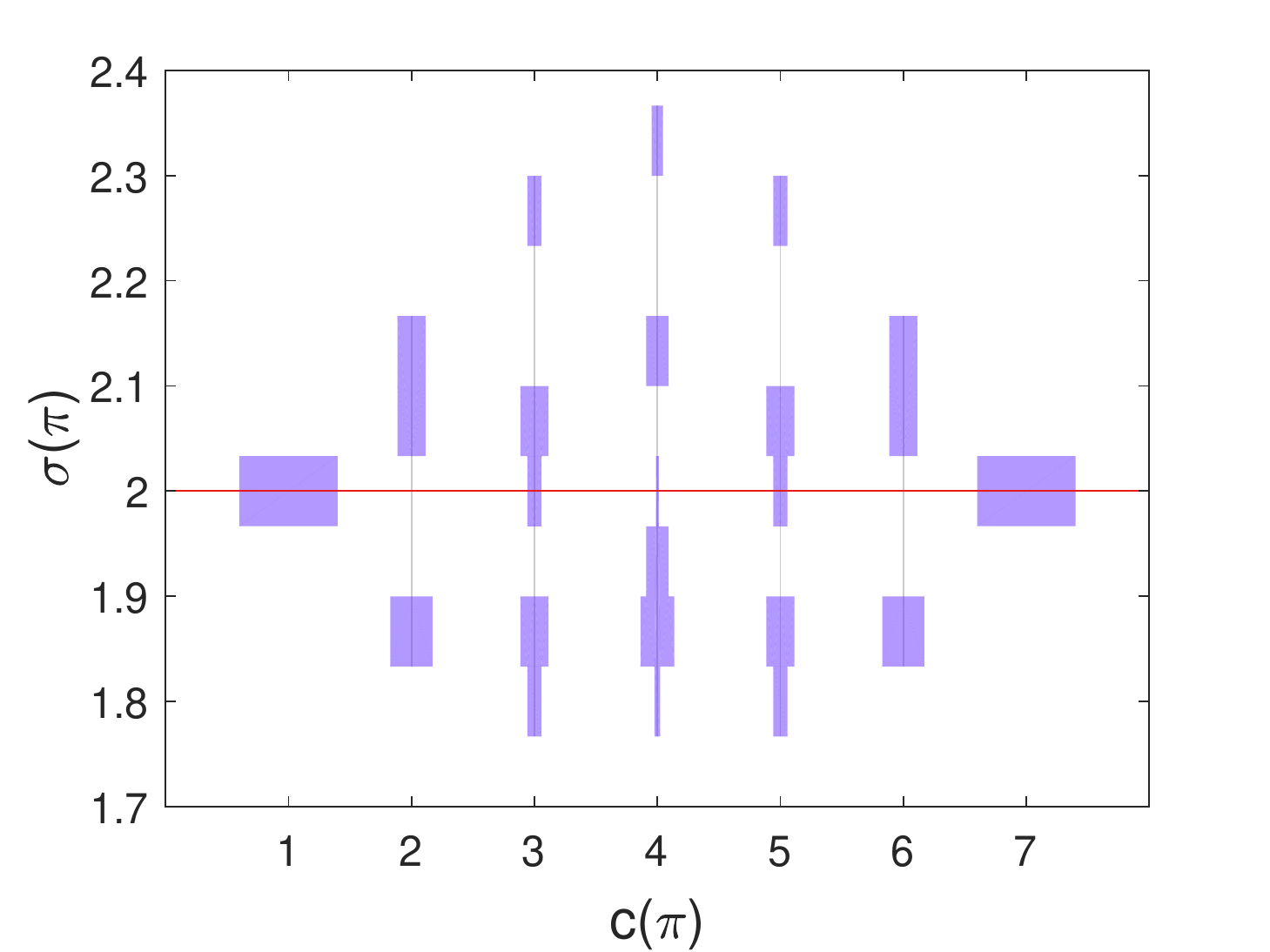}
\includegraphics[trim = 0mm 0mm 0mm 0mm,clip,width=8.25cm, height=5.9cm]{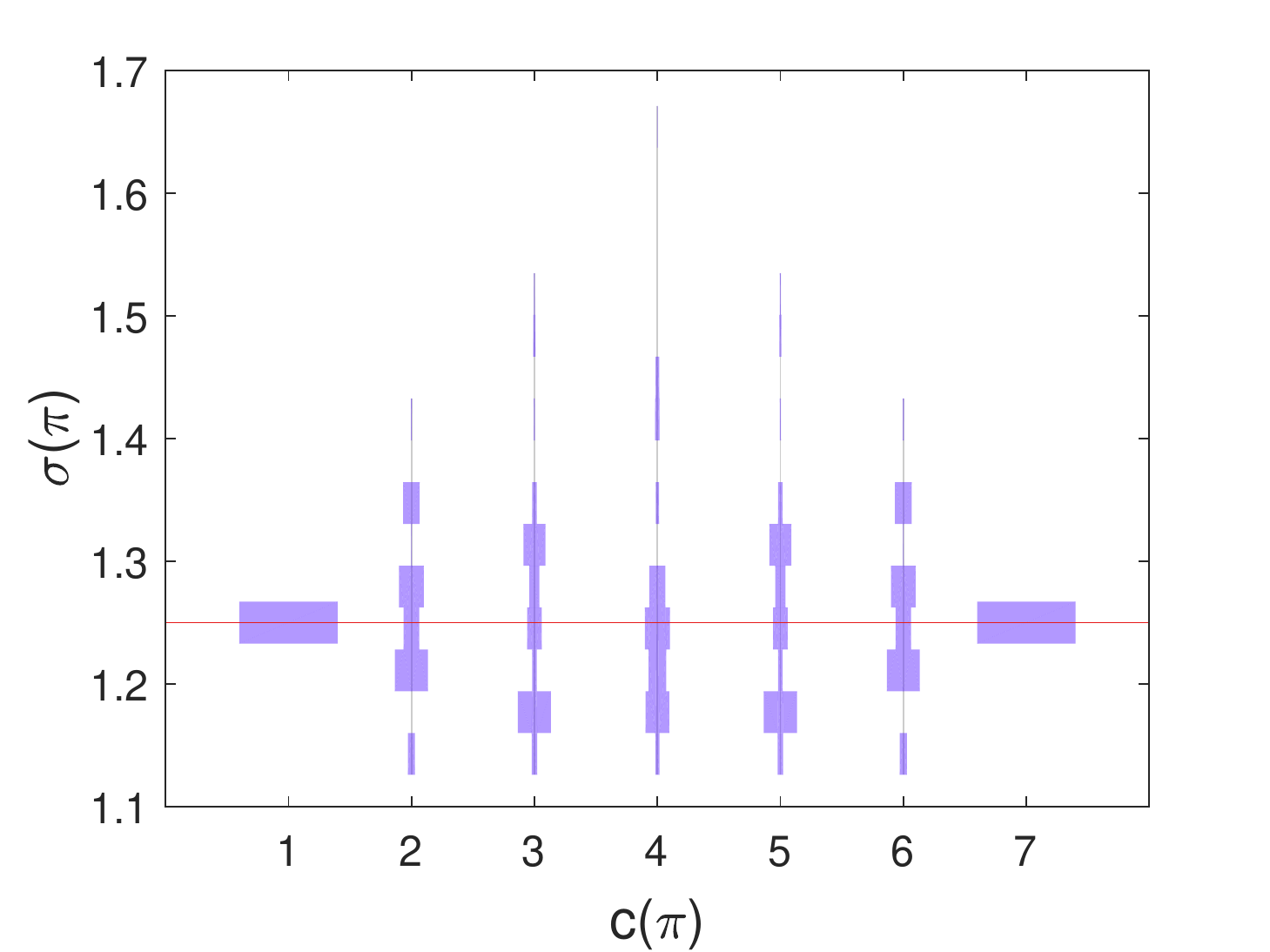}

\hspace{1cm}a) $d=2$  \hspace{7cm} b) $d=3$

\includegraphics[trim = 0mm 0mm 0mm 0mm,clip,width=8.25cm, height=5.9cm]{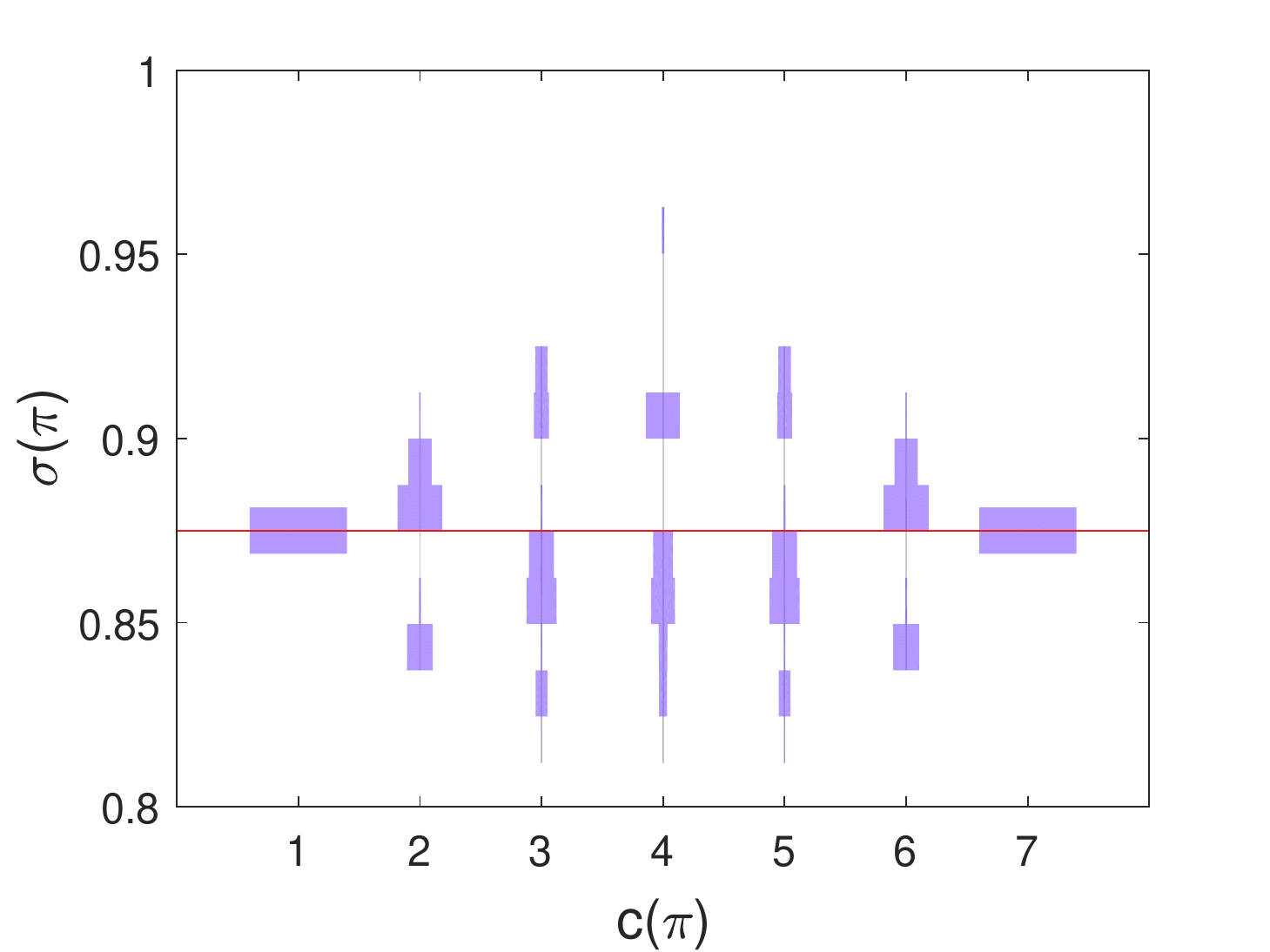}
\includegraphics[trim = 0mm 0mm 0mm 0mm,clip,width=8.25cm, height=5.9cm]{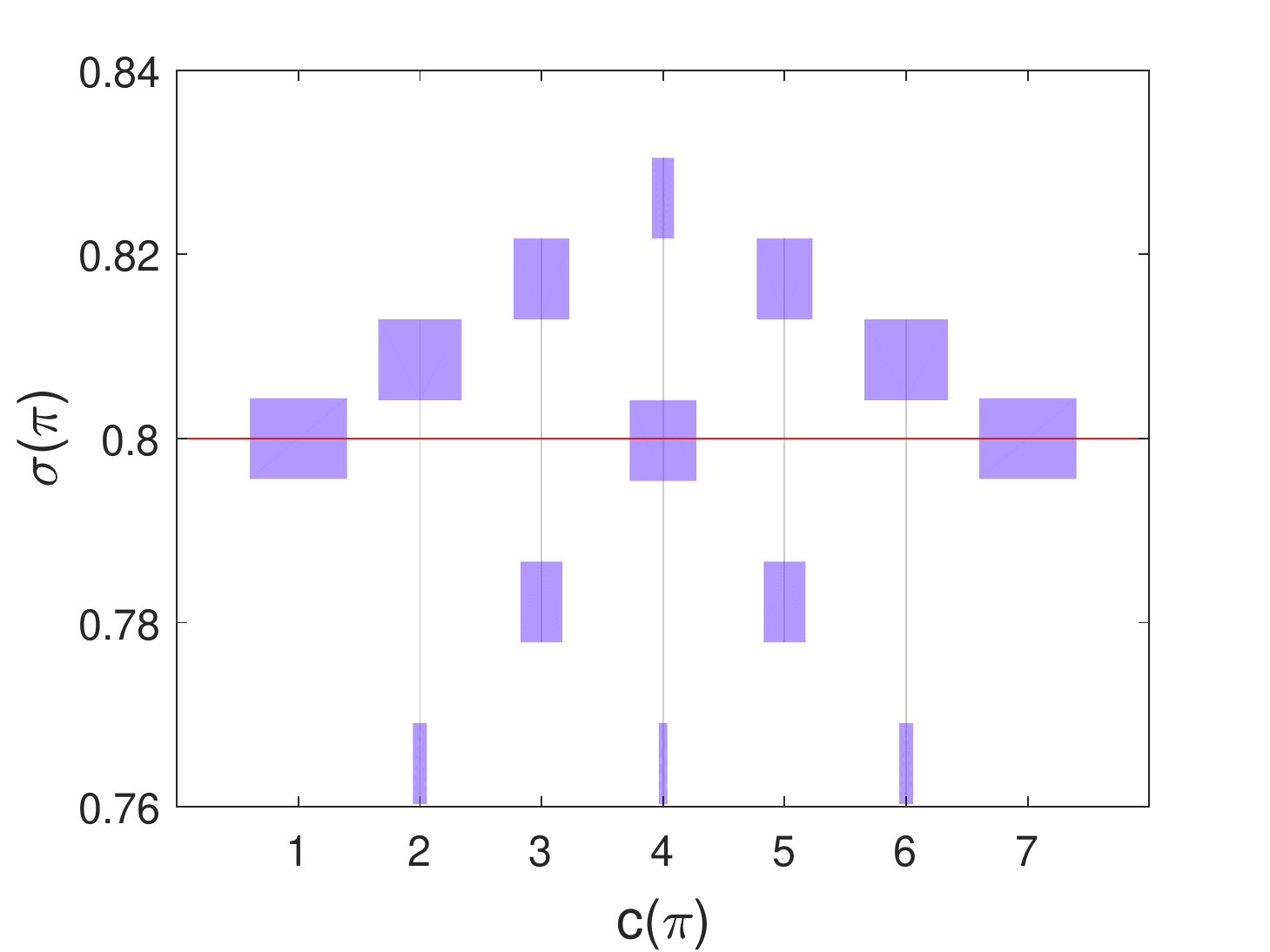}

\hspace{1cm}c) $d=5$  \hspace{7cm} d) $d=6$

\caption{Histograms of the distribution of $\sigma(\pi)$  over interaction networks as violin plots for different number of cooperators $c(\pi)$ and $N=8$.  The red line gives the generic $\sigma$ according to Eq. (\ref{eq:sigm_gen}). The coefficient $\sigma$ is an intermediate value intersecting  the distribution $\sigma(\pi)$ where different interaction networks  may result in both larger and smaller values. The largest range of $\sigma(\pi)$ is generally obtained for $c(\pi)=N/2$, which means for a balance  between cooperators and defectors. }
\label{fig:shape1}
\end{figure}

\begin{figure}[tb]
\includegraphics[trim = 0mm 0mm 0mm 0mm,clip,width=8.25cm, height=5.9cm]{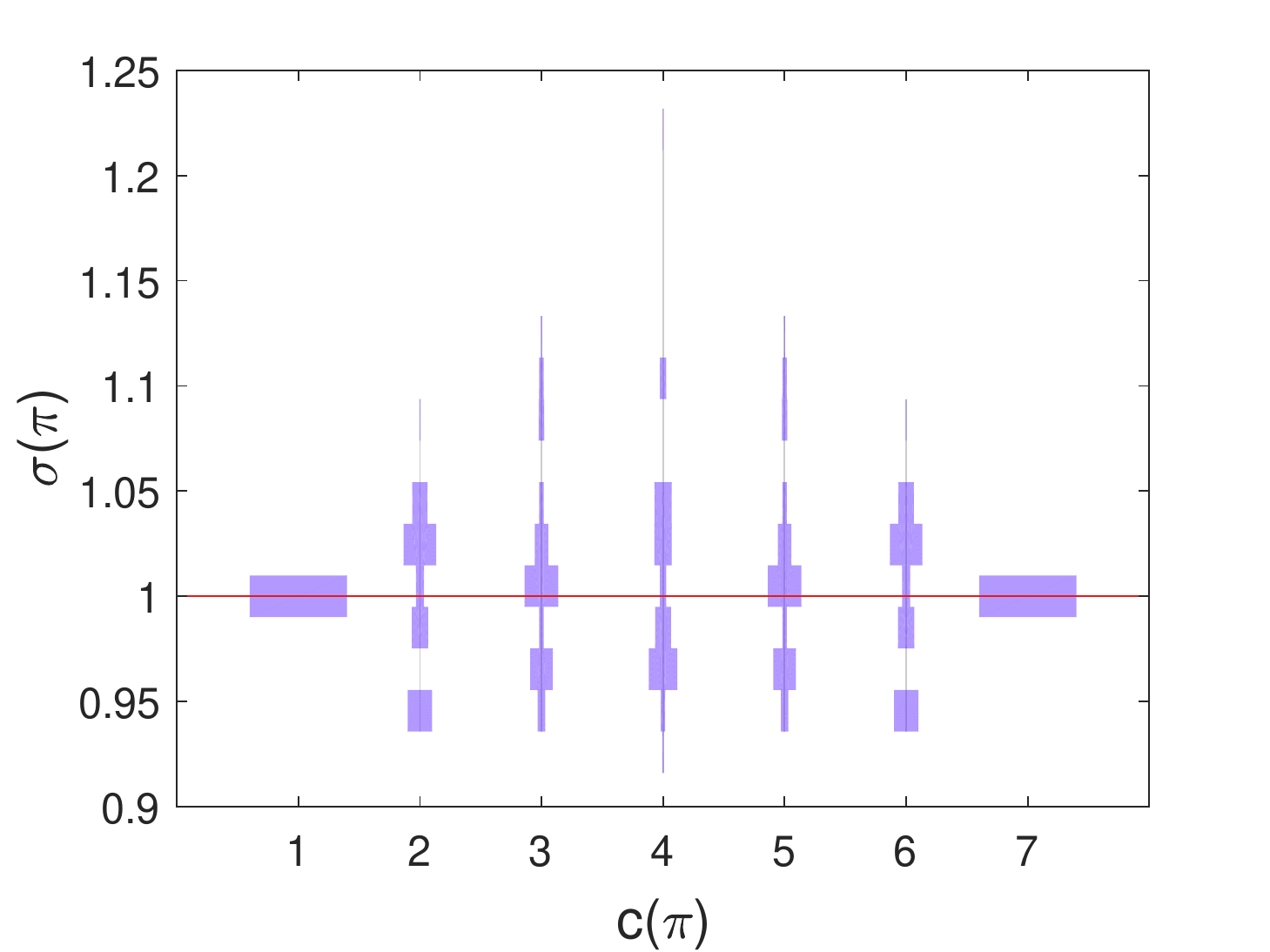}
\includegraphics[trim = 0mm 0mm 0mm 0mm,clip,width=8.25cm, height=5.9cm]{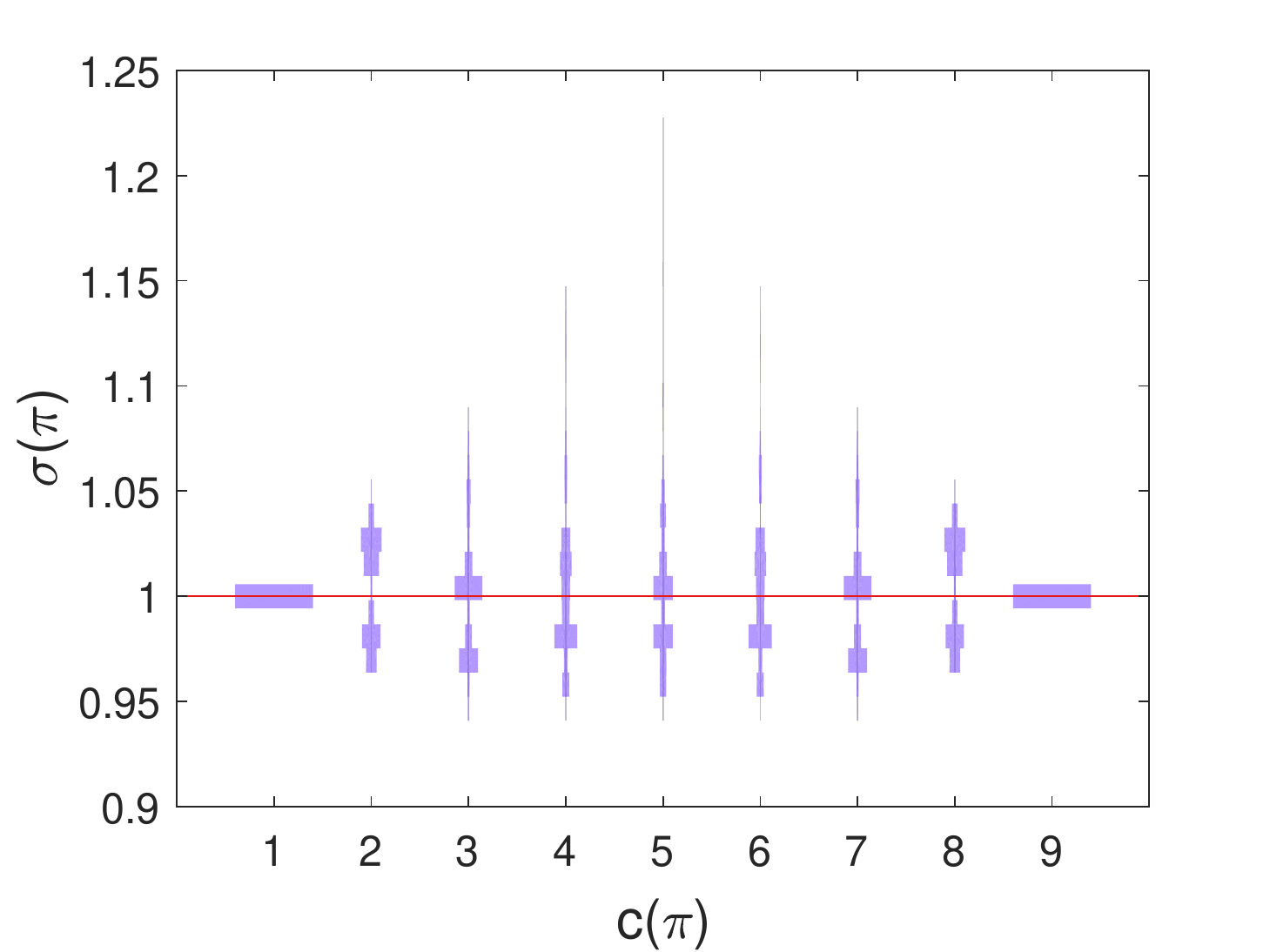}

\hspace{1cm} a) $N=8$ \hspace{7cm} b) $N=10$

\includegraphics[trim = 0mm 0mm 0mm 0mm,clip,width=8.25cm, height=5.9cm]{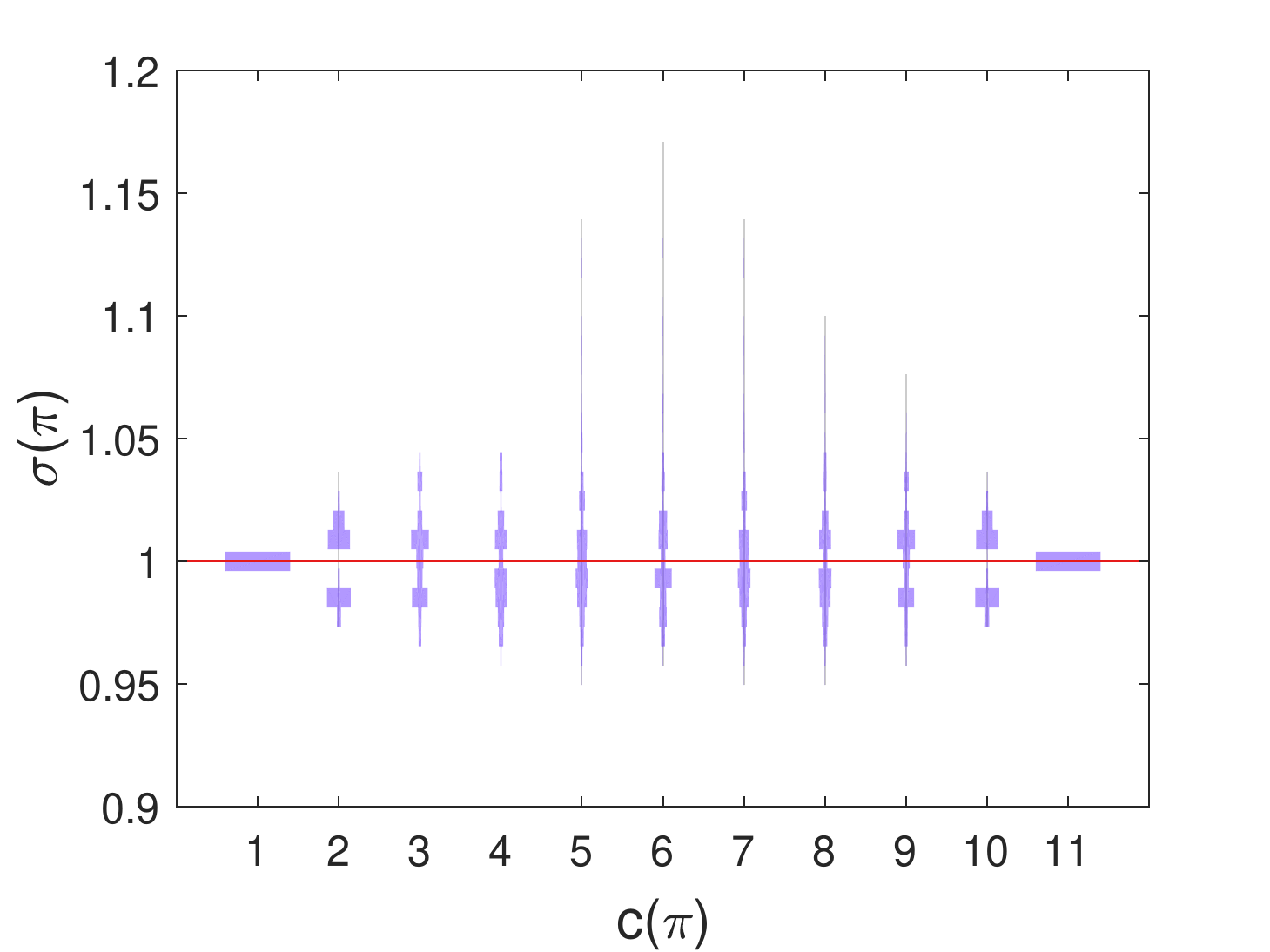}
\includegraphics[trim = 0mm 0mm 0mm 0mm,clip,width=8.25cm, height=5.9cm]{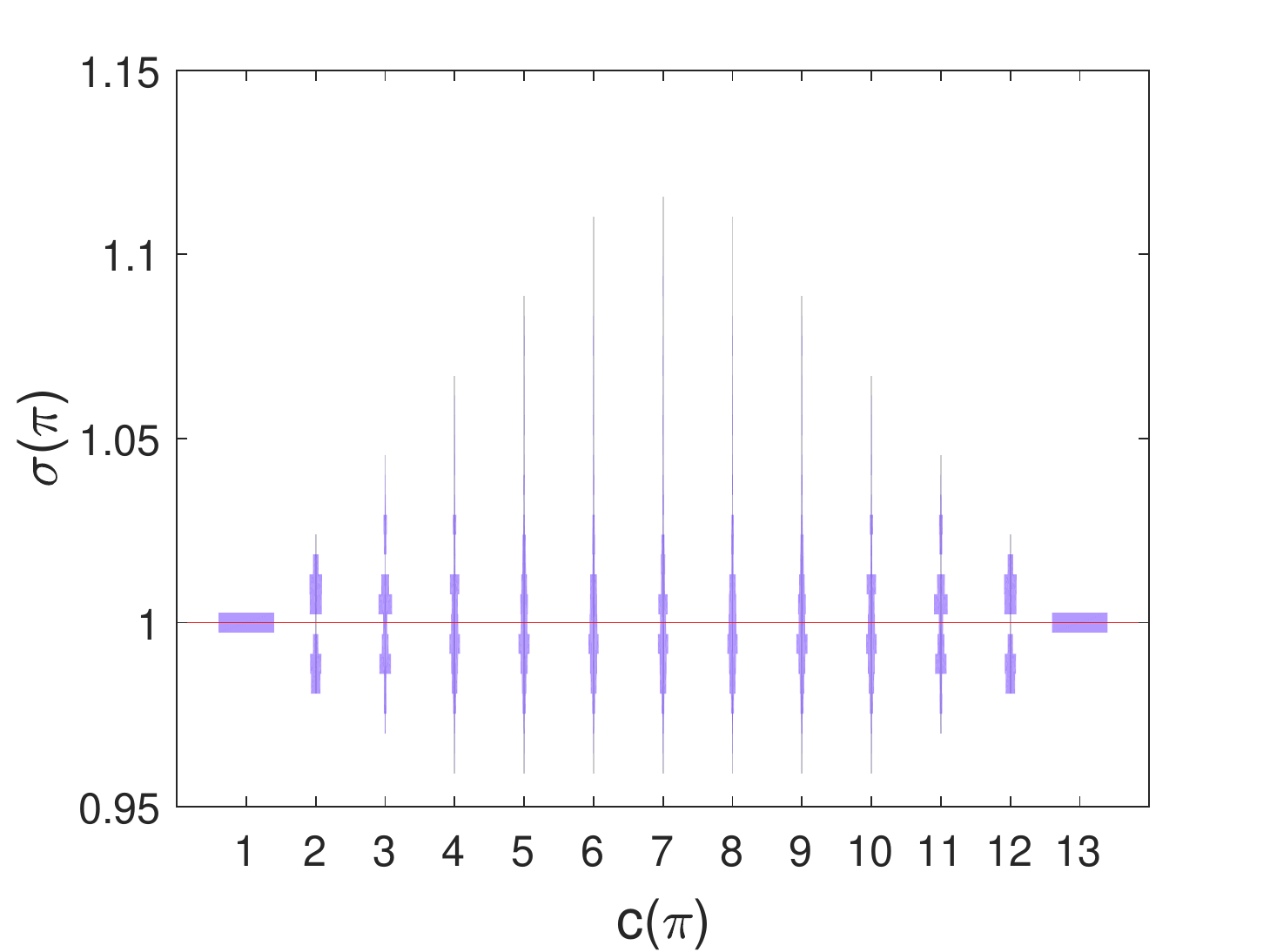}

\hspace{1cm} c) $N=12$ \hspace{7cm} d)  $N=14$

\caption{Histograms of the distribution of $\sigma(\pi)$  over interaction networks as violin plots for different number of cooperators $c(\pi)$ and $d=N/2$.  The red line gives the generic $\sigma$ according to Eq. (\ref{eq:sigm_gen}), which is $\sigma=1$ for $d=N/2$ and all $N$. The range between the smallest and the largest value of the distributions gets slowly smaller for $N$ getting larger. }
\label{fig:shape2}
\end{figure}

We next look at the shape of the distributions, see Figs. \ref{fig:shape1} and \ref{fig:shape2}, which gives histograms of $\sigma(\pi)$  as violin plots over the number of cooperators $c(\pi)$.  The number of bins of the histograms is calculated according to the Freedman--Diaconis rule. The central axis of the violin plots gives the range between the largest and the smallest value of $\sigma(\pi)$. To show general trends of how $\sigma(\pi)$ is distributed over $A_I$, Fig.  \ref{fig:shape1}  gives results for $N=8$ and $d=\{2,3,5,6\}$, while Fig.  \ref{fig:shape2} shows results for $d=\frac{N}{2}$ and $N=\{8,10,12,14\}$. For other combinations of $N$ and $d$, the results are similar and the properties discussed next apply likewise. A first property is that for a given $N$ and $d$ the single value of $\sigma(\pi)$ for $c(\pi)=1$ and $c(\pi)=N-1$ reproduces the generic $\sigma$ for $N$ and $d$:~(\cite{taylor07,lehmann07,tarnita09}) \begin{equation} \sigma=\frac{(d+1)N-4d}{(d-1)N}, \label{eq:sigm_gen}\end{equation} which is shown as red line in the Figs. \ref{fig:shape1} and \ref{fig:shape2}. In other words, the $2N$ configurations $\pi$ with Hamming weight $\text{hw}_1(\pi)=1$ and $\text{hw}_1(\pi)=N-1$ are invariant over interaction networks and thus for these configurations the specific coefficients $\sigma(\pi)$ of Eq. (\ref{eq:sigma}) reproduce the generic coefficients $\sigma$ of Eq. (\ref{eq:sigm_gen}).
For $2 \leq c(\pi) \leq N-2$. we find a distribution of values that may be both smaller and larger than $\sigma$ in Eq. (\ref{eq:sigm_gen}). Put differently, the generic coefficient $\sigma$ intersects the distribution of specific coefficients $\sigma(\pi)$.

If we compare the values over $d$, we notice that for increasing $d$ the range between largest and smallest value of the distribution slightly decreases. 
Generally, it can be seen that the values of $\sigma(\pi)$ are sparsely distributed. Over the range between the largest and the smallest value,  the values of $\sigma(\pi)$ only group within a small subset. Frequently, values are very close to each other, which means they are counted within the same bin of the histogram. In most cases do the distributions have tails where smallest and largest values of $\sigma(\pi)$ have a very small frequency. A exception from this rule are the distributions for $d=N-2$, see Fig.  \ref{fig:shape1}d for $N=8$ (for the other $N$ we get similar results), for which we find more heavy tails with smallest and largest values of $\sigma(\pi)$ having  a rather substantial frequency. 
The largest range  we obtain for intermediate values of cooperators $c(\pi)$, mostly for $c(\pi)=\frac{N}{2}$. The distributions are symmetric with respect to $c(\pi)$. For $N$ increasing, it can be noticed that the distribution remains within  a similar interval with the range between largest and smallest value of the distribution decreases slightly. Along with the number of different structure coefficients increasing with $N$, the distributions for $N=\{12,14\}$ get less sparse and assume a  more dense  shape.

\begin{figure}[tb]
\includegraphics[trim = 0mm 0mm 0mm 0mm,clip,width=8.25cm, height=5.9cm]{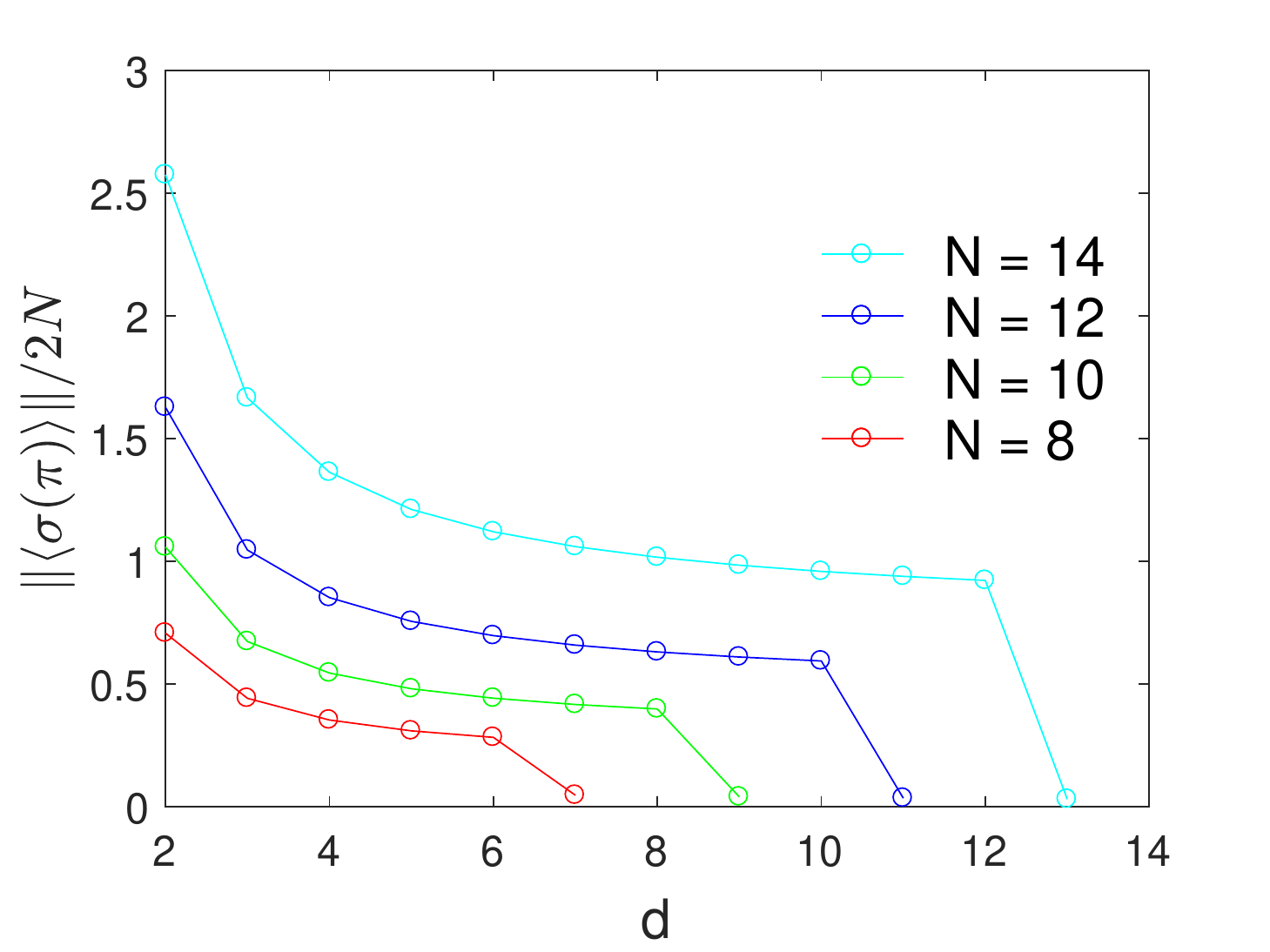}
\includegraphics[trim = 0mm 0mm 0mm 0mm,clip,width=8.25cm, height=5.9cm]{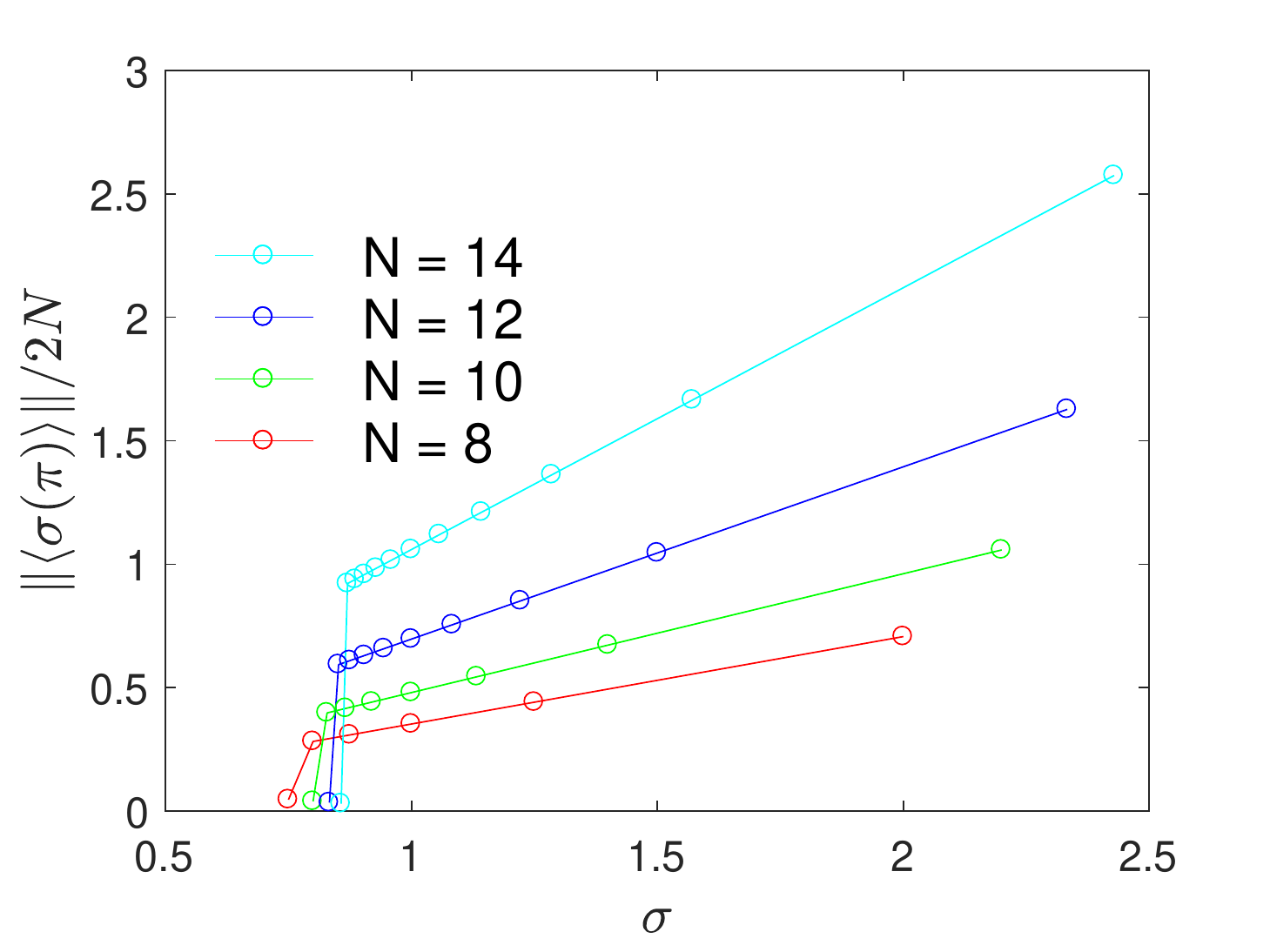}

\hspace{1cm} a) \hspace{7cm} b) 
\caption{The quantity $\| \langle \sigma(\pi) \rangle\| /2N$ representing an average network for $N=\{8,10,12,14 \}$ and $2\leq d \leq N-1$: a) $\| \langle \sigma(\pi) \rangle\| /2N$ versus the number of coplayers $d$; b) $\| \langle \sigma(\pi) \rangle\| /2N$ versus the generic structure coefficient $\sigma(N,d)$ according to Eq. (\ref{eq:sigm_gen}). We see that $\| \langle \sigma(\pi) \rangle\| /2N$ is linearly related to $\sigma$, except for $d=N-1$.}
\label{fig:sigma}
\end{figure}

\begin{figure}[tb]
\includegraphics[trim = 0mm 0mm 0mm 0mm,clip,width=8.25cm, height=4.1cm]{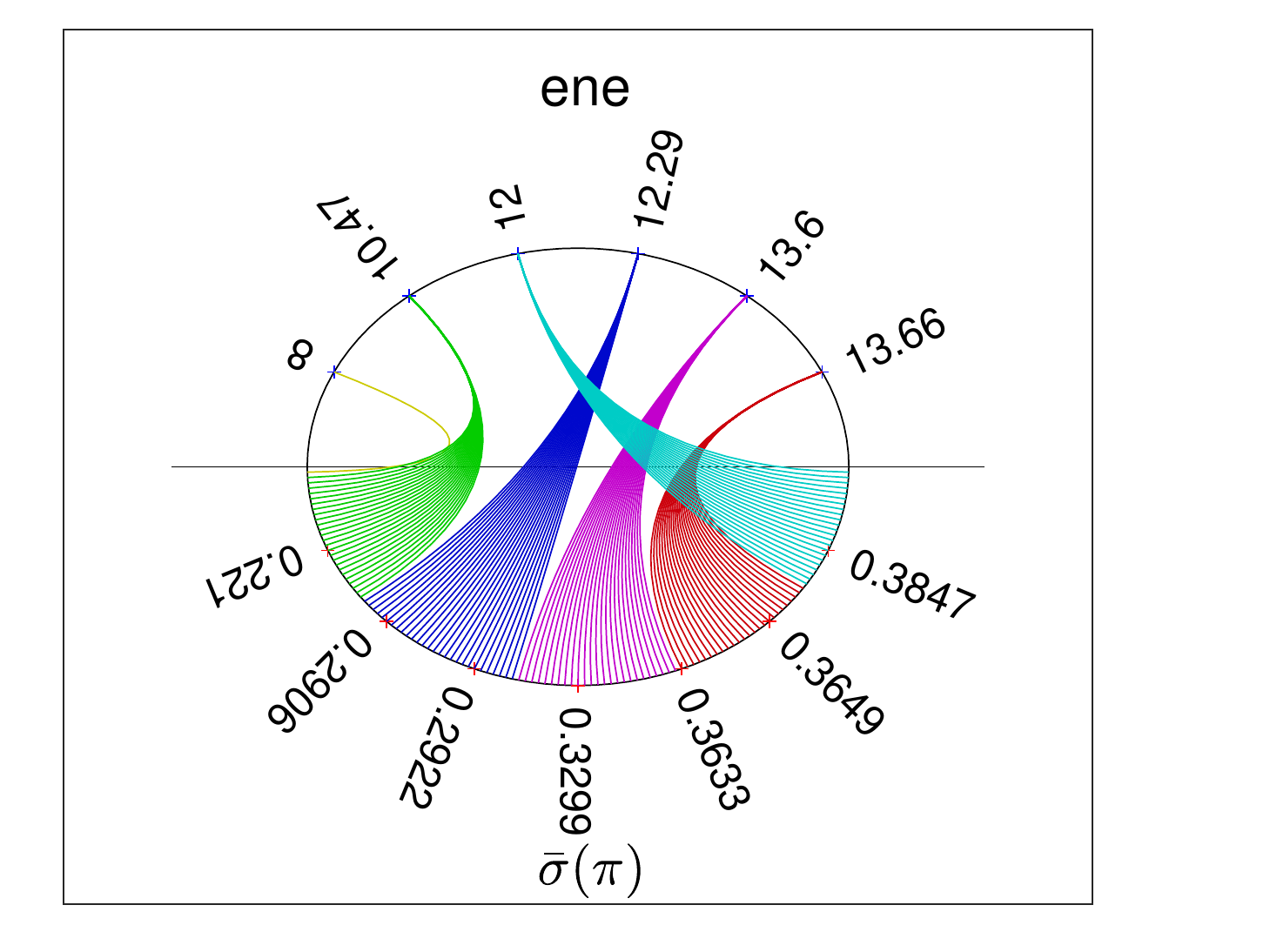}
\includegraphics[trim = 0mm 0mm 0mm 0mm,clip,width=8.25cm, height=4.1cm]{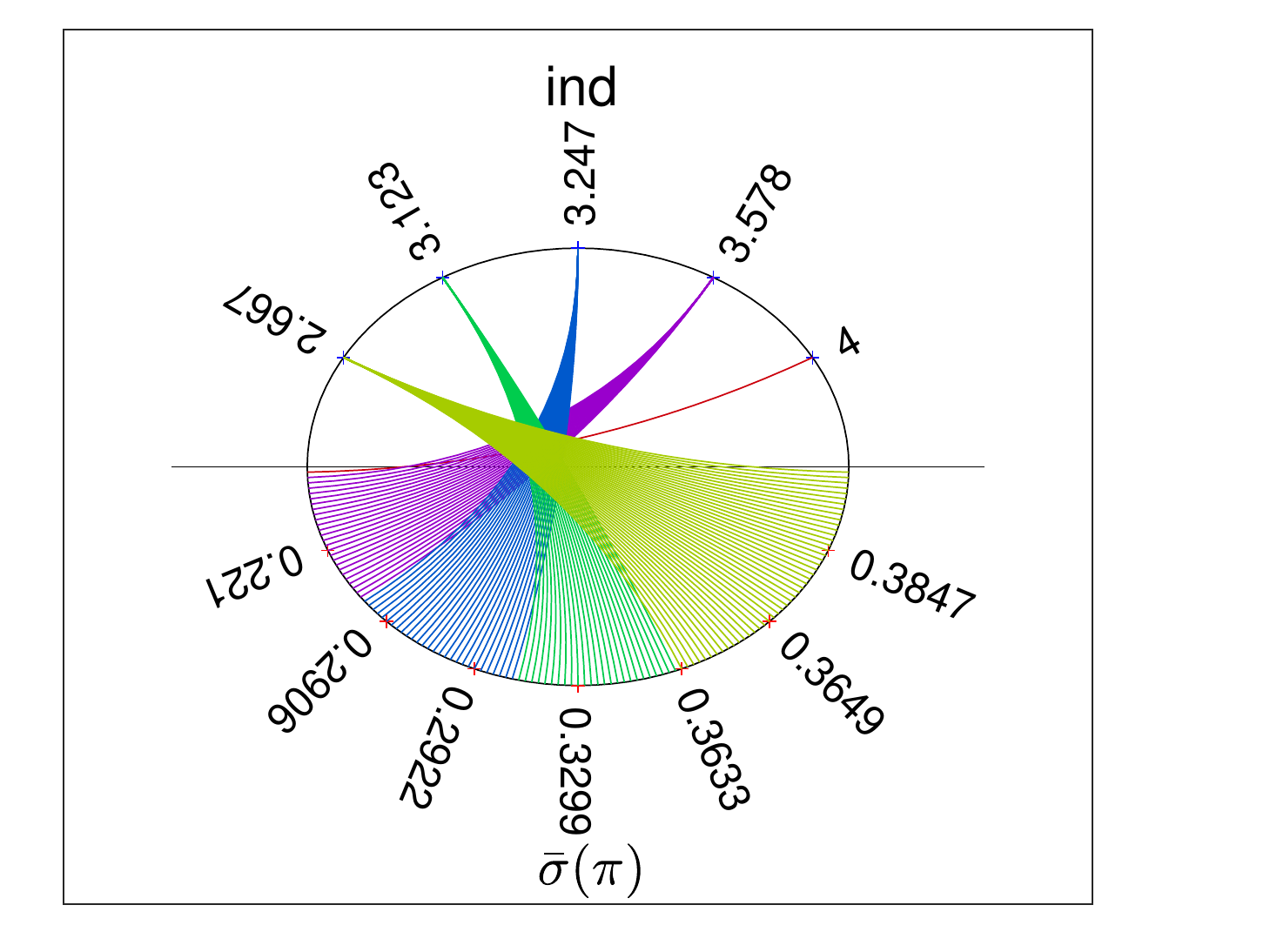}

\hspace{1cm} a) graph energy \hspace{5cm} b) independence number

\includegraphics[trim = 0mm 0mm 0mm 0mm,clip,width=8.25cm, height=4.1cm]{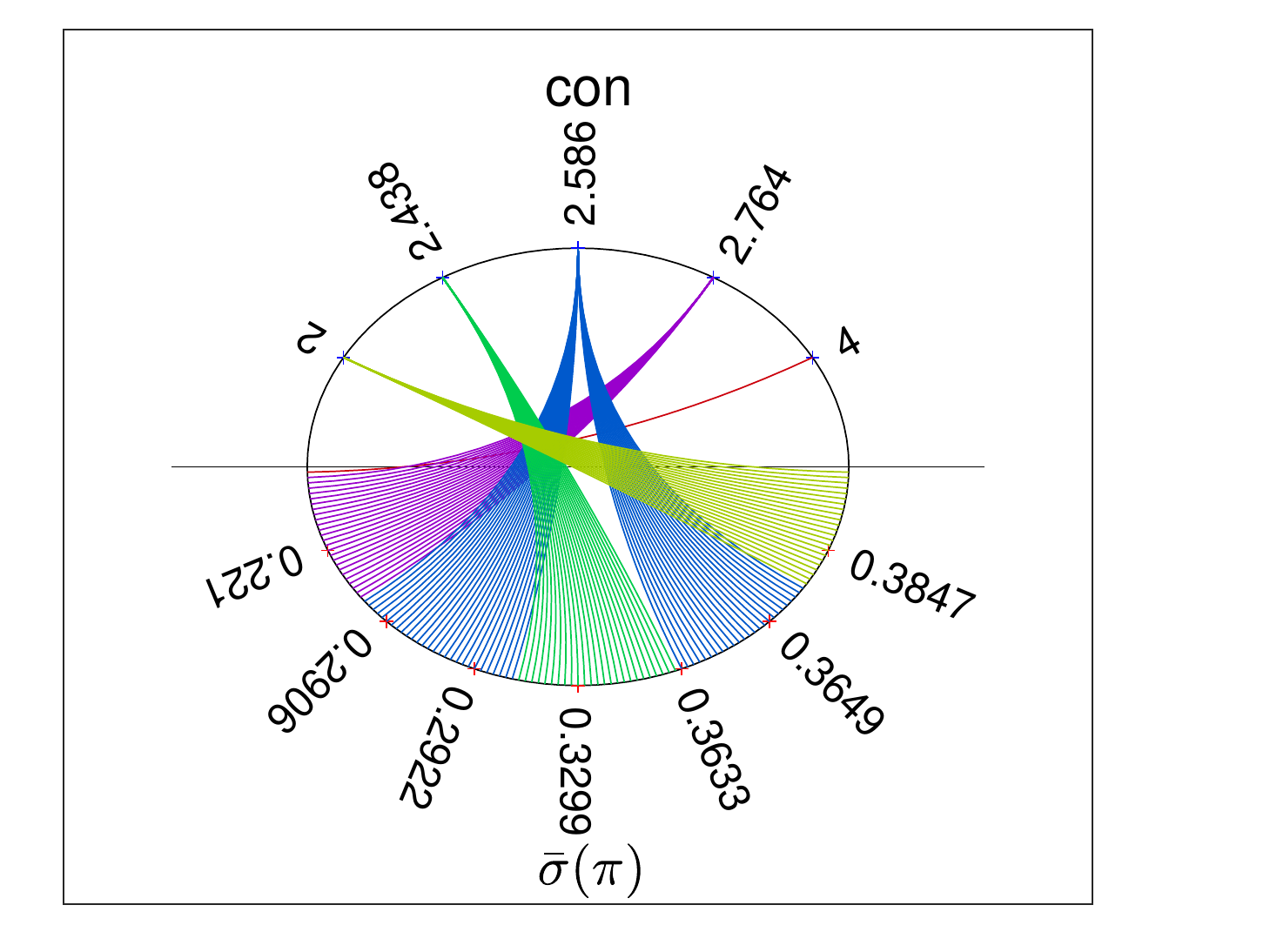}
\includegraphics[trim = 0mm 0mm 0mm 0mm,clip,width=8.25cm, height=4.1cm]{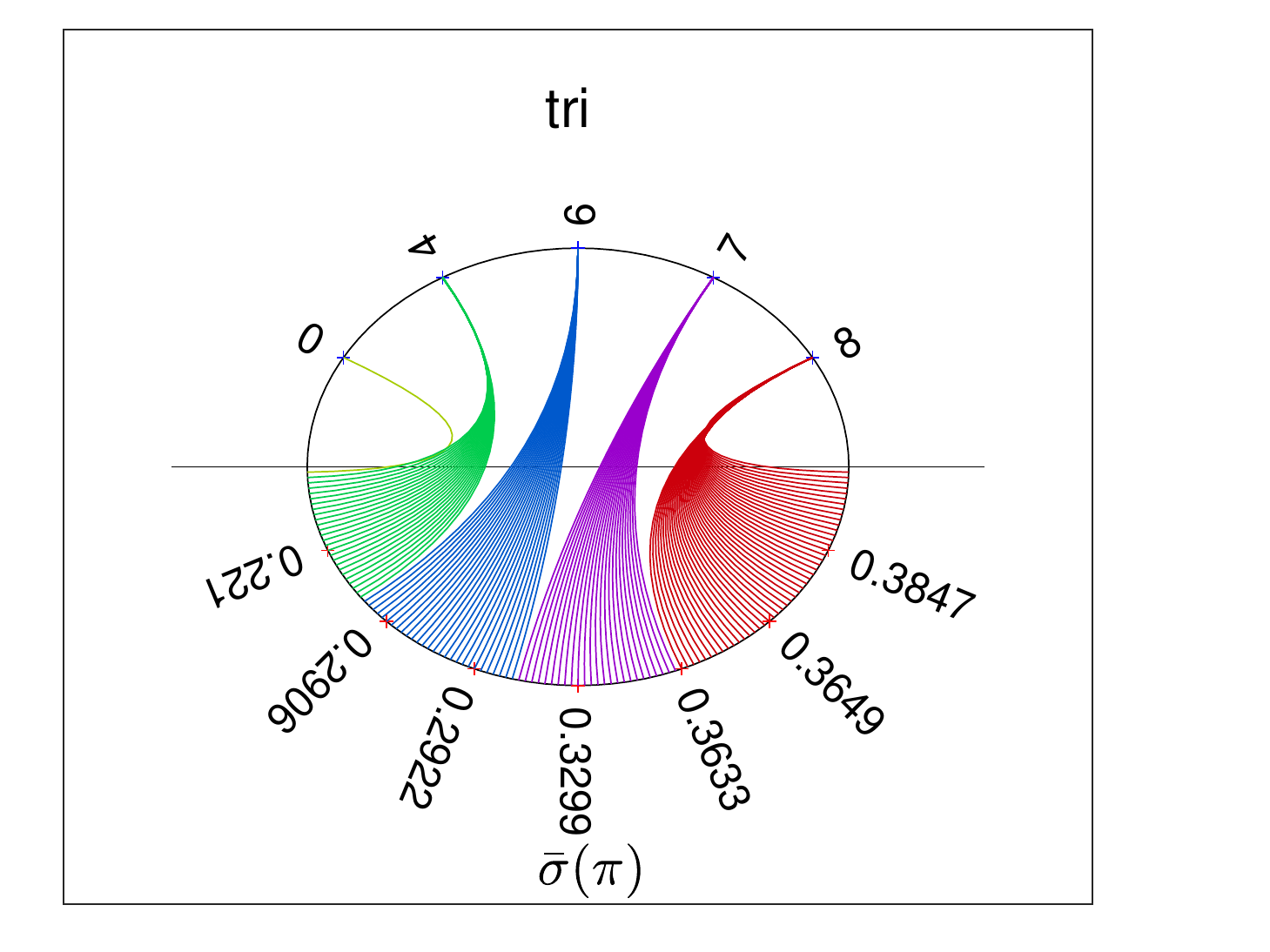}

\hspace{1cm} c) algebraic connectivity \hspace{3.6cm} d)  triangular number

\caption{Schemaballs visualizing the relations between the distance measure $\bar{\sigma}$ and spectral graph measures for $\{N,d,c(\pi)\}=\{8,4,3\}$. a) graph energy $\text{ene}$,  Eq. (\ref{eq:ene}), b) independence number $\text{ind}$, Eq. (\ref{eq:ind}), c)  algebraic connectivity $\text{con}$, Eq. (\ref{eq:con}),  d) triangular number $\text{tri}$, Eq. (\ref{eq:tri}). We draw Bezier curves connecting the values of the spectral graph measures of the interaction matrix $A_I$ with the values of the distance measure $\bar{\sigma}(\pi)$ belonging to this interaction network. The curves are colored in such a way that the same value of the graph measure has only connections of the same color. We see that each value of the spectral graph measures is connected to an interval of the distance measure, but the connection are not linear. }
\label{fig:schemball_gm}
\end{figure} 
A second series of experiments addresses relationships between the structure coefficients $\sigma(\pi)$ and graph measures of the interaction networks.  As shown with Figs. \ref{fig:sigma_abs_8_14}--\ref{fig:shape2}, there is variety in $\sigma(\pi)$ over interaction networks and it may be interesting to ask if these differences are reflected by graph--theoretical properties of the networks.  
We here describe the interaction networks by instances of $d$--regular graphs and each graph can be characterized by spectral graph measures, see \ref{sec:meth}. Methods. 
A first result is that the spectral graph measures tested in this paper, graph energy (ene),  independence number (ind), algebraic connectivity (con) and triangle number (tri), form discrete sets where the count of different values is much lower than the count of the set. In fact, the count for graph measures is even lower than the count for structure coefficients.  We now study how the distributions of the values over the sets relate from structure coefficients to spectral graph measures.  

Therefore, we consider  $\bar{\sigma}(\pi)=\|\sigma(\pi)- \langle \sigma(\pi) \rangle\|$, which is the distance between the distribution $\sigma(\pi)$ from an average over interaction networks $\langle \sigma(\pi) \rangle=1/I\sum_{i=1}^I \sigma_i(\pi)$. The quantity $\langle \sigma(\pi) \rangle$ can be interpreted as  representing an average network, which is underlined by the result that $\| \langle \sigma(\pi) \rangle\|$  is well--related to the generic structure coefficient $\sigma$, see Fig. \ref{fig:sigma}, which shows $\| \langle \sigma(\pi) \rangle\|$ for $N=\{8,10,12,14\}$ and $2 \leq d \leq N-1$ over $d$ and over $\sigma$ calculated by Eq. (\ref{eq:sigm_gen}).  We see that  $\| \langle \sigma(\pi) \rangle\|$ linearly scales to $\sigma$,  except for $d=N-1$.

\begin{figure}[htb]
\includegraphics[trim = 0mm 0mm 0mm 0mm,clip,width=8.25cm, height=4.1cm]{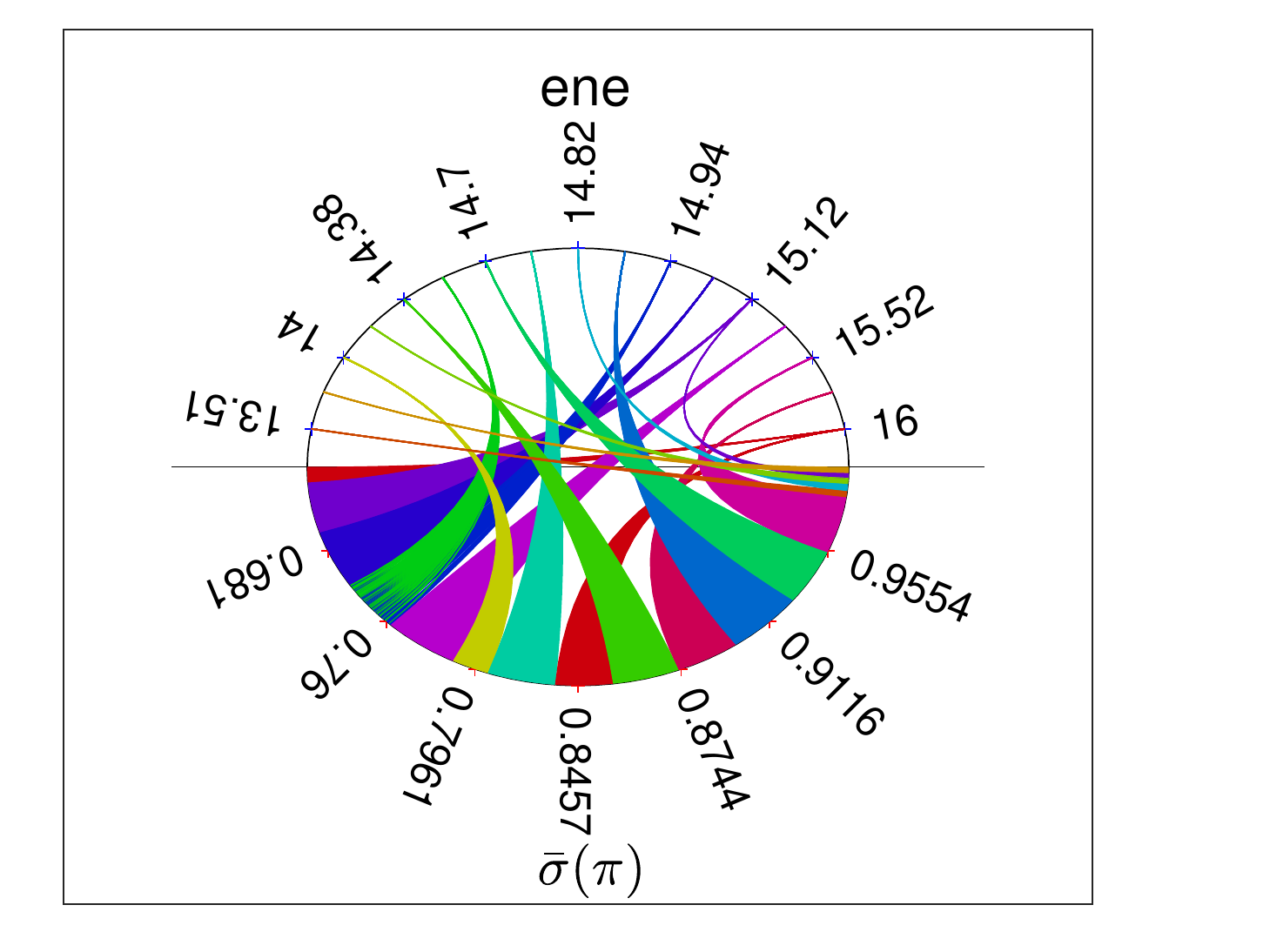}
\includegraphics[trim = 0mm 0mm 0mm 0mm,clip,width=8.25cm, height=4.1cm]{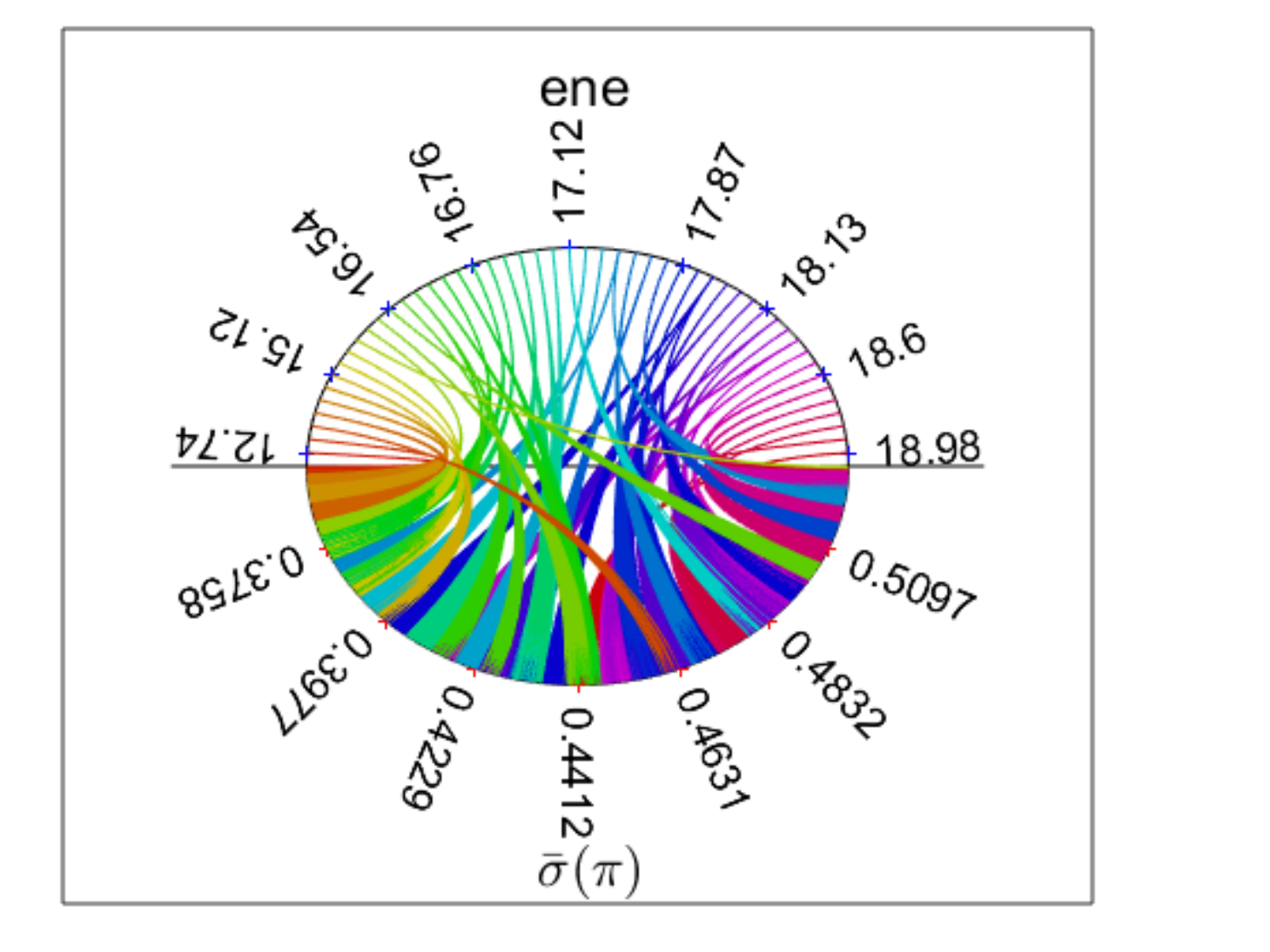}

\hspace{1cm} a) $\{N,d,c(\pi)\}=\{10,3,6\}$ \hspace{3cm} b) $\{N,d,c(\pi)\}=\{10,5,5\}$

\includegraphics[trim = 0mm 0mm 0mm 0mm,clip,width=8.25cm, height=4.1cm]{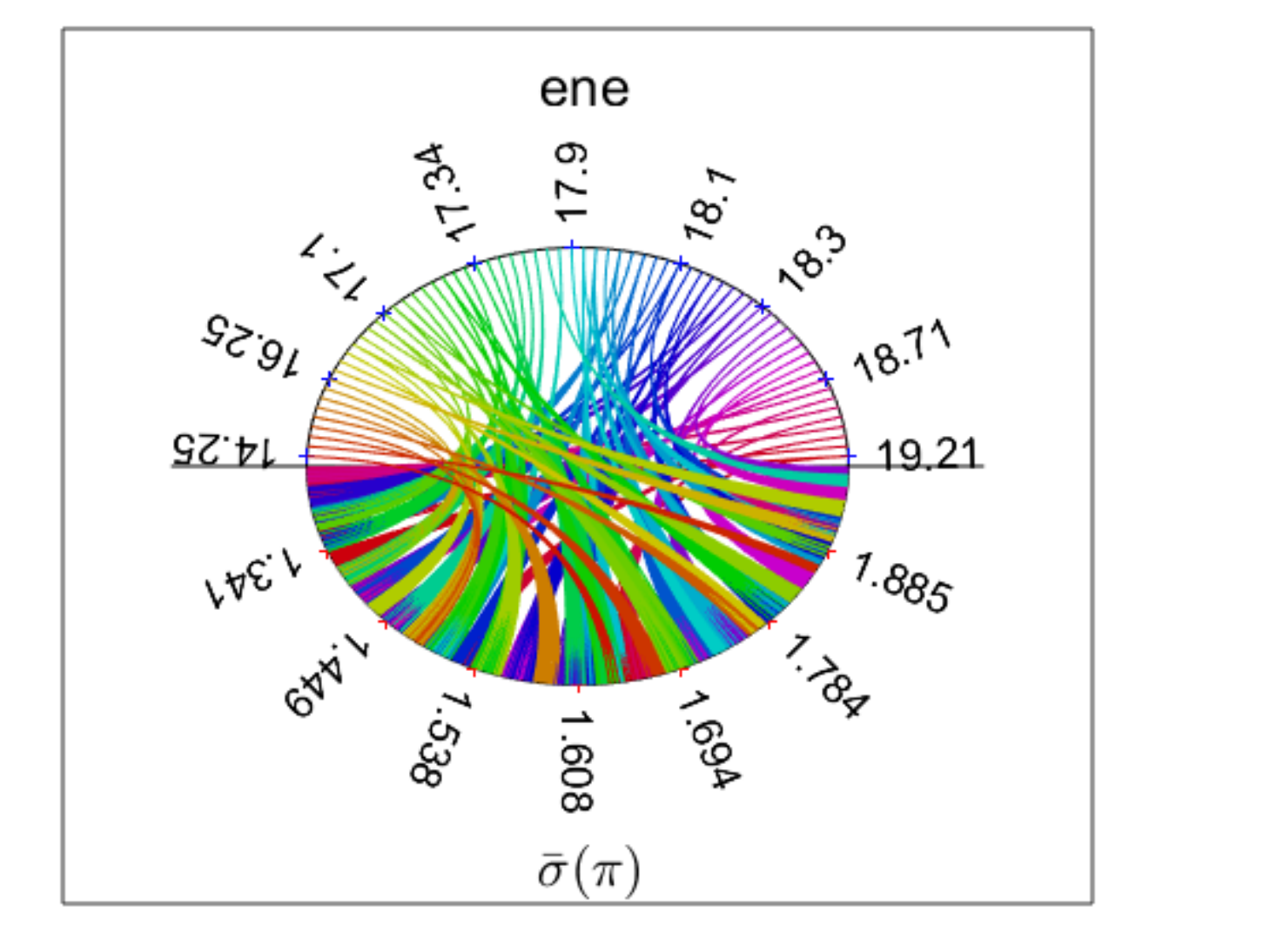}
\includegraphics[trim = 0mm 0mm 0mm 0mm,clip,width=8.25cm, height=4.1cm]{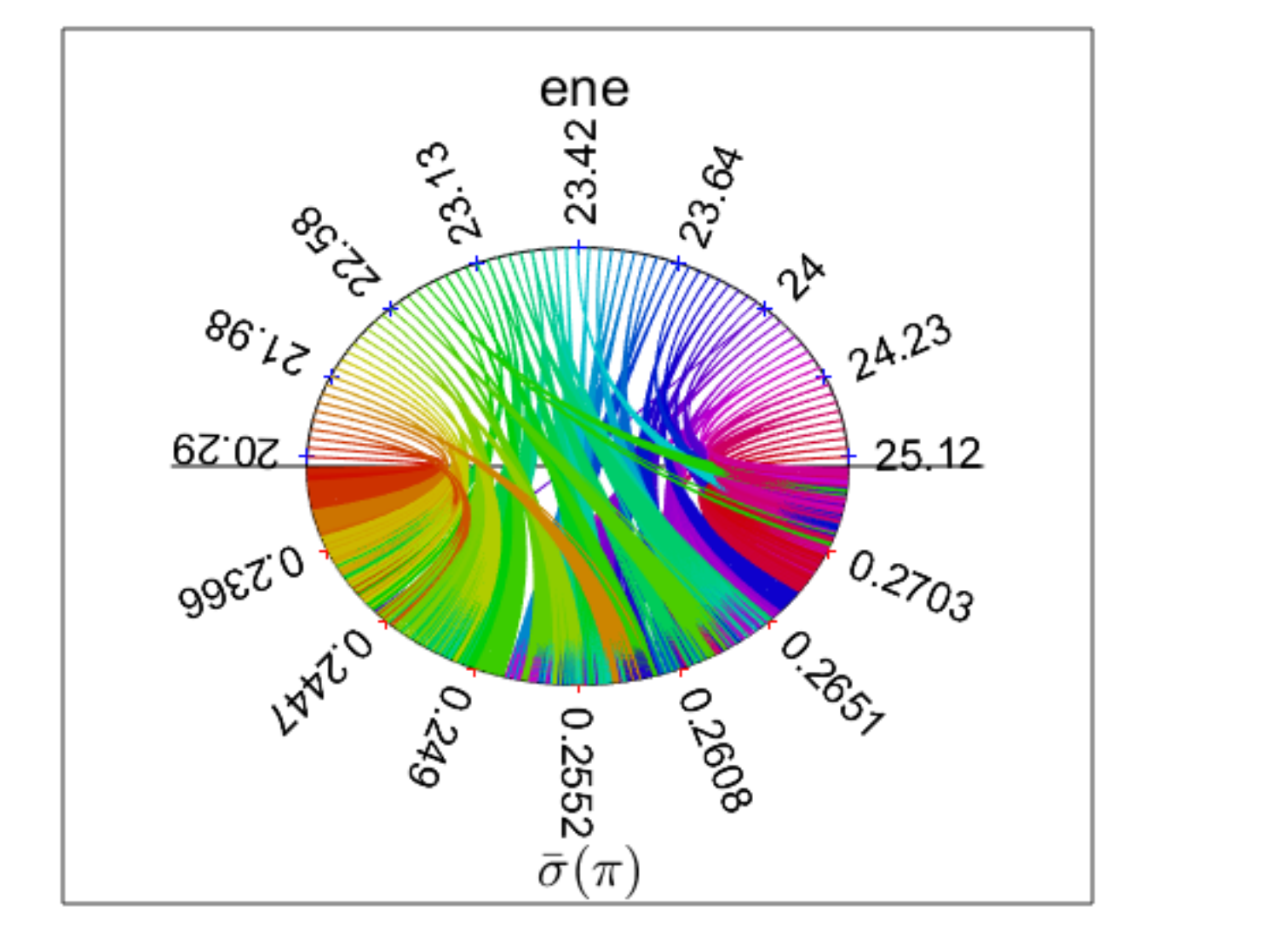}

\hspace{1cm} c) $\{N,d,c(\pi)\}=\{12,3,6\}$ \hspace{3cm} d)  $\{N,d,c(\pi)\}=\{12,8,4\}$

\includegraphics[trim = 0mm 0mm 0mm 0mm,clip,width=8.25cm, height=4.1cm]{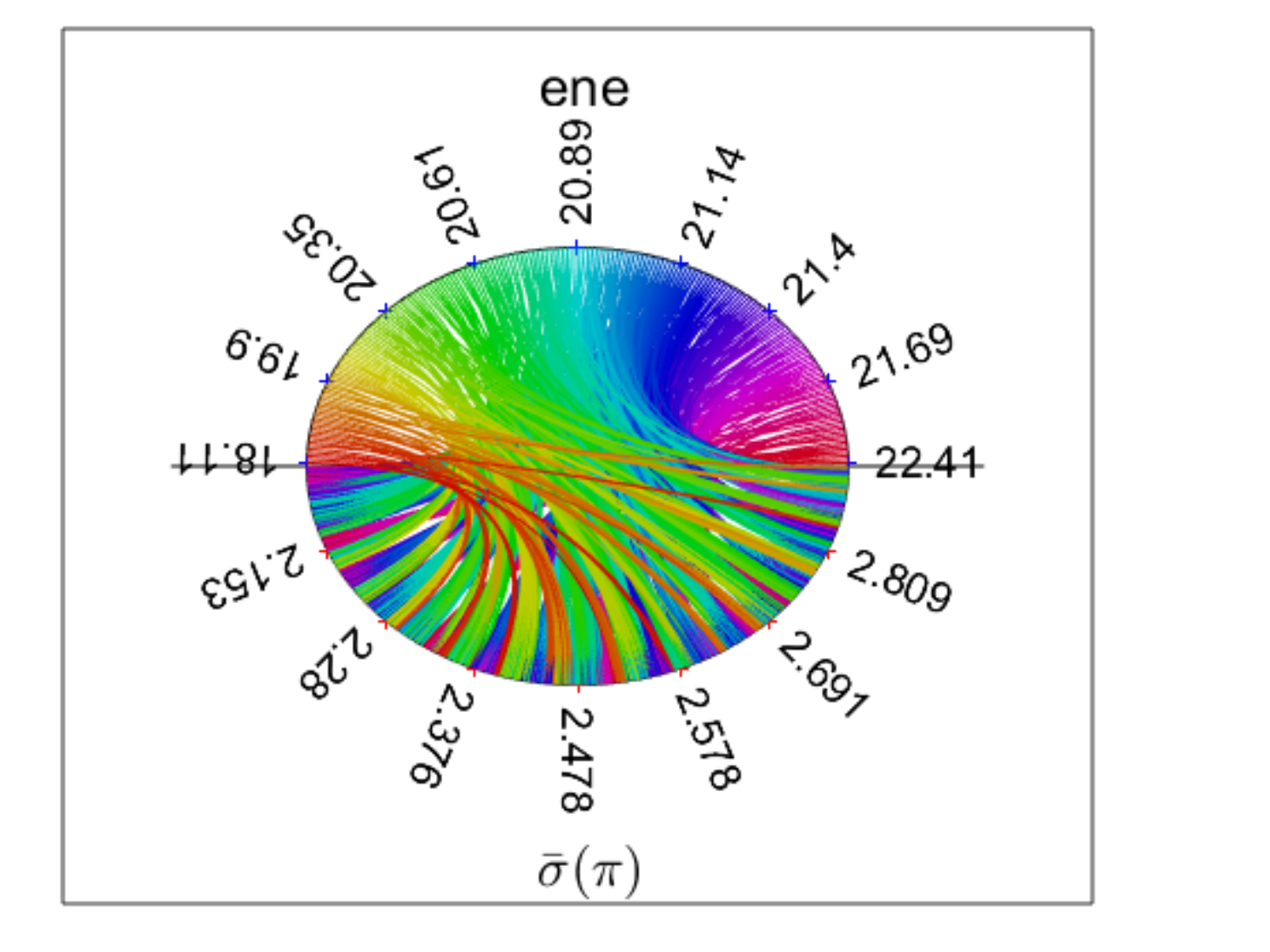}
\includegraphics[trim = 0mm 0mm 0mm 0mm,clip,width=8.25cm, height=4.1cm]{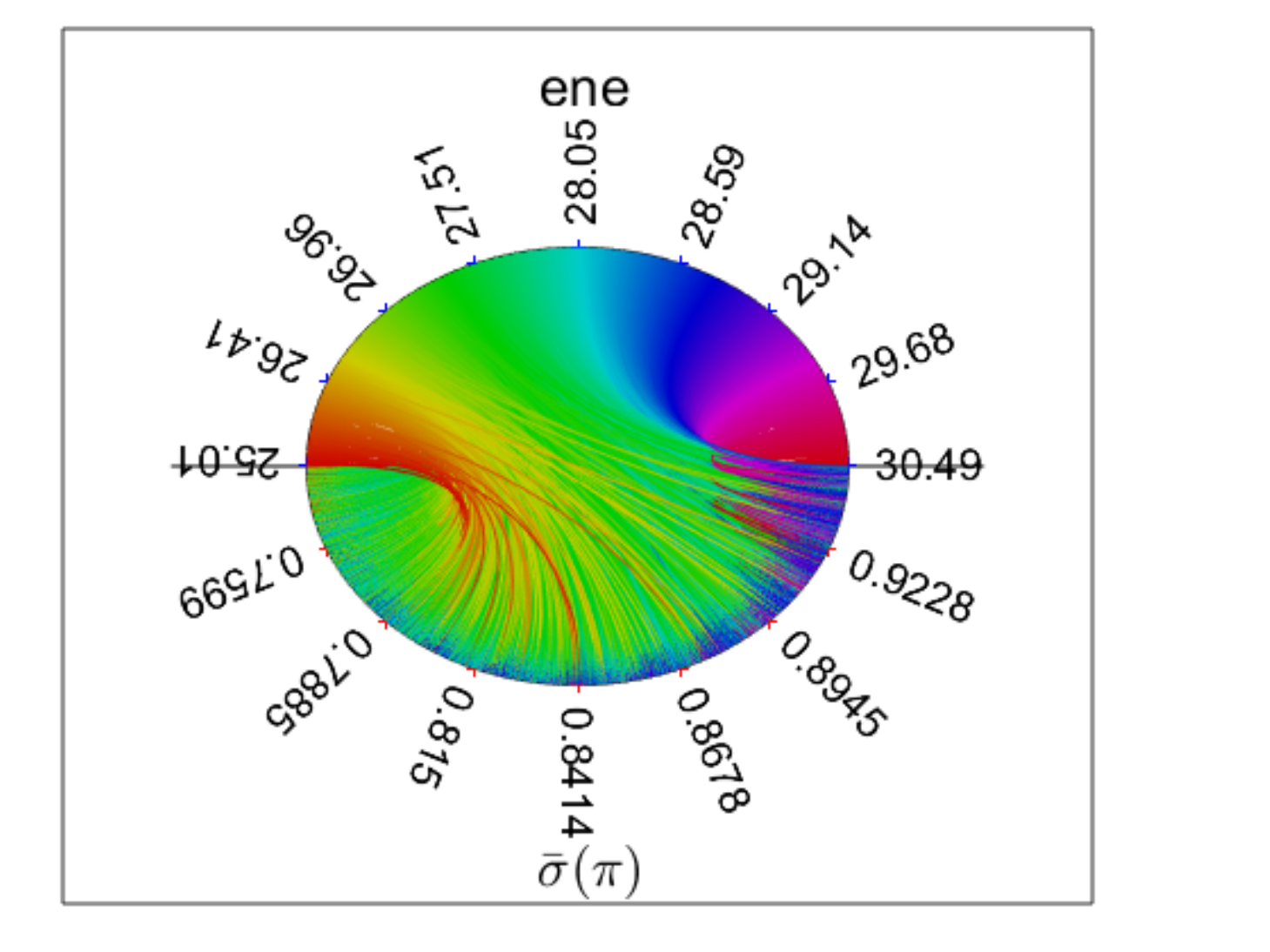}

\hspace{1cm} e) $\{N,d,c(\pi)\}=\{14,3,8\}$ \hspace{3cm} f)  $\{N,d,c(\pi)\}=\{14,7,7\}$

\caption{Schemaballs for $N=\{10,12,14\}$,  which also show connections between the distance measure $\bar{\sigma}$ and graph energy $\text{ene}$ similar to $N=8$ in Fig. \ref{fig:schemball_gm}, but the number of intervals and the fragmentation increases. }
\label{fig:schemball}
\end{figure}

\begin{figure}[h]
\includegraphics[trim = 0mm 0mm 0mm 0mm,clip,width=8.25cm, height=5.9cm]{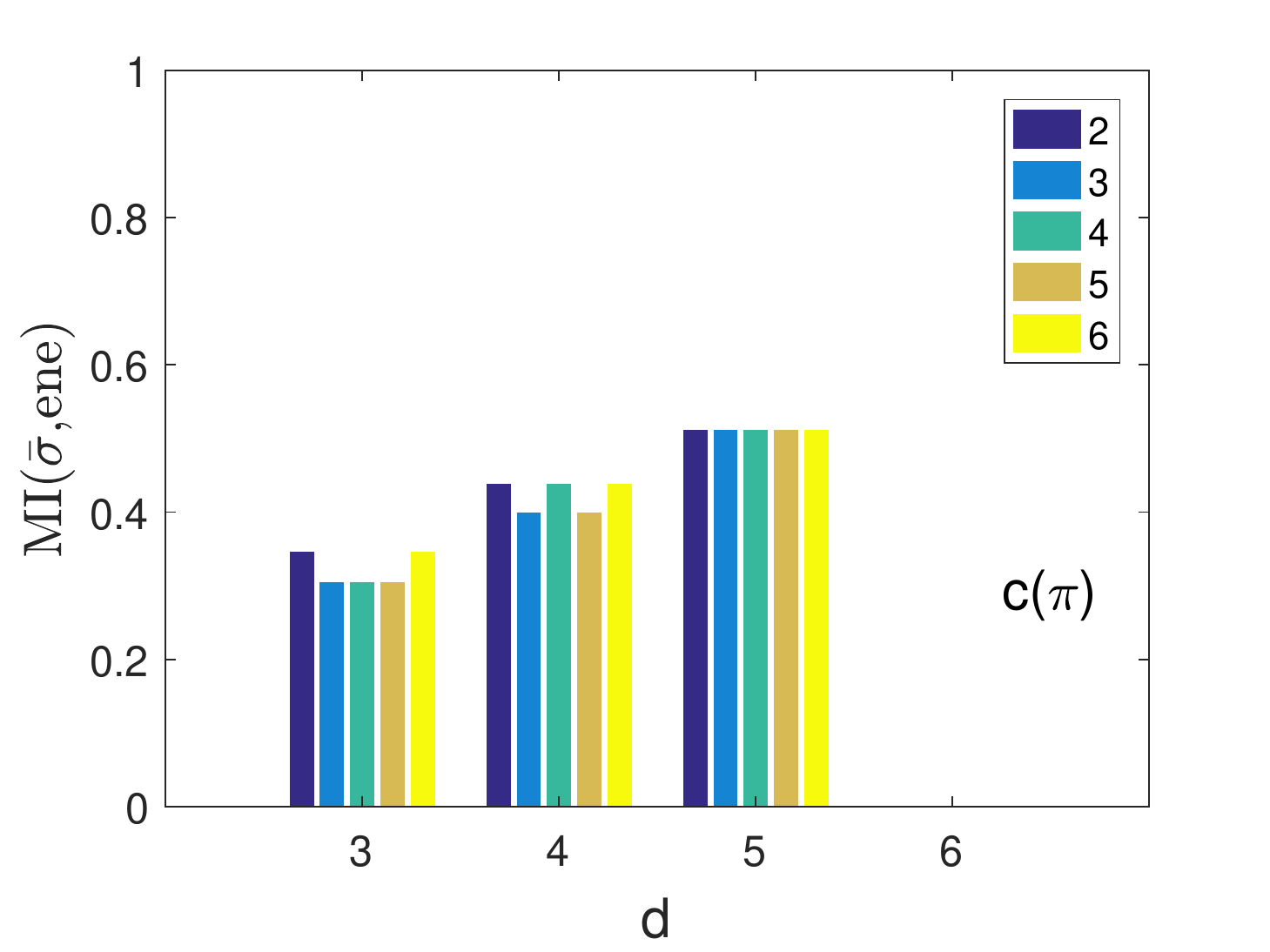}
\includegraphics[trim = 0mm 0mm 0mm 0mm,clip,width=8.25cm, height=5.9cm]{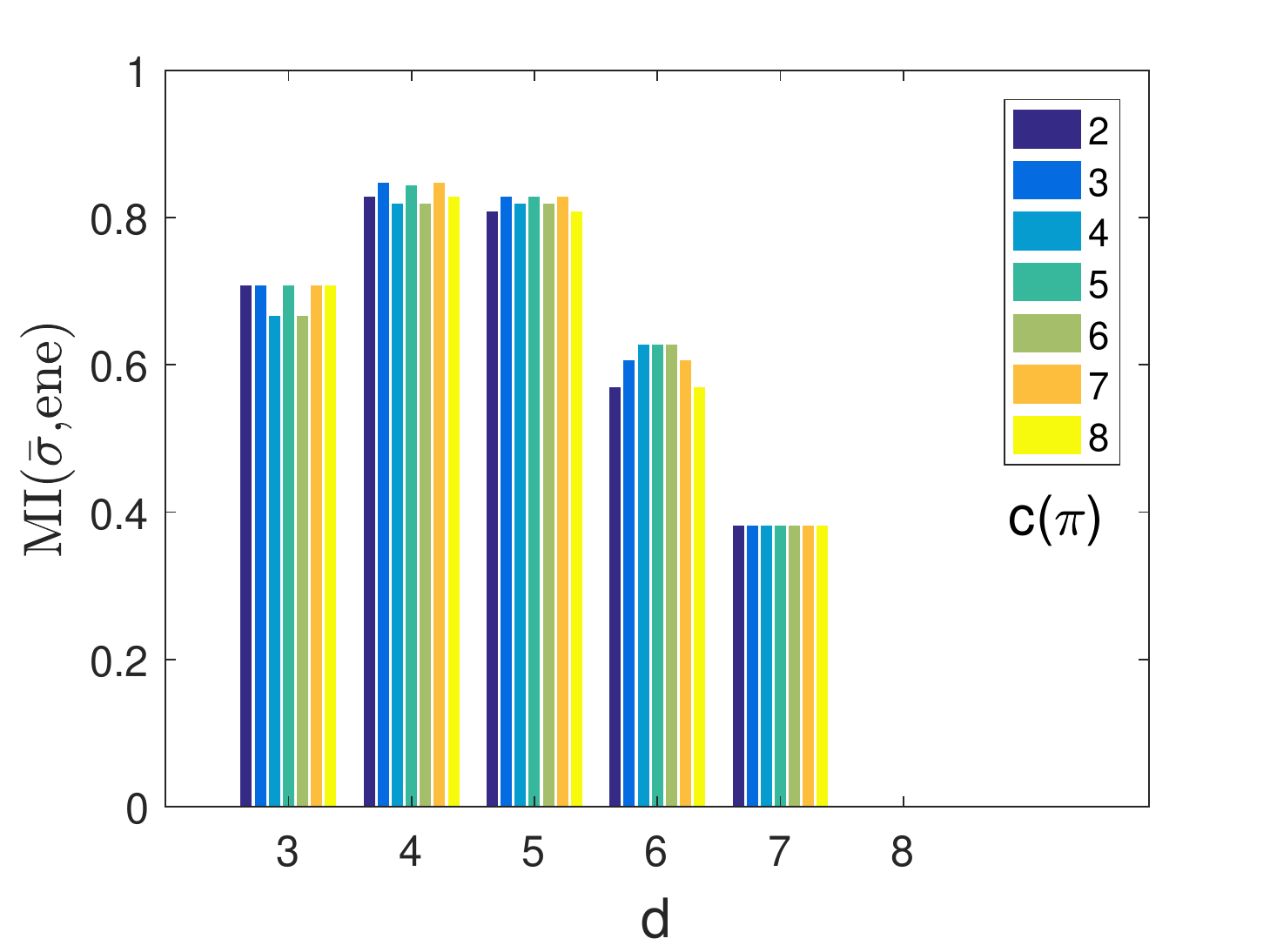}

\hspace{1cm} a) $N=8$ \hspace{7cm} b) $N=10$

\includegraphics[trim = 0mm 0mm 0mm 0mm,clip,width=8.25cm, height=5.9cm]{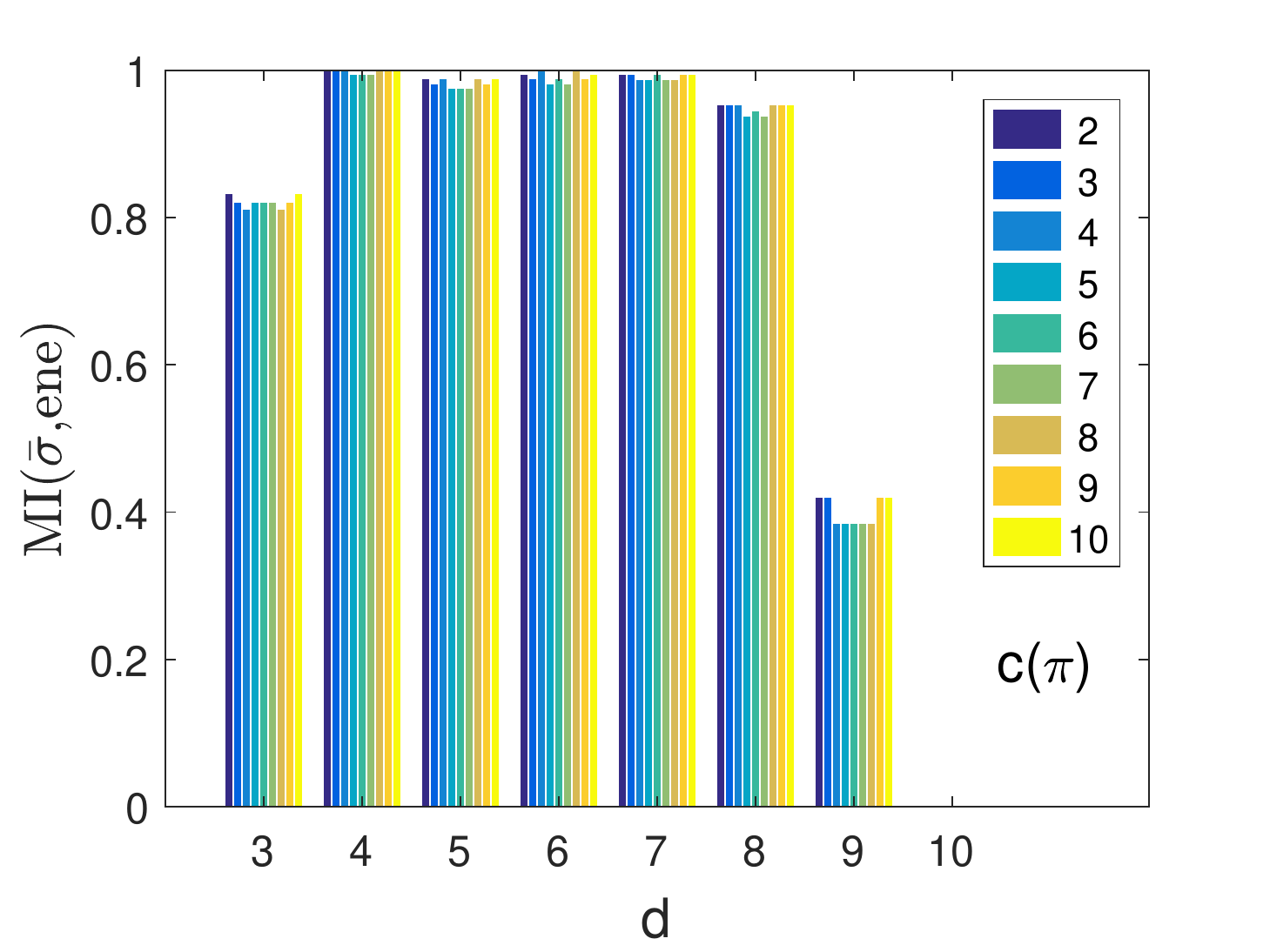}
\includegraphics[trim = 0mm 0mm 0mm 0mm,clip,width=8.25cm, height=5.9cm]{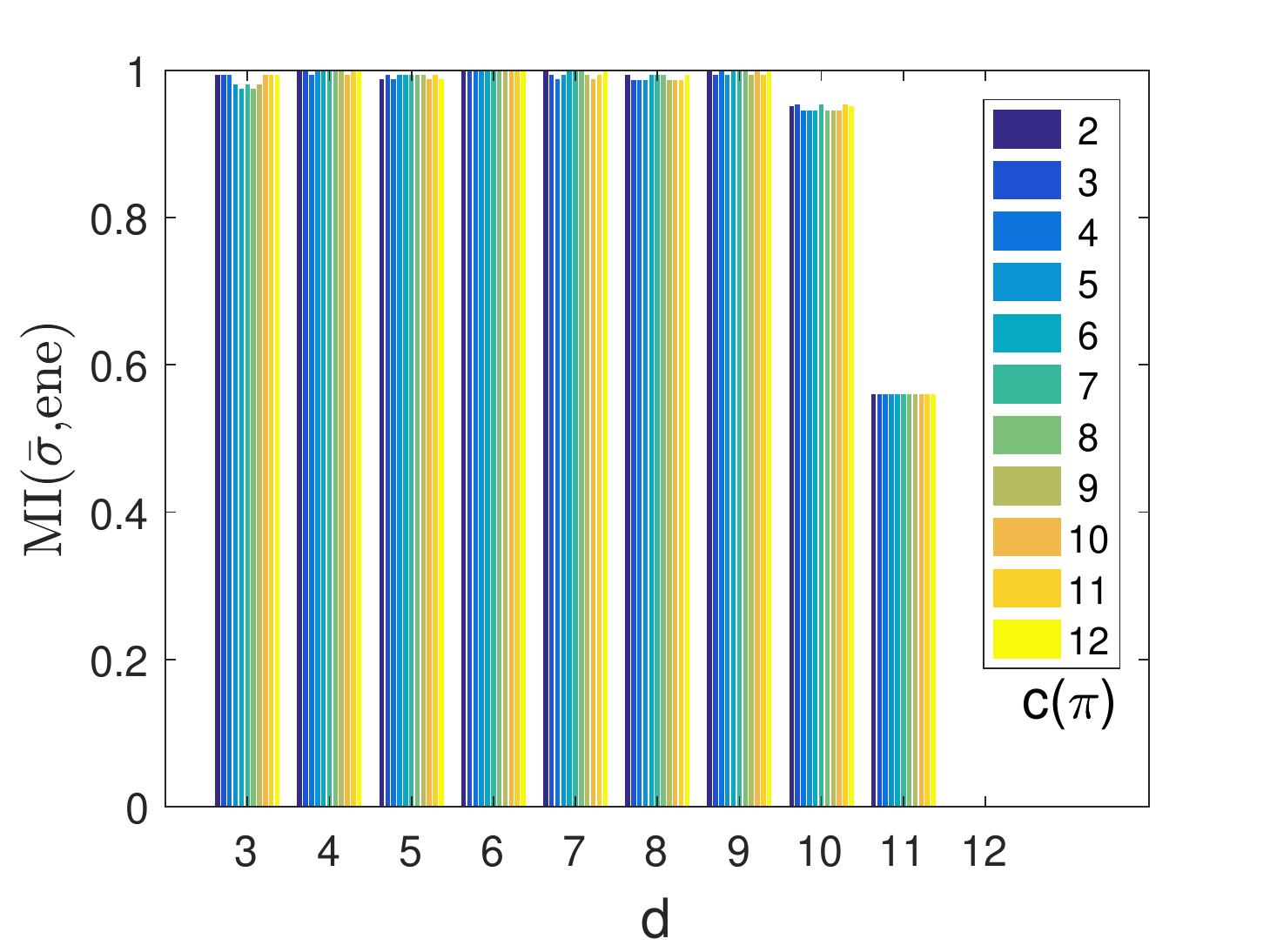}

\hspace{1cm} c) $N=12$ \hspace{7cm} d)  $N=14$

\caption{Normalized mutual information between the distance measure $\bar{\sigma}$ and graph energy $\text{ene}$ for $N$ and $3 \leq d \
\leq N-3$ and $2 \leq c(\pi) \leq N-2$. We generally see high values of $\text{MI}(\bar{\sigma},\text{ene})$ which indicate a strong dependency between the quantities. A major exception is $N=8$, but also for $d=N-3$ the dependency is weaker.  }
\label{fig:MI_energy}
\end{figure} 

\begin{figure}[tb]
\includegraphics[trim = 0mm 0mm 0mm 0mm,clip,width=8.25cm, height=5.9cm]{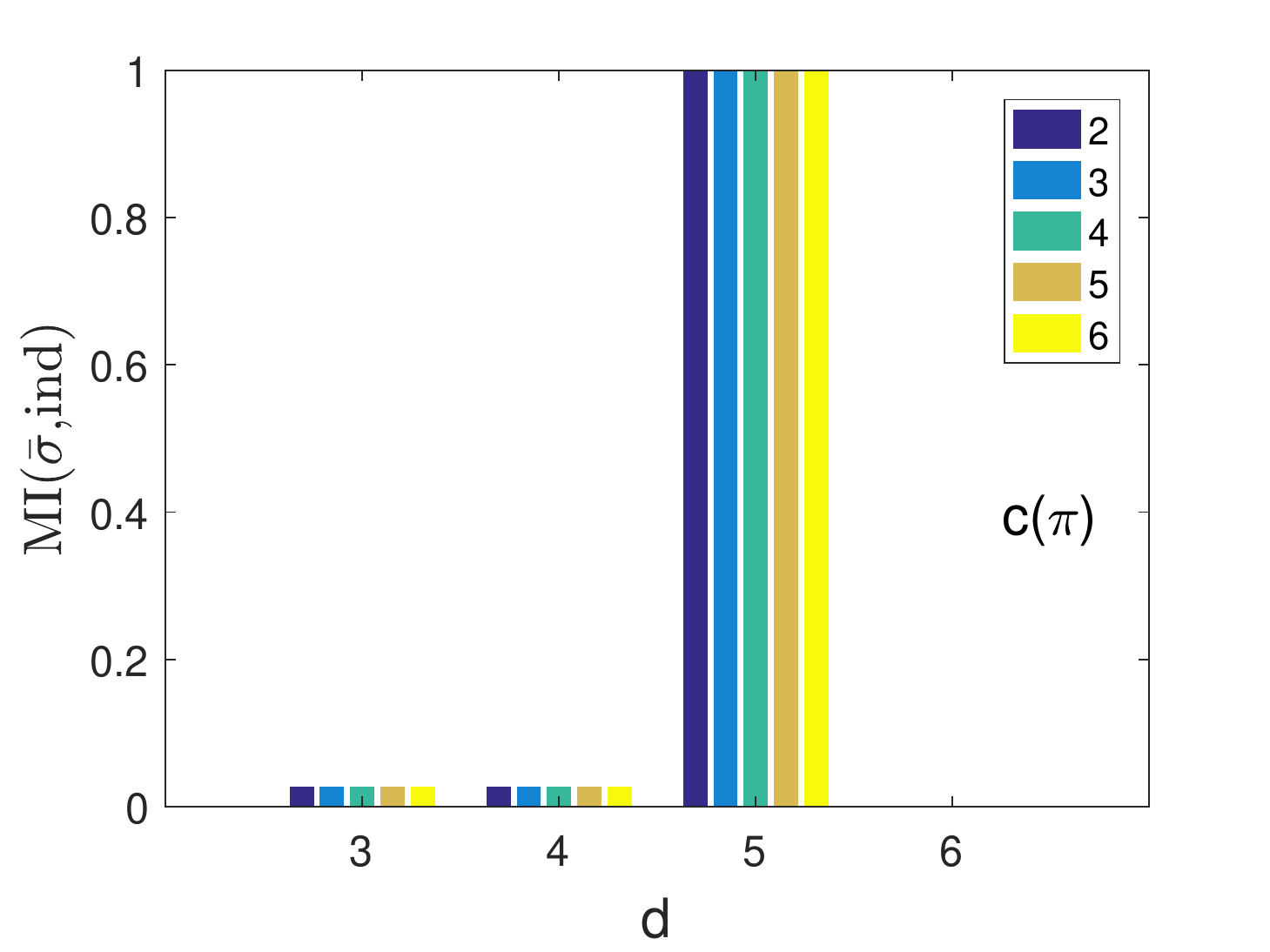}
\includegraphics[trim = 0mm 0mm 0mm 0mm,clip,width=8.25cm, height=5.9cm]{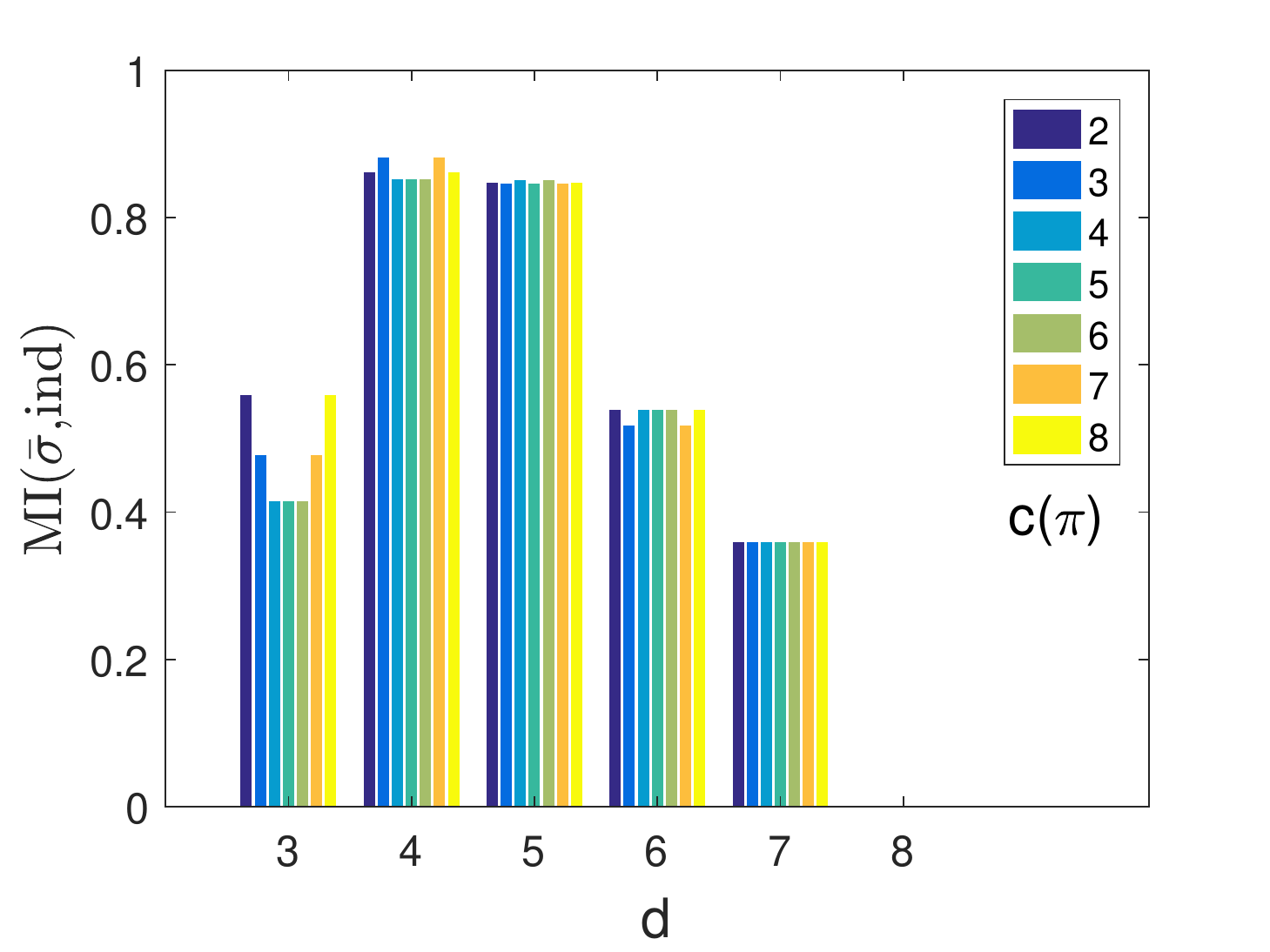}

\hspace{1cm} a) $N=8$ \hspace{7cm} b) $N=10$

\includegraphics[trim = 0mm 0mm 0mm 0mm,clip,width=8.25cm, height=5.9cm]{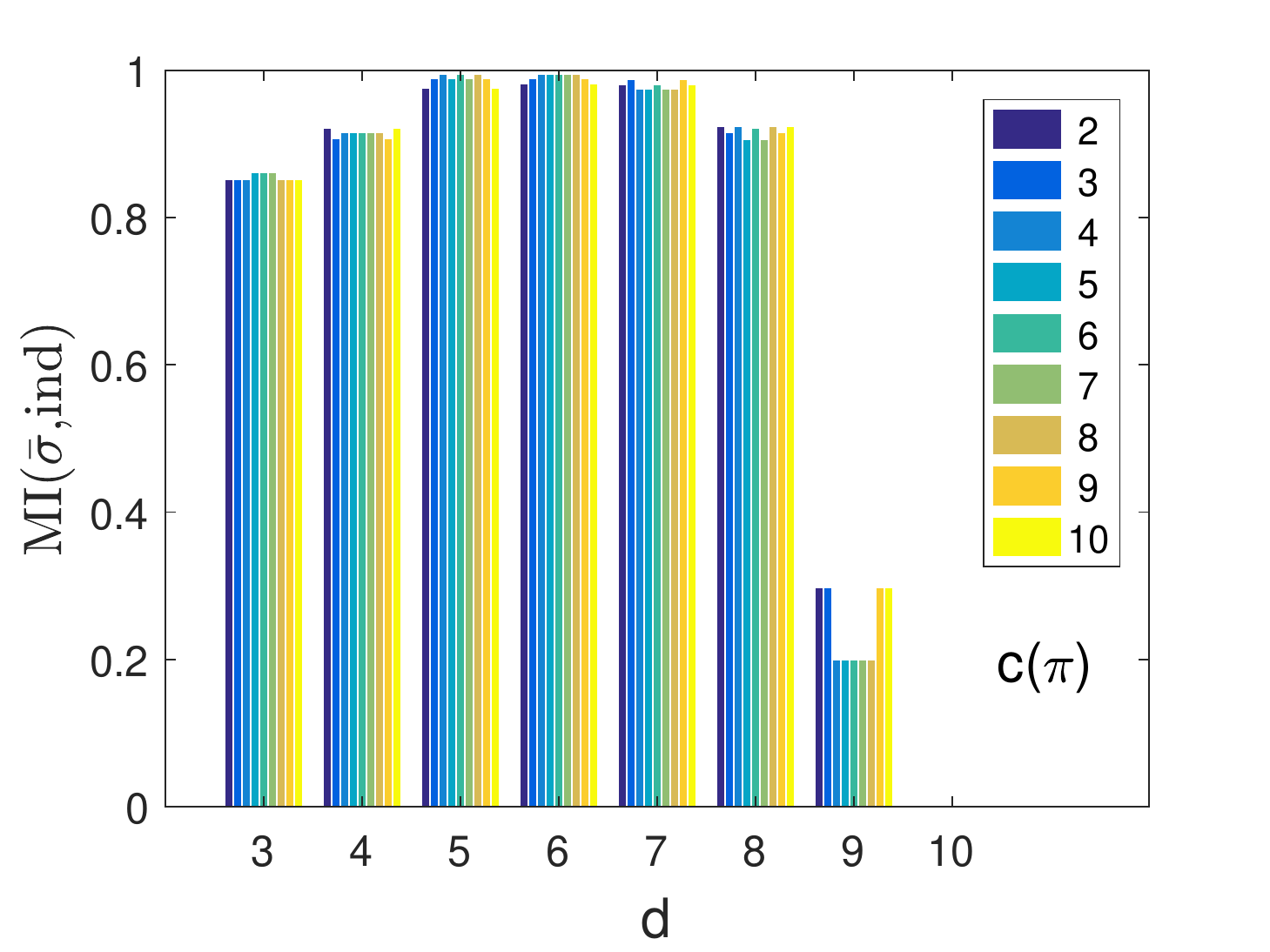}
\includegraphics[trim = 0mm 0mm 0mm 0mm,clip,width=8.25cm, height=5.9cm]{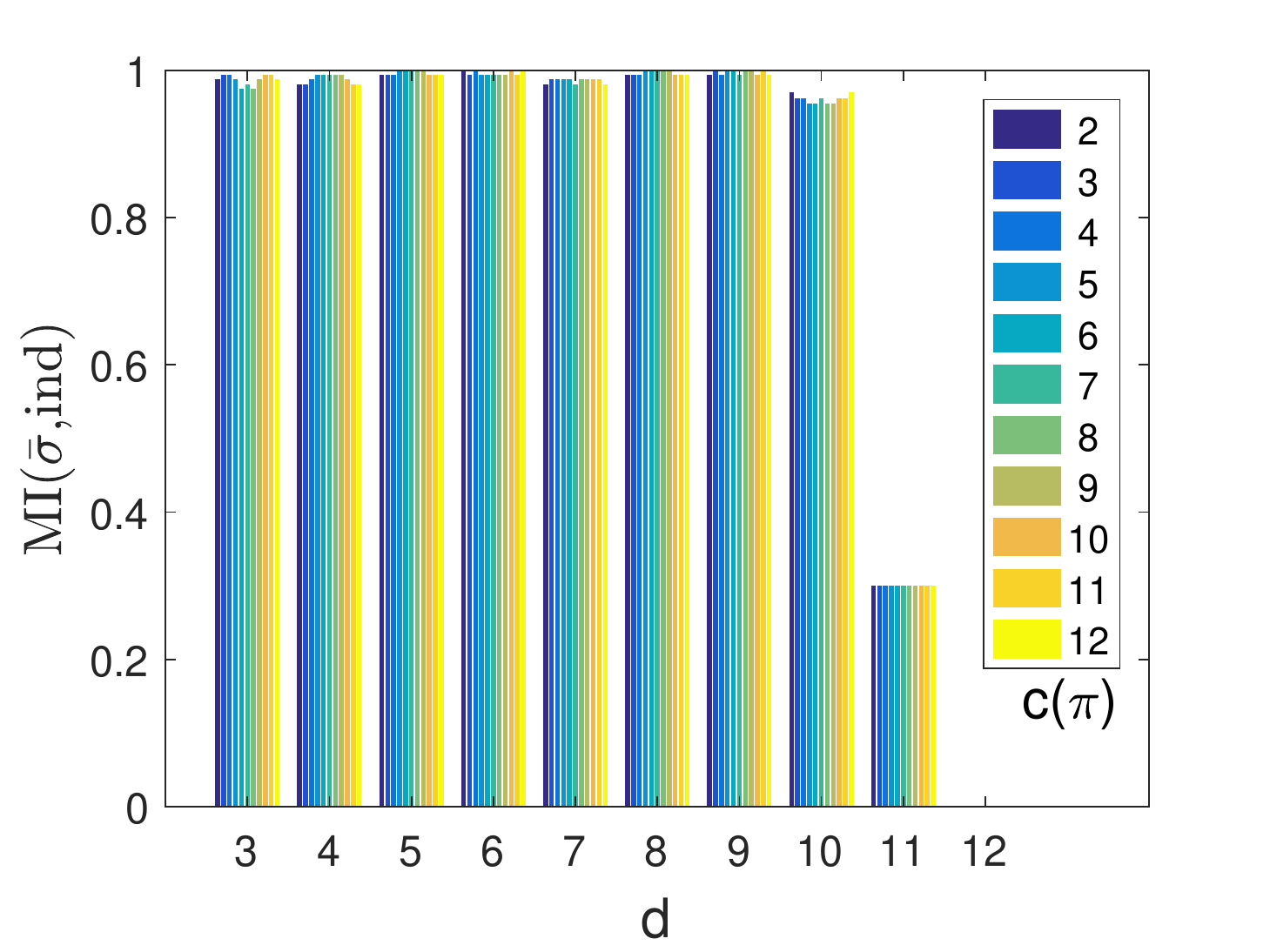}

\hspace{1cm} c) $N=12$ \hspace{7cm} d)  $N=14$

\caption{Normalized mutual information between the distance measure $\bar{\sigma}$ and independence number $\text{ind}$ for $N$ and $3 \leq d \
\leq N-3$ and $2 \leq c(\pi) \leq N-2$. We also see rather high values of $\text{MI}(\bar{\sigma},\text{ind})$. The exceptions found for  $\text{MI}(\bar{\sigma},\text{ene})$ apply likewise. }
\label{fig:MI_ind}
\end{figure}

A possible visualization of the relationships between the distance $\bar{\sigma}(\pi)$ and spectral graph measures is based on schemaballs,~\cite {krz04}.    We again sort $\bar{\sigma}(\pi)$  
according to the same number of cooperators $c(\pi)$ to obtain a distribution over interaction networks  for the number of players, coplayers and cooperators,   $\{N,d,c(\pi)\}$. In the schemaballs we draw Bezier curves connecting the values of the spectral graph measures of the interaction matrix $A_I$ with the values of $\bar{\sigma}(\pi)$ belonging to this interaction network, see Figs. \ref{fig:schemball_gm} and \ref{fig:schemball}. The upper half of the ball gives the sorted values of the graph measure, the lower half  the sorted values of the distance measure $\bar{\sigma}(\pi)$. As both the spectral graph measures as well as the distance measure $\bar{\sigma}$ are sparsely distributions, the gaps between values are cut out to have left over only unique values.  Thus, in the balls the values are distanced such that only unique values are placed equidistant. 
 The curves are colored in such a way that the same value of the graph measure has only connections of the same color. The color are selected  equidistant from a HSV color circle. 
As an example, Fig. \ref{fig:schemball_gm} shows the schemaballs of $\bar{\sigma}$ and the spectral graph measures according to
(\ref{eq:ene})--(\ref{eq:tri}) for $\{N,d,c(\pi)\}=\{8,4,3\}$. We see that certain values of the graph measures imply certain distance measures $\bar{\sigma}(\pi)$. For instance, in Fig. \ref{fig:schemball_gm}a showing the relation between graph energy (\ref{eq:ene}) and $\bar{\sigma(\pi)}$, we see that $\text{ene}=12.29$ implies the interval $0.28 \leq \bar{\sigma}(\pi) \leq 0.31$. Note that between the spectral graph measures and the distance measure $\bar{\sigma}$ there is generally no simple dependence which could be expressed as a correlation implying a linear (or piece--wise linear) relationship between the variables. 
For instance, the schemaball for algebraic connectivity, Fig.   \ref{fig:schemball_gm}c, shows that $\text{con}=2.586$  
is related to two intervals at $\bar{\sigma} \approx 0.36$ and  at $\bar{\sigma} \approx 0.29$, but not to the intermediate interval $\bar{\sigma} \approx 0.32$. In some ways, the scale of $\bar{\sigma}$ is fragmented, while each of the fragments is connected to a value of the graph measure. 
On the other hand, between $\text{tri}$ and $\bar{\sigma}$ there seems to be (at least for the example of  $\{N,d,c(\pi)\}=\{8,4,3\}$) an almost linear relation, see Fig. \ref{fig:schemball_gm}d.
Furthermore, it can be seen by comparing the schemaballs that different spectral graph measures relate differently to the distance measure $\bar{\sigma}$.
For instance $\text{ene}$ and $\text{tri}$ as well as $\text{con}$ and $\text{ind}$ appear to be more related than $\text{ene}$ and $\text{con}$ or  $\text{ind}$ and $\text{tri}$. Looking at the results in the schemaballs it may be interesting to ask if there is a link to the graph--theoretical interpretation of the spectral measures, see
 \ref{sec:meas}. Methods. 
For the example in Fig. \ref{fig:schemball_gm}, it can be seen that small values of the independence number $\text{ind}$, which indicate a small independent set of vertices, small values of the algebraic connectivity $\text{con}$, which characterize more path--like graphs with low girth, and large values of the triangular number $\text{tri}$, which means a larger number of triangles in the graph, all imply large values of the distance measure $\bar{\sigma}$. It might be interesting to explore these links in detail in future work. 

Fig. \ref{fig:schemball} shows schemaballs for $N=\{10,12,14\}$. In principle, we can see that the relationships between the  graph energy $\text{ene}$ and intervals of the distance measure $\bar{\sigma}$ can be found similarly to $N=8$. Most importantly, it seems that the substantial amount of structure in the data is preserved. However, as the number of unique values of the related quantities increases, also the number of the intervals goes up and also the degree of fragmentation. 
Clearly, for $N=14$ the usefulness of schemaballs as tool for visualizing the relationships approaches its limits.

To go beyond visualization we need  a measures of the strength of the relationships between structure coefficients and spectral graphs measures. As argued using the schemaballs in Figs.  \ref{fig:schemball_gm} and \ref{fig:schemball}, most likely there are no linear or piece--wise linear relationships which excludes correlation coefficients mostly accounting for the strength of linear relationships. Therefore, we measure the strength of the relationships by normalized mutual information ($\text{MI}$),~\cite{cov91}, between the distribution $\bar{\sigma}$ and distributions of the graph measures graph energy $\text{ene}$  and independence number $\text{ind}$. Roughly speaking, mutual information of two series measures the amount of information that can be obtained from one through the other, see, e.g.~\cite{mis12,nem17} for application in biological sciences.  Thus, mutual information can be interpreted as a generalized measure of the dependency which does not assume any specific algebraic form of the relationship. The mutual information considered here is normalize to the interval $[0,1]$. Similarly to correlation coefficients, $\text{MI}=0$ means that the data sets tested are statistically independent. 

The normalized mutual information $\text{MI}(\bar{\sigma},\text{ene})$ and $\text{MI}(\bar{\sigma},\text{ind})$ of the distance measure $\bar{\sigma}$ and the spectral graph measures  is given in Figs. \ref{fig:MI_energy} and \ref{fig:MI_ind}. Only for $3 \leq d \leq N-3$ we get $\text{MI}>0$, as for $d=1$ and $d=N-1$, there is only one value of $\sigma(\pi)$ and thus also only one value of $\bar{\sigma}$, while for $d=2$ and $d=N-2$ the spectral graph measures considered here have the same value for all $2$--regular and $(N-2)$--regular graphs. Also, for $c(\pi)=1$ and $c(\pi)=N-1$, the mutual information $\text{MI}=0$, as for a single cooperator (and a single defector), there is no variance in $\bar{\sigma}$ over interaction networks.  The results for the remaining $d$ and $c(\pi)$ show that $\bar{\sigma}$ and $\{\text{ene},\text{ind} \}$ have a considerable degree of dependency, which increases with $N$. For $N=14$ we have almost the maximal value of $\text{MI}=1$. Also, there are just small differences over  the number of cooperators $c(\pi)$ (except $N=8$), which confirms results obtained by visually inspecting the schemaball in Figs.  \ref{fig:schemball_gm} and \ref{fig:schemball}. Also, it can be seen that the dependency by mutual information is not very different from graph energy  to independence number (again except $n=8$), which similarly can also be obtained if considering the other two spectral graph measures $\{\text{con},\text{tri} \}$.    The results in  Figs. \ref{fig:MI_energy} and \ref{fig:MI_ind}  show clearly that there are strong relationships between the structure coefficients obtained for different interaction networks and spectral graph measure of these networks.

\section{Discussion} \label{sec:dis}

\begin{figure}[tb]
\includegraphics[trim = 0mm 0mm 0mm 0mm,clip,width=8.25cm, height=5.9cm]{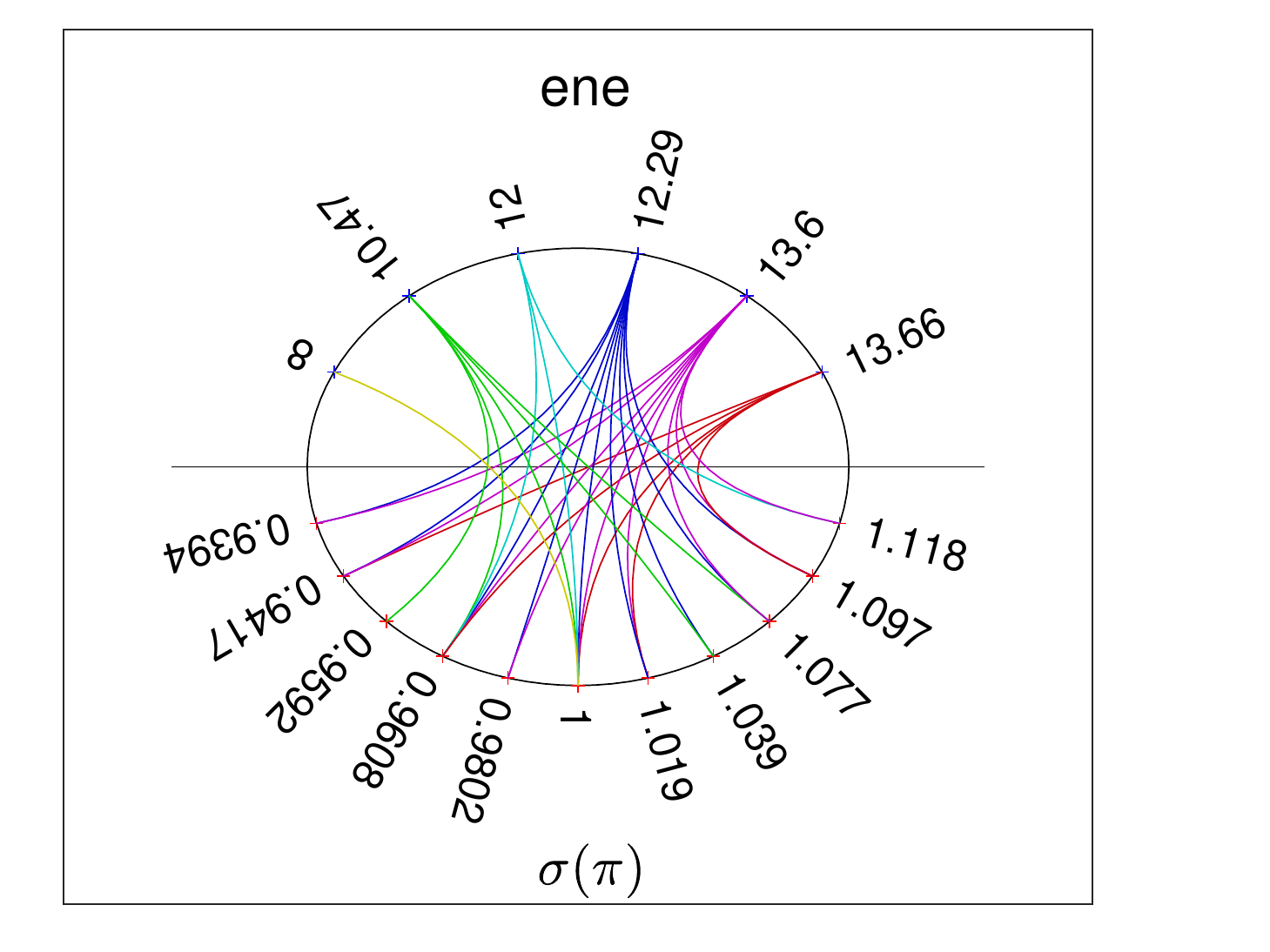}
\includegraphics[trim = 0mm 0mm 0mm 0mm,clip,width=8.25cm, height=5.9cm]{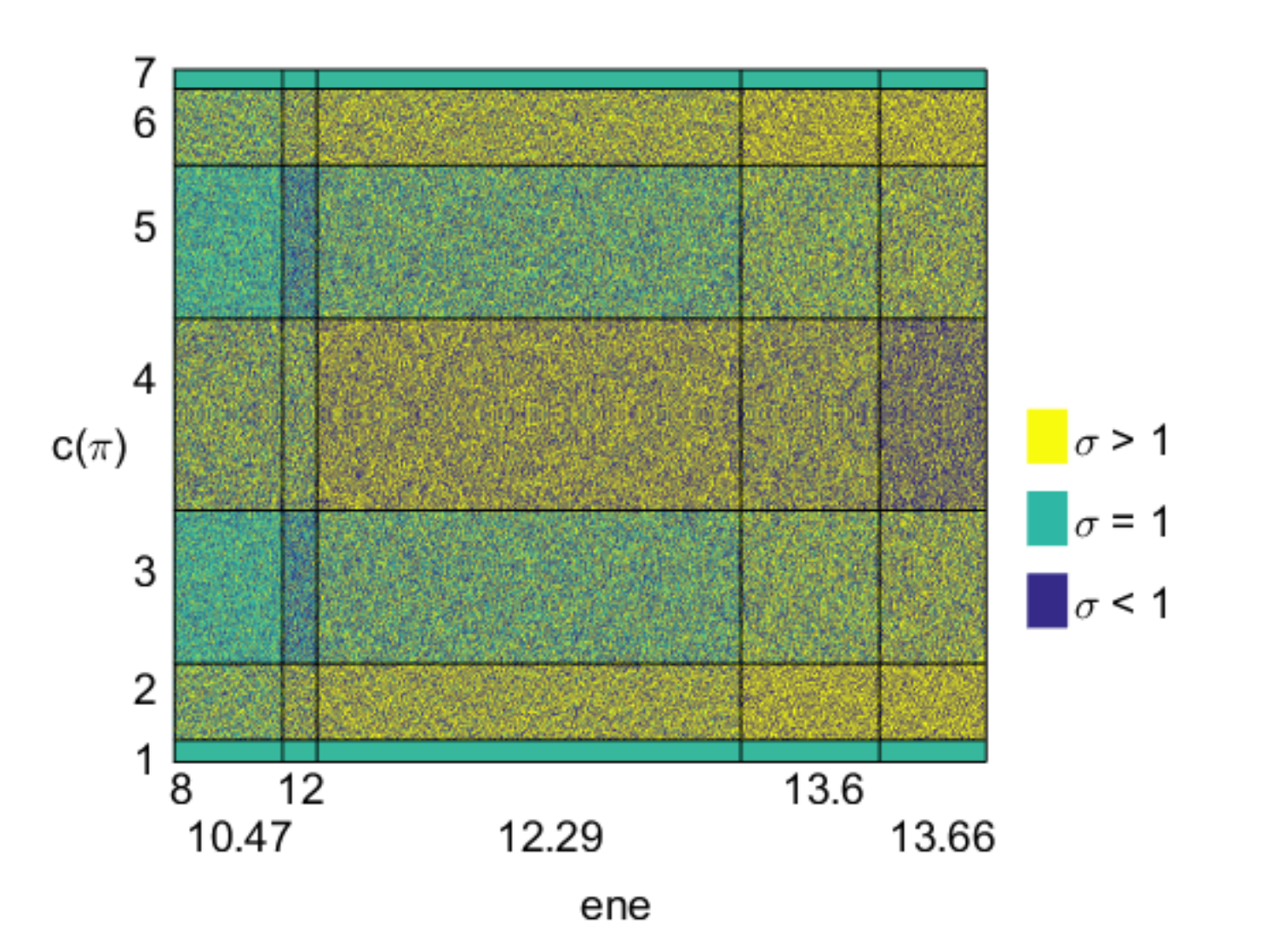}

\hspace{1cm} a)  \hspace{7cm} b)

\caption{a) Schemaball for graph energy and the structure coefficient $\sigma(\pi)$, which combines the schemaball in Fig. \ref{fig:schemball_gm}a with the violin plot in Fig. \ref{fig:shape2}a, each for $N=8$, $d=4$ and $c(\pi)=3$. The structure coefficient $\sigma(\pi)$ has values $0.9394 \leq \sigma(\pi) \leq 1.1180$, but unlike the results in Fig. \ref{fig:schemball_gm}a, all graph measures $\text{ene}$ are connected to all unique values of $\sigma(\pi)$, except $\text{ene}=8$ is only connected to $\sigma(\pi)=1$. Note that the  $11$ values of $\sigma(\pi)$ in the lower half of the schemaball correspond to  $\#_\sigma=11$ in Fig. \ref{fig:sigma_abs_8_14}a for $d=4$ and $c(\pi)=3$. b)  The sign of the distance to $\sigma(\pi)=1$  as different colors over configurations and graph energy, also for $N=8$ and $d=4$. The configurations are sorted according to the number of cooperators $c(\pi)$, while the graph measure $\text{ene}$ is sorted according to unique values of $\text{ene}$.  There is a kind of tricolor ``Cantorian dust'' with intermixed all three signs of the distance to $\sigma(\pi)=1$ for almost all fields of same graph energy and same number of cooperators.  }
\label{fig:sign}
\end{figure}

In \cite{chen16} is was shown that  for  $2 \times 2$ games with $N$ players on $d$--regular graphs with payoff matrix (\ref{eq:payoff}) strategy $C$ is favored over $D$ if \begin{equation} \sigma(\pi)> \frac{c-b}{a-d}. \label{eq:cond} \end{equation}
The  condition (\ref{eq:cond}) generalizes the same condition with $\sigma(\pi)=\sigma=\frac{(d+1)N-4d}{(d-1)N}$,
which applies to a single cooperator (and a single defector),~\cite{taylor07,lehmann07,tarnita09}. It is now valid for any configuration $\pi$ of cooperators and defectors and for any interaction network modeled by a $d$--regular graph.

In~\ref{sec:results}.~Results experimental findings are reported showing that and how the structure coefficients $\sigma(\pi)$ vary over interaction networks $A_I$. It is further demonstrated that the variability of the structure coefficients can be linked to varying values of spectral graph measures obtained for the interaction networks. 
Thus, if we were to insert into Eq. (\ref{eq:cond}) the distributions over interaction networks of $\sigma(\pi)$ as discussed above, we may obtain different ratios $\frac{c-b}{a-d}$ that satisfy the condition  (\ref{eq:cond}). This explains how different parametrizations of the payoff matrix  (\ref{eq:payoff}) may produce different game dynamics (and possibly varying fixation properties) for the same structure coefficient, that is for the same arrangement of cooperators and defectors on the same evolutionary graph.  To give an example of this effect, consider the donation game (also simplified Prisoner's Dilemma), which is
a special case of the payoff matrix (\ref{eq:payoff}) with
\begin{equation} 
\bordermatrix{~ & C_j & D_j \cr
                  C_i & b-c & -c \cr
                  D_i & b & 0 \cr}, \label{eq:donation}
\end{equation}
where $b$ is interpreted as benefit and $c$ as cost,~\cite{allen14,sha12}. For the donation game, the  condition  (\ref{eq:cond}) simplifies to the critical benefit--to--cost ratio,~\cite{taylor07,tarnita09,chen16} \begin{equation} \left( \frac{b}{c} \right)^* = \frac{\sigma(\pi)+1}{\sigma(\pi)-1}. \label{eq:bene_cost}\end{equation}
Condition (\ref{eq:bene_cost}) implies that  
properties of the distribution of the structure coefficients over interaction networks and the relationships between the structure coefficients and graph measures can be immediately 
transferred to benefit--to--cost ratios.
From condition (\ref{eq:bene_cost})  we further see that
if $\sigma(\pi)>1$, cooperation is favored as benefit exceeds cost.
If $\sigma(\pi)=1$, then $\left( \frac{b}{c} \right)^* \rightarrow \infty$, which means cooperation cannot be favored, while for $\sigma<1$, we have a spite game in which  player pay cost to harm another. The experimental findings given above show that there are cases where it depends on the interaction network and the configuration whether we have $\sigma(\pi)>1$ or $\sigma(\pi)=1$ or $\sigma(\pi)<1$, even if the number of cooperators is the same. See for instance Fig. \ref{fig:shape2} which shows this result for $d=N/2$ and any number of cooperators $c(\pi)$ with $2 \leq c(\pi) \leq N-2$.

This observation leads to the question of an optimal configuration and/or an optimal interaction network. For a donation game optimal means best promotion of cooperation, which is for a minimized critical benefit--to--cost ratio (\ref{eq:bene_cost}), or a maximized $\sigma(\pi)$. An interesting result of the numerical finding reported above is that there is not an optimal configuration or an optimal network, but there are combinations of configuration and network that are better than others. In other words, the question of optimality 
cannot be answered for configurations independent from networks, and vice versa. To illustrate this point, let us take another look at the findings given above showing the distributions of $\sigma(\pi)$ over interaction networks, see Fig.  \ref{fig:sign}. Fig. \ref{fig:sign}a shows a schemaball as in Figs. \ref{fig:schemball_gm} and \ref{fig:schemball} for $N=8$, $d=4$ and $c(\pi)=3$,
but with the distribution of the structure coefficient $\sigma(\pi)$ instead of the distance $\bar{\sigma}$.
In fact, Fig. \ref{fig:sign}a combines the schemaball in Fig. \ref{fig:schemball_gm}a with the violin plot in Fig. \ref{fig:shape2}a, each for $N=8$ and $d=4$. The structure coefficient $\sigma(\pi)$ has values $0.9394 \leq \sigma(\pi) \leq 1.1180$, but unlike the results in Fig. \ref{fig:schemball_gm}a, all graph measures $\text{ene}$ are connected to all unique values of $\sigma(\pi)$, except for $\text{ene}=8$ which is only connected to $\sigma(\pi)=1$. This result is general for all tested $d$ and $A_I$ with $N=\{8,10,12,14 \}$. 
Certain values of the graph energy (or any of the other spectral graph measures tested) of an interaction network  alone do not specify a structure coefficient $\sigma(\pi)$ and thus game dynamics and fixation properties.

To explore this point a little further, see the results in Fig.    \ref{fig:sign}b, which shows the sign of the distance to $\sigma(\pi)=1$, ($\sigma(\pi)<1$, $\sigma(\pi)=1$, $\sigma(\pi)>1$) as different colors over configurations and graph measures, again for $N=8$ and $d=4$. The configurations are sorted according to the number of cooperators $c(\pi)$, while the graph measure $\text{ene}$ is sorted according to unique values of $\text{ene}$. As there are $7$ different numbers of cooperators $c(\pi)$ with $N=8$ players ($c(\pi)=0$ and $c(\pi)=8$ are absorbing) and $6$ different values of $\text{ene}$, we get $42$ different fields in Fig. \ref{fig:sign}b. (However, only $5$ different value of $\text{ene}$ are clearly visible as $\text{ene}=8$ is only a tiny portion with $19$ out of the $10.000$ tested networks.) There are two main observations to be made. A first is that for every field (except $c(\pi)=1$ and $c(\pi)=7$, which is for a single cooperator and a single defector), there is a kind of tricolor ``Cantorian dust'' with intermixed $\sigma(\pi)<1$, $\sigma(\pi)=1$ and $\sigma(\pi)>1$. The other observation is that the mix is not always the same. There are at least some fields with more values $\sigma(\pi)<1$, for instance $c(\pi)=4$ and $\text{ene}=13.66$ or more values of $\sigma=1$, for instance $c(\pi)=3$ and $c(\pi)=5$ and $\text{ene}=10.47$, and so on.  Similar results as in Fig. \ref{fig:sign}b can be obtained for other combinations of $N$ and $d$ as well. However,  only for $d=N/2$
 the critical value $\sigma(\pi)=1$ coincides with the generic structure coefficient $\sigma$ according to Eq. (\ref{eq:sigm_gen}), which intersects the distribution of $\sigma(\pi)$ over interaction networks.  Thus, to obtain the   ``Cantorian dust''  as in  Fig. \ref{fig:sign}b, the  sign of the distance should be calculated  against the threshold  $\sigma(\pi)=\sigma$. Admittedly, as $\sigma>0$ for $d<N/2$ and $\sigma>1$ for $d>N/2$, contrary to $d=N/2$ such an image would therefore not directly be related to the critical benefit--to--cost ratio. 
 All
 these findings once again support that game dynamics and fixation is generally defined by the interplay of configuration and properties of the interaction network. For interaction networks modeled as $d$--regular graphs the relevant network properties can be expressed as spectral graph measures, which may open up spectral analysis of evolutionary graphs.

Closely related to optimality is the
design problem which addresses the question of which configuration and which network of interaction should be taken to promote certain fixation properties, for instance the emergence of cooperation. As configurations describe on which vertices of the evolutionary graph there are cooperators and defectors, this question can be answered by an optimal initial configurations, from which the game starts. Again, designing the initial configuration cannot be undertaken without designing the interaction network. In both cases, doing so by enumeration is only possible for a small number of players. Even then
we could only be sure to have the ``optimal configuration'' if we checked all $\ell=2^N$ possible configurations.  In contrast, 
whether or not we have found the  ``optimal network''  is far from being clear, even for the rather small number of players $N=\{8,10,12,14\}$ as considered here. It is not yet known how many different $d$--regular graphs on $N$ vertices
there are for a given $N$ and $3 \leq d \leq N-2$. There is an asymptotic approximation for the total number of $d$--regular graphs $\mathcal{L}_d(N)$ that assigns for $d=\hbox{o}(\sqrt{N})$ and $dN$ even a magnitude    $\mathcal{L}_d(N)=\mathcal{O}(N^N)$, which is rather huge,~\cite{worm99,rich17}. The number of different interaction networks grows even faster than the number of configurations.  On the other hand, the results given above also show that for $N$ getting large, the range of the distribution over interaction networks gets smaller, shrinking to $\sigma$ for $N\rightarrow \infty$. Thus, the design problem appears to be most interesting for intermediate values of $N$ such that with some likelihood a numerical treatment might remain feasible.

Moreover, the results given in this paper suggest another approach to the design problem. At least for  $N=\{8,10,12,14\}$, interaction networks with certain spectral graph measures propose to lead to network--and--configuration pairs with enhanced likelihood for certain structure coefficients. Thus, we may potentially prescribe interaction networks with particular spectral graph measures and hence reduce the design space of interaction networks.
Naturally, this would require a better understanding of the relationships between the structure coefficients $\sigma(\pi)$, the distance measure $\bar{\sigma}$ and the spectral graph measures. 
Mutual information has shown to be sufficient to establish that there are strong relationships and hence it seems reasonable that by more sophisticated methods a more precise description of the relationships could be obtained, for instance by nonlinear statistical models or machine learning algorithms. Also, 
the conclusions given rely upon the assumption that the set of interaction networks tested represents well all existing interaction networks with given $N$ and $d$. To question this assumption experiments with smaller subsets of the set  of up to $G=10.000$ different  interaction networks have been carried out. The results obtained for the smaller subsets are consistent with the results for the whole set, which may be seen as an argument in favor of the assumption.  However, if these results hold for $N>14$ needs to be addressed by future work. 

Calculating structure coefficients by Eq.~(\ref{eq:sigma}) is for regular graphs and all conclusions drawn from this method are as well. 
A final remark can be made about how the framework presented may be a staring point for going beyond regular graphs.  
Recently,~\cite{allen17} proposed a method to calculate the benefit--to--cost ratios for any population structure based on coalescence times of random walks, which can be done for any graph structure.  The simulation results show a substantial degree of variety in the ratios and it could be interesting to see if this variety stems from similar relationships to spectral graph measures  as those reported here.  If so, further understanding could be reached about how network properties contribute to differences in the fixation properties of games.

\section{Methods} \label{sec:meth}
\subsection{Coevolutionary games and configurations}

Coevolutionary games of $N$ players can be described by: (i) the payoff matrix, (ii) the interaction network,  and (iii) the strategy of each player,~\cite{szabo07,perc10,rich16,rich17}. We assume the game dynamics be caused by both updating the interaction network and updating the strategies. The former is meant to imply dynamic interaction graphs, while for the latter there is a rule for updating strategies, see e.g.~\cite{allen14,sha12,patt15}, for instance death--birth (DB) or birth--death (BD) updating. As both the interaction network and the strategies may change, the game is called coevolutionary.     For a game with two strategies $C$ and $D$, and pairwise interactions between two players $\mathcal{I}_i$ and $\mathcal{I}_j$ (which consequently are mutual coplayers), there is the $2 \times 2$ payoff matrix      
\begin{equation} 
\bordermatrix{~ & C_j & D_j \cr
                  C_i & a & b \cr
                  D_i & c & d \cr}. \label{eq:payoff}
\end{equation}
We understand a player that uses strategy $C_i$ (or $D_i$) to be a cooperator (or defector). In fact, payoff matrix (\ref{eq:payoff}) may apply to any two strategies (say $A_i$ and $B_i$). However, as variability of the elements of  (\ref{eq:payoff})  is not our topic here, cooperate and defect will suffice.
 
The interaction network of the coevolutionary game specifies who--plays--whom. According to evolutionary graph theory~(\cite{lieb05,allen14,sha12,ohts07,patt15}), each player $\mathcal{I}_i$ belongs to a vertex $i$ of an interaction graph, while an edge connecting vertex $i$ and vertex $j$ shows that the players   $\mathcal{I}_i$ and  $\mathcal{I}_j$ are mutual coplayers. Algebraically,  an interaction  graph  is specified by the adjacency matrix $A_I \in [\text{o},1]^{N \times N}$.  The interaction graph having an edge between vertex 
$i$ and vertex $j$ equals the matrix $A_I$ having an element $a_{ij}=1$, while  $a_{ij}=\text{o}$ shows that the players   $\mathcal{I}_i$ and  $\mathcal{I}_j$ are no coplayers.  In this paper, we consider interaction networks in which each player has the same number of coplayers. Thus, the interaction network can be described by a  random $d$--regular graph,~\cite{hinder15,rich16,rich17}. There are $\mathcal{L}_d(N)$ different instances of $d$--regular graphs on $N$ vertices that can serve as interaction networks for $N$ players with $d$ coplayers, thus offering to study the effect of changing the setting as to who--plays--whom. Furthermore, there are fast and efficient algorithms available for generating such graphs,~\cite{bay10,blitz11}. As $d$--regular graphs with $N$ vertices have $dN/2$ edges, they are only defined for $dN$ even. In the experiments we set $N$ even to have graphs for all $2 \leq d \leq N-1$. 

The strategy vector $\pi(k)$ comprises the $N$ strategies $\pi_i(k)$ that players may select and execute in  a given round $k$ of the game. The strategy vector can 
be understood as a configuration of the game,~\cite{chen13,chen16,rich16,rich17}.~A configuration stands for the spatial arrangement of cooperators and defectors on the interaction graph. For any finite number of players there is a finite number of configurations. For a game with two strategies, there are $\ell=2^N$ configurations, which equals the number of words with length $N$ over a two--letter alphabet $\mathcal{A}$, for instance  the strategies cooperate and defect   $\mathcal{A}=\{C_i,D_i\}$, for which we use the binary code $\mathcal{A}=\{1,0\}$. Consider the example of $N=4$ players. There are $2^4=16$ configurations. For instance, the configuration 
$\pi=(\pi_1\pi_2\pi_3\pi_4)=(0110)$   means that players $\mathcal{I}_2$ and $\mathcal{I}_3$ cooperate, while    $\mathcal{I}_1$ and $\mathcal{I}_4$ defect. To see how many players cooperate, we  count how often the symbol $1$ appears in the string $\pi$, which is known as Hamming weight (also bitcount) $\text{hw}_1(\pi)$. For instance, $\text{hw}_1(0110)=2$.

\subsection{Structure coefficients and local frequencies}
 \begin{table}
\caption{Calculation of the local frequencies $\overline{\omega^1}$, $\overline{\omega^0}$, $\overline{\omega^{10}}$ and 	$\overline{\omega^1 \omega^0}$ for $N=4$, $d=3$ (well--mixed population), $d=2$ (structured population). There are $\ell=2^4=16$ configurations with the number of cooperators $c(\pi)$. We have the number of configurations with the same number of coplayers $\#_{c(\pi)}=\left (\#_0, \#_1, \#_2, \#_3, \#_4 \right)=\left (1, 4, 6, 4, 1 \right)$, which  are the binomials $\#_{c(\pi)}=\frac{4!}{c(\pi)! (4-c(\pi))!}$. 
The local frequencies $\overline{\omega^1}$ and $\overline{\omega^0}$ are characteristic for a configuration $\pi$ and remain invariant over the number of coplayers ($d=2$ and $d=3$) and over different interaction networks $A_I$. For the structured population ($d=2$), however, there is variety in $\overline{\omega^{10}}$ and  $\overline{\omega^1 \omega^0}$ over interaction networks. For instance, consider the configuration $\pi=(1100)$. For $A_I(1)$ and according to Eq. (\ref{eq:loc_fre_play}), we obtain $\omega_1^1=\text{hw}_1\left( (1100) \circ (\text{o}\text{o}11)\right)/2=\omega_2^1=0$ and $\omega_3^1=\text{hw}_1\left( (1100) \circ (11\text{o}\text{o})\right)/2=\omega_4^1=1$, while for  $A_I(2)$, there is $\omega_1^1=\text{hw}_1\left( (1100) \circ (\text{o}11\text{o})\right)/2=\omega_4^1=1/2$ and $\omega_2^1=\text{hw}_1\left( (1100) \circ (1\text{o}\text{o}1)\right)/2=\omega_3^1=1/2$.   Likewise, $\omega_1^0=\omega_2^0=1$ and $\omega_3^0=\omega_4^0=0$  for
$A_I(1)$, but  $\omega_1^0=\omega_2^0=\omega_3^0=\omega_4^0=1/2$ for $A_I(2)$. Thus, $\overline{\omega^1}=\overline{\omega^0}=1/2$ for both   $A_I(1)$ and $A_I(2)$, but $\overline{\omega^1 \omega^0}=0$ for $A_I(1)$  and $\overline{\omega^1 \omega^0}=1/4$  for  $A_I(2)$.
 For variety in  $\overline{\omega^{10}}$, see Fig. \ref{fig:loc_fre}.  } \label{tab:N4}

\begin{tabular}{|cc||c||c|c|c|}
\hline \hspace{1.77cm} &  \hspace{1.44cm} & \hspace{1.75cm} & \hspace{1.68cm} &  \hspace{1.85cm} &\\
& \hspace{1cm}&$A_I(0)$&$A_I(1)$&$A_I(2)$&$A_I(3)$ \\
&&$=$&$=$&$=$&$=$ \\
	&		&	$\left(\begin{smallmatrix} \text{o} & 1 & 1 & 1 \\ 1 & \text{o} & 1 & 1 \\ 1 &1 & \text{o} & 1\\ 1 & 1 & 1 & \text{o} \end{smallmatrix} \right)$	&			 $\left(\begin{smallmatrix} \text{o} & \text{o} & 1 &1 \\ \text{o} &  \text{o} & 1 & 1 \\ 1 & 1 & \text{o} & \text{o}\\ 1 & 1 & \text{o} & \text{o} \end{smallmatrix} \right)$		&		$\left( \begin{smallmatrix}\text{o}  & 1 & 1 & \text{o} \\ 1 &  \text{o} & \text{o} & 1 \\ 1 & \text{o} & \text{o} & 1\\ \text{o} & 1 & 1 & \text{o} \end{smallmatrix} \right)$ &	$\left(\begin{smallmatrix} \text{o} & 1 & \text{o} & 1 \\ 1 &  \text{o} & 1 & \text{o} \\ \text{o} & 1 & \text{o} & 1\\ 1 & \text{o} & 1 & \text{o} \end{smallmatrix} \right)$		\\   
	&&&&& \\
	\hline
&&$d=3$&$d=2$&$d=2$&$d=2$ \\ 
\hline
 \end{tabular}

\begin{tabular}{|c|c|cc||cc||cc|cc|cc|}
\hline
	&&&&&&&&&&& \\
$\pi$	& $c(\pi)$	&$\overline{\omega^1}$	&	$\overline{\omega^0}$	&	 $\overline{\omega^{10}}$ 	&	$\overline{\omega^1 \omega^0}$ &	$\overline{\omega^{10}}$	&	$\overline{\omega^1 \omega^0}$	&	$\overline{\omega^{10}}$	&	$\overline{\omega^1 \omega^0}$&	$\overline{\omega^{10}}$	&	$\overline{\omega^1 \omega^0}$	\\ \hline
0000	&	0&         0	&	    1	&	         0	&	         0	&	         0	&	         0	&	0	&	         0	&	         0	&	         0	\\
1000	& 1& 	    1/4	&	    3/4	&	    1/4	&	    1/6	&	    1/4	&	    1/8	&	         1/4	&	    1/8	&	    1/4	&	    1/8	\\
0100	&1& 	    1/4	&	    3/4	&	    1/4	&	    1/6	&	    1/4	&	    1/8	&	    1/4	&	    1/8	&	    1/4	&	    1/8	\\
1100	&2& 	    1/2	&	    1/2	&	    1/3	&	    2/9	&	    1/2	&	    0	&	    1/4	&	   1/4	&	    1/4	&	    1/4	\\
0010	&1& 	    1/4	&	    3/4	&	    1/4	&	    1/6	&	    1/4	&	    1/8	&	    1/4	&	    1/8	&	    1/4	&	    1/8	\\
1010	&2& 	    1/2	&	    1/2	&	    1/3	&	    2/9	&	    1/4	&	    1/4	&	    1/4	&	    1/4	&	    1/2	&	         0	\\
0110	&2& 	    1/2	&	    1/2	&	    1/3	&	    2/9	&	    1/4	&	    1/4	&	    1/2	&	    0	&	    1/4	&	    1/4	\\
1110	&3& 	    3/4	&	    1/4	&	    1/4	&	    1/6	&	    1/4	&	    1/8	&	    1/4	&	    1/8	&	    1/4	&	    1/8	\\
0001	&1& 	    1/4	&	    3/4	&	    1/4	&	    1/6	&	    1/4	&	    1/8	&	    1/4	&	    1/8	&	    1/4	&	    1/8	\\
1001	&2& 	    1/2	&	    1/2	&	    1/3	&	    2/9	&	    1/4	&	    1/4	&	    1/2	&	         0	&	    1/4	&	    1/4	\\
0101	&2& 	    1/2	&	    1/2	&	    1/3	&	    2/9	&	    1/4	&	    1/4	&	    1/4	&	    1/4	&	    1/2	&	         0	\\
1101	&3& 	    3/4	&	    1/4	&	    1/4	&	    1/6	&	    1/4	&	    1/8	&	    1/4	&	    1/8	&	    1/4	&	    1/8	\\
0011	&2& 	    1/2	&	    1/2	&	    1/3	&	    2/9	&	    1/2	&	         0	&	    1/4	&	  1/4	&	    1/4	&	    1/4	\\
1011	&3& 	    3/4	&	    1/4	&	    1/4	&	    1/6	&	    1/4	&	    1/8	&	    1/4	&	    1/8	&	    1/4	&	    1/8	\\
0111	&3& 	    3/4	&	    1/4	&	    1/4	&	    1/6	&	    1/4	&	    1/8	&	    1/4	&	    1/8	&	    1/4	&	    1/8	\\
1111	&4& 	    1	&	         0	&	         0	&	         0	&	         0	&	         0	&	    0	&	         0	&	         0	&	         0	\\
 \hline

\end{tabular}

\end{table}
Whether or not weak selection favors the fixation of a strategy $C_i$ over $D_i$ from $\mathcal{A}=\{C_i,D_i\}=\{1,0\}$  in a game described by the payoff matrix  (\ref{eq:payoff}) can be expressed by the condition \begin{equation}
\sigma(\pi) (a-d)+(b-c)>0
\end{equation}
where $\sigma(\pi)$ is the structure coefficient of configuration $\pi$,~\cite{nowak10,ohtsnow06,traulsen08,antal09,chen16}. For any $N$, $2 \leq d \leq N-1$, and for a single cooperator (and a single defector) the structure coefficient $\sigma(\pi)$ does not depend on the actual configuration and can be calculated as (\cite{taylor07,lehmann07,tarnita09}):  \begin{equation} \sigma=\frac{(d+1)N-4d}{(d-1)N}. \label{eq:sigm_gen_methods}\end{equation}
In~\cite{chen16} an explicit formula for calculating the structure coefficient has been derived from local frequencies for both DB and BD updating, any configuration and any interaction network described by $d$--regular graphs. We here focus on DB updating as BD updating always opposes the emergence of cooperation.  
For DB updating the structure coefficient is
\begin{equation} 
\sigma(\pi)=\frac{N\left(1+1/d \right) \overline{\omega^1} \cdot \overline{\omega^0}-2\overline{\omega^{10}}-\overline{\omega^1 \omega^0} }{N\left(1-1/d \right) \overline{\omega^1} \cdot \overline{\omega^0}+\overline{\omega^1 \omega^0}}. \label{eq:struc_methods}
\end{equation}
The four local frequencies  in Eq. (\ref{eq:struc_methods}), $\overline{\omega^1}$, $\overline{\omega^0}$, $\overline{\omega^{10}}$ and $\overline{\omega^1 \omega^0}$, can be interpreted probabilistically considering random walks on the interaction graph. According to this interpretation, the local frequency  $\overline{\omega^1}$ (or $\overline{\omega^0}=1-\overline{\omega^1})$ is the probability that starting from a vertex chosen uniformly--at--random and for a given configuration $\pi$, the player at the first step of the walk is a cooperator (or defector). Moreover, $\overline{\omega^{10}}$ is the probability for a cooperator at the first step and a defector at the second, and  $\overline{\omega^1 \omega^0}$  assumes two random walks independent of each other and assigns the probability that the player at the first step is a cooperator of the first walk, but a defector of the second.   As pointed out above Eq. (\ref{eq:struc_methods}) reproduces Eq. (\ref{eq:sigm_gen_methods}) for the $2N$ configurations $\pi$ with $\text{hw}_1(\pi)=1$ and  $\text{hw}_1(\pi)=N-1$

The local frequencies in Eq. (\ref{eq:struc_methods}) are defined  as follows,~\cite{chen16}. For every player $\mathcal{I}_i$, all configurations $\pi$ and a given network of interaction described by $A_I$, there are local frequencies that the coplayers cooperate $\omega_i^1(\pi)$ or defect $\omega_i^0(\pi)$. We define $\omega_i^1(\pi)$ ($\omega_i^0(\pi)$) by the fraction of cooperators (defectors) among the coplayers that a player 
$\mathcal{I}_i$ has in configuration $\pi$. To calculate these quantities, we take the element--wise product  $\pi \circ A_I^{(i)}$ of configuration $\pi$ and $i$--th row of the adjacency matrix  $A_I^{(i)}$,  observe  $1\text{o}=0\text{o}=\text{o}$ and have
\begin{equation} \omega_i^1(\pi)=\frac{\text{hw}_1\left(\pi \circ A_I^{(i)}\right)} {d} \label{eq:loc_fre_play} \end{equation}
where $\text{hw}_1$ is the Hamming weight of a  string.  For  $\omega_i^0(\pi)$ we take the Hamming weight $\text{hw}_0$  counting the number of $0$ symbols. Both Hamming weight discard the $\text{o}$ symbol, which indicates no edge in the adjacency matrix $A_I$.
Consider $N=4$ players with $d=2$ coplayers. The network of interaction is described by a $2$--regular graph on $4$ vertices with $\mathcal{L}_2(4)=3$ instances:
\begin{equation}  	A_I(1)=\left( \begin{smallmatrix}\text{o}  & 1 & 1 & \text{o} \\ 1 &  \text{o} & \text{o} & 1 \\ 1 & \text{o} & \text{o} & 1\\ \text{o} & 1 & 1 & \text{o} \end{smallmatrix} \right) \quad	A_I(1)=\left(\begin{smallmatrix} \text{o} & \text{o} & 1 &1 \\ \text{o} &  \text{o} & 1 & 1 \\ 1 & 1 & \text{o} & \text{o}\\ 1 & 1 & \text{o} & \text{o} \end{smallmatrix} \right)	 \quad A_I(3)=\left(\begin{smallmatrix} \text{o} & 1 & \text{o} & 1 \\ 1 &  \text{o} & 1 & \text{o} \\ \text{o} & 1 & \text{o} & 1\\ 1 & \text{o} & 1 & \text{o} \end{smallmatrix} \right). \label{eq:AI_4}	 \end{equation}
   For the configuration 
$\pi=(1110)$, player $\mathcal{I}_1$ and $A_I(1)$, we get $(1100) \circ (\text{o}11\text{o})=(\text{o}10\text{o})$ which gives $\omega_1^1(\pi)=\omega_1^0(\pi)=1/4$.

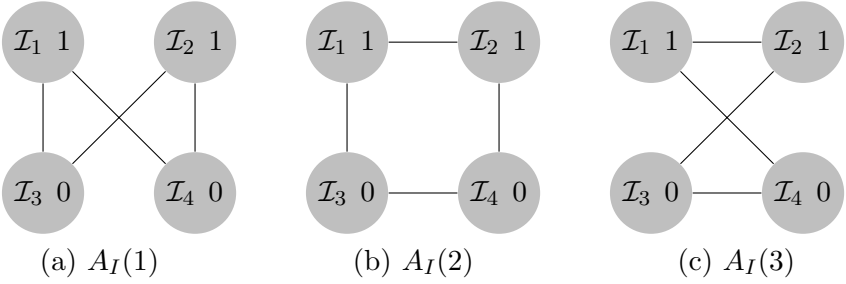
\begin{figure}

\begin{tikzpicture}[shorten >=2pt,->]
  \tikzstyle{vertex}=[circle,fill=black!25,minimum size=12pt,inner sep=3pt]

  \node[vertex] (G_1) at (3,1) {$\mathcal{I}_1 \: \: 1$};
  \node[vertex] (G_2) at (5,1)   {$\mathcal{I}_2 \: \: 1$};
  \node[vertex] (G_3) at (5,-1)  {$\mathcal{I}_4 \: \: 0$};
 \node[vertex] (G_4) at (3,-1)  {$\mathcal{I}_3 \: \: 0$};
  \draw (G_1) -- (G_2) -- (G_3) -- (G_4) -- (G_1) -- cycle;

  \node[vertex] (G_1) at (-1,1) {$\mathcal{I}_1 \: \: 1$};
  \node[vertex] (G_2) at (1,1)   {$\mathcal{I}_2 \: \: 1$};
  \node[vertex] (G_3) at (1,-1)  {$\mathcal{I}_4 \: \: 0$};
 \node[vertex] (G_4) at (-1,-1)  {$\mathcal{I}_3 \: \: 0$};
  \draw (G_1) -- (G_4) -- (G_2) -- (G_3) -- (G_1) -- cycle;

  \node[vertex] (G_1) at (7,1) {$\mathcal{I}_1 \: \: 1$};
  \node[vertex] (G_2) at (9,1)   {$\mathcal{I}_2 \: \: 1$};
  \node[vertex] (G_3) at (9,-1)  {$\mathcal{I}_4 \: \: 0$};
 \node[vertex] (G_4) at (7,-1)  {$\mathcal{I}_3 \: \: 0$};
  \draw (G_1) -- (G_2) -- (G_4) -- (G_3) -- (G_1) -- cycle;

\end{tikzpicture}

\hspace{0.5cm}(a) $A_I(1) $\hspace{2.55cm}(b) $A_I(2)$ \hspace {2.55cm}(c) $A_I(3)$ 
\caption{Calculation of the local frequencies $\overline{\omega^{10}}$  for $N=4$ and $d=2$ with $\mathcal{L}_2(4)=3$ possible networks of interaction. Each of them is a circle. As there are $1/d^2$ paths of length $2$ for each vertex in a $d$--regular graphs, for each player we need to check $4$ path where a cooperator ($\pi_i=1$) is followed by a defector ($\pi_j=0)$, thus the property $(10)$. For all these networks the configuration is $\pi=(1100)$, which means that the players $\mathcal{I}_1$ and  $\mathcal{I}_2$ cooperate ($\pi_1=\pi_2=1$) and the players   $\mathcal{I}_3$ and  $\mathcal{I}_4$ defect ($\pi_3=\pi_4=0$).  For   (a), we find no path of length 2 and property $(10)$ for players  $\mathcal{I}_1$ and $\mathcal{I}_2$ ($\omega_1^{10}=\omega_2^{10}=0$), but 4 paths each for  players  $\mathcal{I}_3$ and $\mathcal{I}_4$: $(\mathcal{I}_3\rightarrow \mathcal{I}_1 \rightarrow \mathcal{I}_3)$, $(\mathcal{I}_3\rightarrow \mathcal{I}_1 \rightarrow \mathcal{I}_4)$, $(\mathcal{I}_3\rightarrow \mathcal{I}_2 \rightarrow \mathcal{I}_3)$, $(\mathcal{I}_3\rightarrow \mathcal{I}_2 \rightarrow \mathcal{I}_4)$  and 
 $(\mathcal{I}_4\rightarrow \mathcal{I}_1 \rightarrow \mathcal{I}_3)$, $(\mathcal{I}_4\rightarrow \mathcal{I}_1 \rightarrow \mathcal{I}_4)$, $(\mathcal{I}_4\rightarrow \mathcal{I}_2 \rightarrow \mathcal{I}_3)$, $(\mathcal{I}_4\rightarrow \mathcal{I}_2 \rightarrow \mathcal{I}_4)$. Thus, $\omega_3^{10}=\omega_4^{10}=1$ and $\overline{\omega^{10}}(1100)=1/2$.  For (b), player $\mathcal{I}_1$ has only 1 path ($\mathcal{I}_1\rightarrow \mathcal{I}_2 \rightarrow \mathcal{I}_4$) out of 4 with the property that a cooperator is followed by a defector $(10)$, from which we get $\omega_1^{10}=1/4$. This is the same for the remaining players. For  player $\mathcal{I}_2$, the path is ($\mathcal{I}_2\rightarrow \mathcal{I}_1 \rightarrow \mathcal{I}_3$), for $\mathcal{I}_3$, it is ($\mathcal{I}_3\rightarrow \mathcal{I}_1 \rightarrow \mathcal{I}_3$), and for $\mathcal{I}_4$, there is ($\mathcal{I}_4\rightarrow \mathcal{I}_2 \rightarrow \mathcal{I}_4$). Thus, $\omega_2^{10}=\omega_3^{10}=\omega_4^{10}=1/4$ and $\overline{\omega^{10}}(1100)=1/4$.
 We may finally check that
for (c), we have the same local frequencies as for (b), and also $\overline{\omega^{10}}(1100)=1/4$. }
\label{fig:loc_fre}
\end{figure}

Similarly, we may define by $\omega_i^{10}(\pi)$ the fraction of paths with length $2$ such that for a configuration $\pi$ and a player  $\mathcal{I}_i$ there is a cooperator followed by a defector on this path of the interaction network $A_I$.   To obtain the frequencies in Eq. (\ref{eq:struc_methods}),
$\omega_i^1(\pi)$,  $\omega_i^0(\pi)$  and  $\omega_i^{10}(\pi)$  are finally averaged over players:
\begin{equation}
\overline{\omega^1} (\pi)=\frac{1}{N} \sum_{i=1}^{N} \omega_i^1(\pi), \label{eq:loc1}
\end{equation}
\begin{equation}
\overline{\omega^0} (\pi)=\frac{1}{N} \sum_{i=1}^{N} \omega_i^0(\pi),
\end{equation}
\begin{equation}
\overline{\omega^1 \omega^0} (\pi)=\frac{1}{N} \sum_{i=1}^{N} \omega_i^1(\pi) \cdot  \omega_i^0(\pi),  
\end{equation}
\begin{equation}
\overline{\omega^{10}} (\pi)=\frac{1}{N} \sum_{i=1}^{N} \omega_i^{10}(\pi). \label{eq:loc4}
\end{equation}
Tab. \ref{tab:N4} and Fig. \ref{fig:loc_fre} exemplify the calculating of the local frequencies (\ref{eq:loc1})--(\ref{eq:loc4}) for $N=4$ and $d=\{2,3\}$. 
 The results show some features that can be observed for all tested $N$, $d$ and $A_I$ and may be assumed to apply generally. The quantity $\overline{\omega^1}$ (and $\overline{\omega^0}=1-\overline{\omega^1}$) is characteristic for a configuration $\pi$ and invariant over networks of interaction specified by regular graphs. 
By contrast,  for the same configuration the frequencies $\overline{\omega^{10}}$ and $\overline{\omega^1 \omega^0}$ may  vary over interaction networks.
For instance, 
for the configuration  $\pi=(1100)$, there is $\overline{\omega^{10}}(1100)=1/4$ and $\overline{\omega^{1}\omega^0}(1100)=1/4$ for $A_I(2)$ and  $A_I(3)$, but $\overline{\omega^{10}}(1100)=1/2$ and $\overline{\omega^{1}\omega^0}(1100)=0$ for $A_I(1)$.

If we calculate the structure coefficients $\sigma(\pi)$ according to Eq. (\ref{eq:struc_methods}) from the local frequencies in Tab.  \ref{tab:N4} for $N=4$ and $d=\{2,3\}$, we obtain that all $\sigma(\pi)$ are the same for all configurations, except
that $\sigma(0000)$ and $\sigma(1111)$ are not defined as the denominator in Eq. (\ref{eq:struc_methods}) is zero and these configurations are absorbing.
We get $\sigma(\pi)=1/2$ for $d=3$ and $\sigma(\pi)=1$ for $d=2$, which reproduces the generic value of $\sigma$ in Eq. (\ref{eq:sigm_gen_methods}).
For $N\geq 5$, the values of $\sigma(\pi)$ may vary from $\sigma$ and give a distribution over interaction networks.  In \ref{sec:results}.~Results, a numerical analysis is given for $N=\{8,10,12,14 \}$ and $2 \leq d \leq N-1$.

\subsection{Spectral graph measures of interaction networks} \label{sec:meas}

As discussed above, the local frequencies  $\overline{\omega^{1}} (\pi)$ and  $\overline{\omega^{0}} (\pi)$ remain invariant over interaction graphs, while  
 $\overline{\omega^{10}} (\pi)$ and $\overline{\omega^1 \omega^0} (\pi)$ may vary. In other words, a varying structure of the interaction graph specifies varying frequencies, which is underlined by the fact that they
 can be calculated directly from configuration $\pi$ and adjacency matrix $A_I$,~\cite{chen16}. Therefore, 
the conjugate configuration $\hat{\pi}$ is defined by $\hat{\pi}=1-\pi$. For instance, for $\pi=(1100)$, there is $\hat{\pi}=1-(1100)=(0011)$. We then obtain \begin{equation}  \overline{\omega^{10}} (\pi)=\frac{1}{dN} \pi^TA_I\hat{\pi}\end{equation}
and
\begin{equation}
\overline{\omega^1 \omega^0} (\pi)=\frac{1}{d^2N} \pi^TA_I^2\hat{\pi} .  
\end{equation}
Thus, if  different $A_I$ may result in different $\overline{\omega^{10}} (\pi)$ and $\overline{\omega^1 \omega^0} (\pi)$ (and subsequently leading to possibly varying structure coefficients $\sigma(\pi)$ as well) for the same configuration $\pi$, there could be quantifiers of the interaction graph that reflect these differences. This approach may open up a spectral analysis of interaction networks of co--evolutionary games. 

 Spectral graph theory defines several quantifiers which capture relationships between the algebraic description of a graph and its structural properties; see for instance~\cite{brou12,li12,spiel09}.  By such a quantification structural differences of the interaction graph are mapped to different values of graph--spectral invariants.
Two quantities considered here are based on the spectra of the adjacency matrix: $eig(A_I)=\left(\alpha_1,\alpha_2,\ldots,\alpha_N \right)$. For connected $d$--regular graphs, the spectra is real and can be sorted: $-d\leq \alpha_1 \leq \alpha_2 \leq \ldots \leq \alpha_N\leq d$.
A first quantity is the (normalized) energy of a graph \begin{equation}\text{ene}= \frac{1}{N}\sum_{i=1}^{N}  |\alpha_i |.\label{eq:ene} \end{equation}
It can be interpreted as the spectral distance between a given graph and an empty graph and as scaling to the degree of difference between graphs. 
A second spectral network measure is the independence number \begin{equation} \text{ind}=\frac{-N \alpha_1}{d-\alpha_1}, \label{eq:ind}\end{equation}
which approximates the size of the largest independent  set of vertices in a graph. An independent set is a set of vertices in a graph such that no two vertices of the set are connected by a edge.

A spectral measure based on the Laplacian matrix $L=dI-A_I$ derives from the eigenvalues $eig(L)=\left(\lambda_1,\lambda_2,\ldots,\lambda_N \right)$, which are non--negative and sorted according to $0=\lambda_1 \leq \lambda_2 \leq\ldots \leq\lambda_N$.
The smallest eigenvalue of $L$ larger than zeros, $\lambda_2$,  is known as (normalized) algebraic connectivity \begin{equation} \text{con}=\frac{\lambda_2}{\lambda_N}, \label{eq:con} \end{equation}  and scales to how well a graph is connected.  Connectivity describes the structural property of a graph that removing  vertices or edges disconnects the graph. The Laplacian eigenvalue $\lambda_2=0$ if the graph is not connected, and $\lambda_2=N$ if the graph is complete (that is fully connected). Larger values of $\lambda_2$  indicate graphs with a  rounder shape, high connectivity and high girth, while for smaller values of $\lambda_2$ the graph is more path--like with low connectivity and low girth. Finally, we consider the
triangular number 
\begin{equation} \text{tri}= \frac{1}{6} \text{tr}\left( A_I^3 \right),\label{eq:tri}\end{equation}
which is calculated from the trace of matrix $A_I^3$ and gives the number of triangles in a graph.

\end{document}